\documentclass[10pt,journal,compsoc]{IEEEtran}
\usepackage{cite}

\usepackage{lipsum,graphicx,multicol}
\usepackage{algorithm, algorithmic}

\usepackage{amsmath,graphicx}
\usepackage{placeins}
\usepackage{amsmath,epsfig}
\usepackage{epstopdf}
\usepackage{amssymb}

\usepackage{dblfloatfix}
\usepackage{amssymb}
\usepackage{amsmath}
\usepackage{comment}
\usepackage{caption}
\usepackage{pstricks}
\usepackage{subfloat}
\usepackage{lipsum}
\usepackage{graphicx}
\usepackage{subcaption}
\usepackage{multirow}

\ifCLASSINFOpdf
\else
\fi
\begin{document}


%\title{Architectures for Detecting Real-time Interleaved Multi-stage Network Attacks Using \\Hidden Markov Model}
\title{Architectures for Detecting Interleaved Multi-stage Network Attacks Using \\Hidden Markov Models}


\author{\IEEEauthorblockN{Tawfeeq Shawly,~\IEEEmembership{Member,~IEEE},
Ali Elghariani,~\IEEEmembership{Member,~IEEE}, \\
Jason Kobes,~\IEEEmembership{Member,~IEEE},  and Arif Ghafoor,~\IEEEmembership{Fellow,~IEEE}}}

%\IEEEauthorblockA{\IEEEauthorrefmark{1}School of Electrical and Computer Engineering,
%Purdue University, West Lafayette, IN 47905 USA},\IEEEauthorrefmark{2}NGC Research Consortium, Northrop Grumman Corporation,McLean, USA}
%\IEEEauthorblockA{\IEEEauthorrefmark{2} Univrsity, City, Libya}

% <-this % stops an unwanted space
%\thanks{Manuscript received December 1, 2012; revised August 26, 2015. 
%Corresponding author: M. Shell (email: http://www.michaelshell.org/contact.html).}}

\markboth{IEEE TRANSACTIONS ON DEPENDABLE AND SECURE COMPUTING,~Vol.~xx, No.~x, Month~20xx,  DOI: 10.1109/TDSC.2019.2948623}%
{Shell \MakeLowercase{\textit{et al.}}: Bare Demo of IEEEtran.cls for IEEE Transactions on DEPENDABLE AND SECURE COMPUTING}




% for Transactions on Magnetics papers, we must declare the abstract and
% index terms PRIOR to the title within the \IEEEtitleabstractindextext
% IEEEtran command as these need to go into the title area created by
\IEEEtitleabstractindextext{%
\begin{abstract}
With the growing amount of cyber threats, the need for development of high-assurance cyber systems is becoming increasingly important. The objective of this paper is to address the challenges of modeling and detecting sophisticated network attacks, such as multiple interleaved attacks. We present the interleaving concept and investigate how interleaving multiple attacks can deceive intrusion detection systems. Using one of the important statistical machine learning (ML) techniques, Hidden Markov Models (HMM), we develop two architectures that take into account the stealth nature of the interleaving attacks, and that can detect and track the progress of these attacks. These architectures deploy a database of HMM templates of known attacks and exhibit varying performance and complexity. For performance evaluation, in the presence of multiple multi-stage attack scenarios, various metrics are proposed which include (1) attack risk probability, (2) detection error rate, and (3) the number of correctly detected stages. Extensive simulation experiments are used to demonstrate the efficacy of the proposed architectures. 
\end{abstract}

\begin{IEEEkeywords}
cyber systems, network security, intrusion detection, Hidden Markov Model, interleaved attacks.
\end{IEEEkeywords}}

\maketitle

\IEEEdisplaynontitleabstractindextext

% \fi
\IEEEpeerreviewmaketitle

\section{Introduction}
\IEEEPARstart{L}arge organizations face a daunting challenge in the provision of security for their cyber-based systems. Modern cyber-based infrastructures typically consist of a large number of interdependent systems and exhibit increasing reliance on the security of such systems.
In the present threat landscape, network attacks have become more advanced, sophisticated and diversified, and the rapid pace of coordinated cyber security crimes has witnessed a massive growth over the past several years. 
For instance, in May 2017, the ``WannaCry'' ransomware attack was detected after it locked up over 200,000 servers in more than 150 countries \cite{Website2017}. A month later, another version of the same attack caused outages of most of the government websites and several companies in Ukraine, and eventually, this attack spread worldwide \cite{Website2017a}. 
With the explosive growth of cyber threats, a dire need exists for the development of high-assurance and resilient cyber-based systems. One of the most important requirements for high-assurance systems is the need for advanced and sophisticated attack detection and prediction systems \cite{Jajodia2017}. 

Security reports reveal that, over time, the type of network intrusions have transformed from the original Trojan horses and viruses into more complex attacks comprised of a myriad of individual attacks. These attacks follow a series of long-term steps and actions referred to as multi-stage attacks, and therefore are hard to predict \cite{luktarhan2012multi, navarro2018systematic}. 
During these attacks, an intruder launches several actions, which may not be performed simultaneously, but are correlated in the sense that each action is part of the execution of previous ones and each multi-stage attack is aimed at a specific target. The detection of multi-stage attacks poses a daunting challenge to the existing threat detection techniques \cite{Jajodia2017}. This challenge is exacerbated when multiple attacks such as these are launched simultaneously in the network, originated by a single or multiple attackers trying to stealth certain attacks among others \cite{Qin2004, navarro2018systematic}.

\subsection{Related Work}
In the past, various approaches have been proposed to address intrusion detection challenges related to multi-stage attacks. These approaches can, in general, be categorized as correlation-based techniques \cite{Valeur2004, Cheng2011, Wang2006} or machine learning (ML) based techniques. Examples of ML techniques include Hidden Markov Models, Bayesian Networks, Clustering and Neural Networks \cite{Fava2008, Du2010, Ramaki2015, Manganiello2011}. Correlation-based techniques, based on cause and effect relationships, mainly utilize attack-graphs when searching the possible stages of the attack \cite{Ning2003, huiying2014real, onwubiko2012situational, zali2012real, alserhani2010mars , Zhang2017}. For example, the work in \cite{huiying2014real} focuses on the causal relationships between attack phases on the basis of security information. Onwubiko et al.\cite{onwubiko2012situational} assesses network security through mining and restoring the attack paths within an attack graph. A causal relations graph presented in \cite{zali2012real}, contains the low-level attack patterns in the form of their prerequisites and consequences. In this approach, during the correlation phase, a new search is performed upon the arrival of a new alert. Several other techniques use similar ideas for analyzing attack scenarios from security alerts \cite{alserhani2010mars , Zhang2017}. However, most of these approaches depend on correlation rules in conjunction with the domain knowledge. Due to increased computational complexity in detecting real time attacks, these techniques pose a limitation. 

%Regular Expressions (RegEx) matching using automata is another approach for IDS that is proved to be efficient for detecting patterns on the packet-level \cite{Meiners2010, Yu2016}. However, the models discussed in \cite{Meiners2010, Yu2016} provide a deterministic pattern matching and do not provide support for the probabilistic inference (decoding) for multi-stage attacks. Instead, Probabilistic Automata (PA) \cite{Dupont2005}, such as PDFA, PNFA, and PDIA \cite{Pfau2010}, can be used to detect this type of probabilistic inference. However, as shown in \cite{Dupont2005}, a HMM can be transformed into an equivalent PNFA with the same number of states, but the converse is not true in general. Further, in general, PA is known to have relatively limited expressivity as compared to HMM \cite{Pfau2010}.%Further, it is shown in \cite{Pfau} that Probabilistic Deterministic Infinite Automata (PDIA).

In the category of ML techniques, HMM is a leading approach for the prediction of multi-stages attacks, \cite{Ourston2003, aarnes2005real, Haslum2008, Farhadi2011, sendi2012real, Zonouz2012, kholidy2014finite, holgado2017real, Ramaki2018}. 
In this approach, stages of an attack are modeled as states of the HMM. The HMM is considered the most suitable detection techniques for such attacks for several reasons \cite{Ourston2003}. First, it has a tractable mathematical formalism in terms of the analysis of input-output relationships, and the generation of transition probability matrices based on a training dataset. Second, because of its specialized capacity to deal with sequential data by exploiting transition probability between states, it can track the progress of a multi-stage attack. Despite the existing research in the use of HMM for intrusion detection in general and multi-stage attacks in particular, none of these approaches considers the problem of interleaving multi-stage attacks and analyzing the impact on the detection performance of such attacks. Moreover, existing approaches address only a single multi-stage attack.

%The existing HMM based detection systems, however, are primarily focused on single multi-stage attacks

\subsection{Contribution}
This research addresses several challenges to the detection of interleaved attacks which include: 
(1) how to model each multi-stage attack in terms of HMM states in the presence of mixed observations, 
(2) how to detect a multi-stage attack when an attacker(s) performs interleaved attacks with the intention to hide an attack (i.e. stealthy attacks),
(3) since no standard public dataset is available that can provide interleaved traffic from simultaneous multiple attacks, the generation of this type of datasets poses a challenge to the research community, 
(4) the design of an efficient architecture that can detect and track the progress of multiple simultaneous attacks, and 
(5) the development of an approach to accurately quantify and measure the detection performance of such an architecture.

To address the above challenges, we propose in this paper two architectures based on HMM formalism. The proposed architectures exhibit varying detection performance and processing complexities.
The architectures can detect the occurrence of multiple organized attacks and provide insights into the dynamics of these attacks such as identifying which attack is progressing and which one is idle at any point of time, how fast or slow each attack is progressing, and in which security state each attack is occurring at any given point in time. Knowledge of this information can assist in designing effective response mechanisms that can mitigate security risks to the network \cite{Albanese2011, Jajodia2017}. The design of the first proposed architecture relies on modifying HMM model parameters to detect multiple multi-stage attacks in the presence of mixed alerts. The design of the second proposed architecture relies on de-interleaving mixed alerts from different attacks prior to the HMM processing subsystem.
We compare the two architectures in terms of their detection performance and design complexity.

The remainder of this paper is organized as follows. In Section 2, we discuss how HMM is used to detect multi-stage attacks. We introduce the system model in Section 3. We present the proposed architectures in Section 4 and evaluation and performance measures in Section 5. We conclude the paper in Section 6.

%%%%%%%%%%%%%%%%%%%%%%%%%%%%%%%%%%%%%%%%%%

\section{The Hidden Markov Model (HMM) For Detecting Multi-stage Attacks}
%\subsection{Background}
An HMM is a double-stochastic process \cite{rabiner1989tutorial} in that it has an underlying stochastic process that is hidden, and can only be observed through another set of stochastic processes that produce the sequence of observed symbols. The observation process corresponds to alerts generated by the IDS. 
Mathematical preliminaries for the discrete HMM in the context of multiple multi-stage attacks are given in Appendix A. 

\subsection{Using HMM to Detect Multi-stage Attacks}
In a multi-stage attack, an intruder launches a series of long-term steps and actions that are sequentially correlated in the sense that each action follows the successful execution of the previous one. In other words, the output of one stage serves as the input to a subsequent stage. 
One example of multi-stage attacks is the DDoS attack in which the attacker starts by scanning the targeted network in order to identify potential vulnerabilities. Subsequently, the attacker tries to break into vulnerable hosts which have been compromised by the attacker. After exploiting these hosts, the attacker installs a software such as a Trojan horse. Eventually, the attacker initiates access to the final target, which could be a server accessible from all the exploited hosts, and subsequently, the DDoS attack is launched \cite{darpa2000}. 

Most detection systems have the capability to detect a single-stage attack or each of the stages of a multi-stage attack independently. However, the detection of multi-stage attacks poses a daunting challenge to the existing intrusion detection techniques due to the lack of an ability to analyze the entire attack activity chain as a whole. This challenge is exacerbated if several of these attacks are launched simultaneously in the network, each attack originated by a single or multiple attackers trying to stealth certain attacks within others. 
The difficulty in detecting interleaved attacks comes from the unrelated observations made of unrelated attacks, observations that conceal the details of the activity chains of multi-stage attacks. 
\begin{figure}
\centering
% 1-Col
\includegraphics[width= 8.90cm, height= 5.5cm]{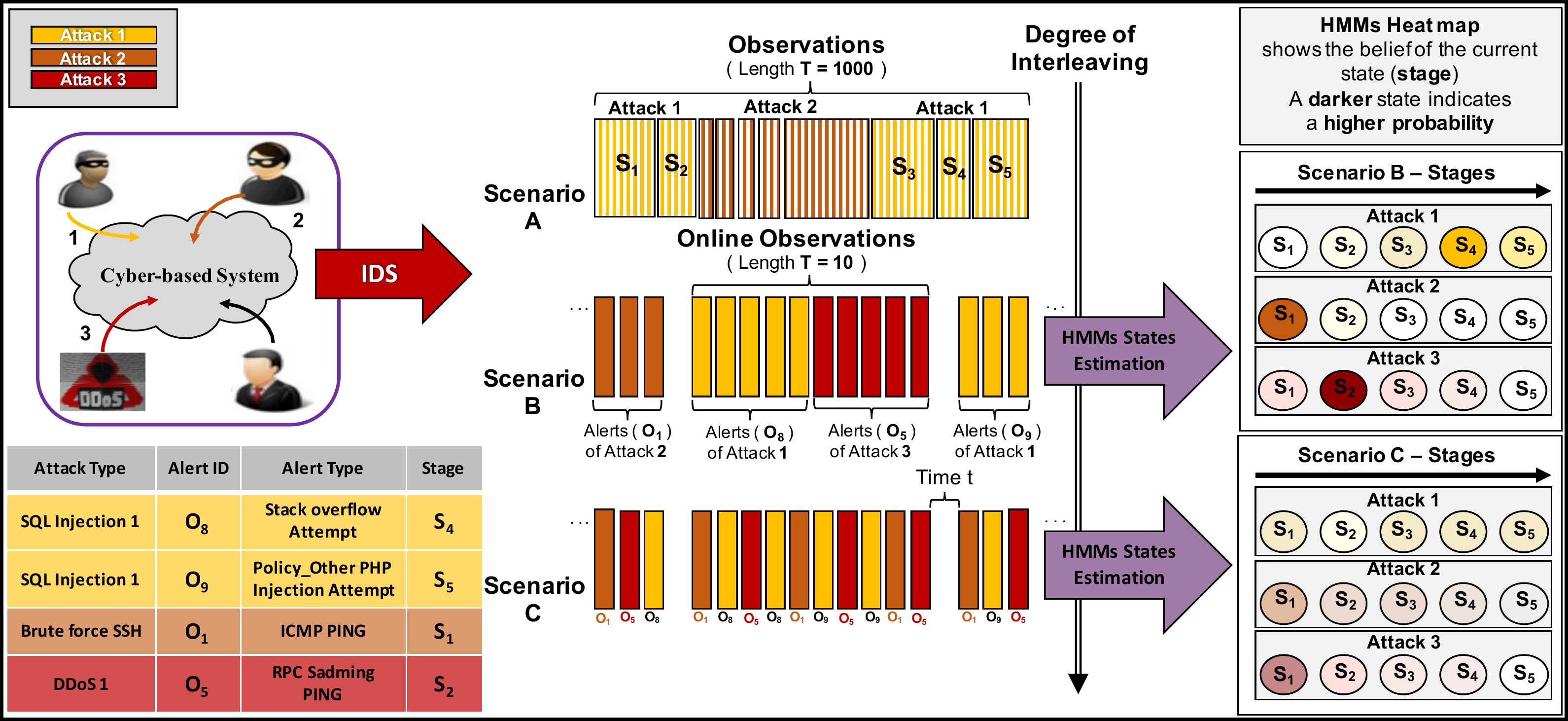}
% 2-Cols
% \includegraphics[width= 8.75cm, height= 5cm]{generalarchitecture2.pdf}
\caption{State Estimation of Multi-stage Attacks with Various Degrees of Interleaving at Time t}
\label{fig:2222}
\end{figure}
	
Fig. \ref{fig:2222} illustrates this challenge by exhibiting three possible scenarios involving the interleaving of three multi-stage attacks that can target a specific or multiple servers. For example, Attack 1, shown in yellow, is an SQL injection attack, wherein Attack 2, in orange, is a Brute force SSH, and Attack 3, in red, is a DDoS attack \cite{darpa2000}.
The table in the lower-left corner of the figure shows the correspondence between the type of attack, Alert ID, Alert type, and stages for some observations of the aforementioned attacks. 
In addition, on the right side of the figure, an HMM heatmap shows the estimation of the belief about the current state assuming there are five stages for each multi-stage attack. A darker color indicates a higher probability and a higher degree of certainty about the current state.

In this example, ICMP PING is a common observation between Attack 1 and Attack 2. Also, assume that the system is in State 4 of Attack 1 and the next alert(s) generated by the IDS is ICMP PING, which is an observation for State 1 in this attack. The ICMP PING observation could be originally generated from Attack 2, thus, in this case, the state estimation can be affected due to the uncertainty regarding the exact current state of the system caused by the unexpected ICMP PING observation(s).

Fig. \ref{fig:2222} also exhibits an example of how the degree of interleaving among the observations of three multi-stage attacks can hypothetically affect the performance of the state estimation over time. For instance, Scenario C in Fig. \ref{fig:2222} has a higher degree of interleaving compared to Scenario B and, consequently, the uncertainty about the current state for each multi-stage attack at time t in Scenario C can be higher than in Scenario B. 
A detailed performance analysis regarding the interleaving is given in Section 5.

In this paper, we use HMM to model and detect possible multi-stage attack scenarios on a targeted cyber-based system. In particular, in order to detect a single multi-stage attack, (say Attack $k$), stages of the attack are modeled as states of the HMM and the observation process corresponds to related alerts generated by the IDS and processed later by a preprocessing component. 
Note, the aforementioned three multi-stage attacks can consist of different types in terms of the order of sequences and the number of stages and corresponding observations. Each attack type ($k$) is modeled using a distinct HMM template $\lambda_k$. In the case of $M$ possible attacks, we have a set of M templates.

Note, the selection of the optimum number of states for each HMM template is a challenge, and no simple theoretical answer exists as to how, in general, this parameter can be selected; this selection depends on the application \cite{rabiner1989tutorial}. In this paper, we model the number of HMM states so that they are similar to the number of stages of the multi-stage attack. The justification for this approach is that the closer the number of states is to the number of stages in the multi-stage attack, the better the details can be provided regarding the progress of the attack; therefore this approach can lead to the development of a more effective response mechanism.

Also note, for each attack type, multiple instances of the same type of attack can be launched by the attacker(s) and consequently, each instance constitutes a distinct attack. The distinction among instances is maintained by a set of observations features such as the source and destination IP addresses and ports. The full description of the attributes and features associated with observations is given in Section 4.

The parameters of the HMM template (i.e. the HMM model $\lambda_k$) for the multi-stage attack $k$ include the number of states of its Markov chain, the number of related IDS observations and aforementioned probability matrices A and B. These parameters are derived offline from a training dataset that contains alerts of a similar multi-stage attack scenario and which can be reestimated and improved online \cite{yin2009}. Specifically, each state is trained based on the observations that belong to the corresponding stage. Subsequently, in the presence of observations related to Attack $k$, HMM estimates the probability of being in each state of the model using Viterbi Algorithm \cite{rabiner1989tutorial}. % More details ? % cite{Online training and Rabiner}.  
However, as mentioned earlier, in the presence of multiple interleaved multi-stage attacks, the performance of the state estimation degrades significantly, especially in a scenario that contains a high degree of interleaving among the observations of multi-stage attacks. %as discussed later in the next sections. %More details will be presented formally in the next sections. 
In the next section, we discuss interleaved multi-stage attacks in detail and present an HMM-based architecture. 

\section{System Model and Architecture}
\begin{figure}
\centering
% 1-Col
\includegraphics[width= 8.90cm, height= 5.5cm]{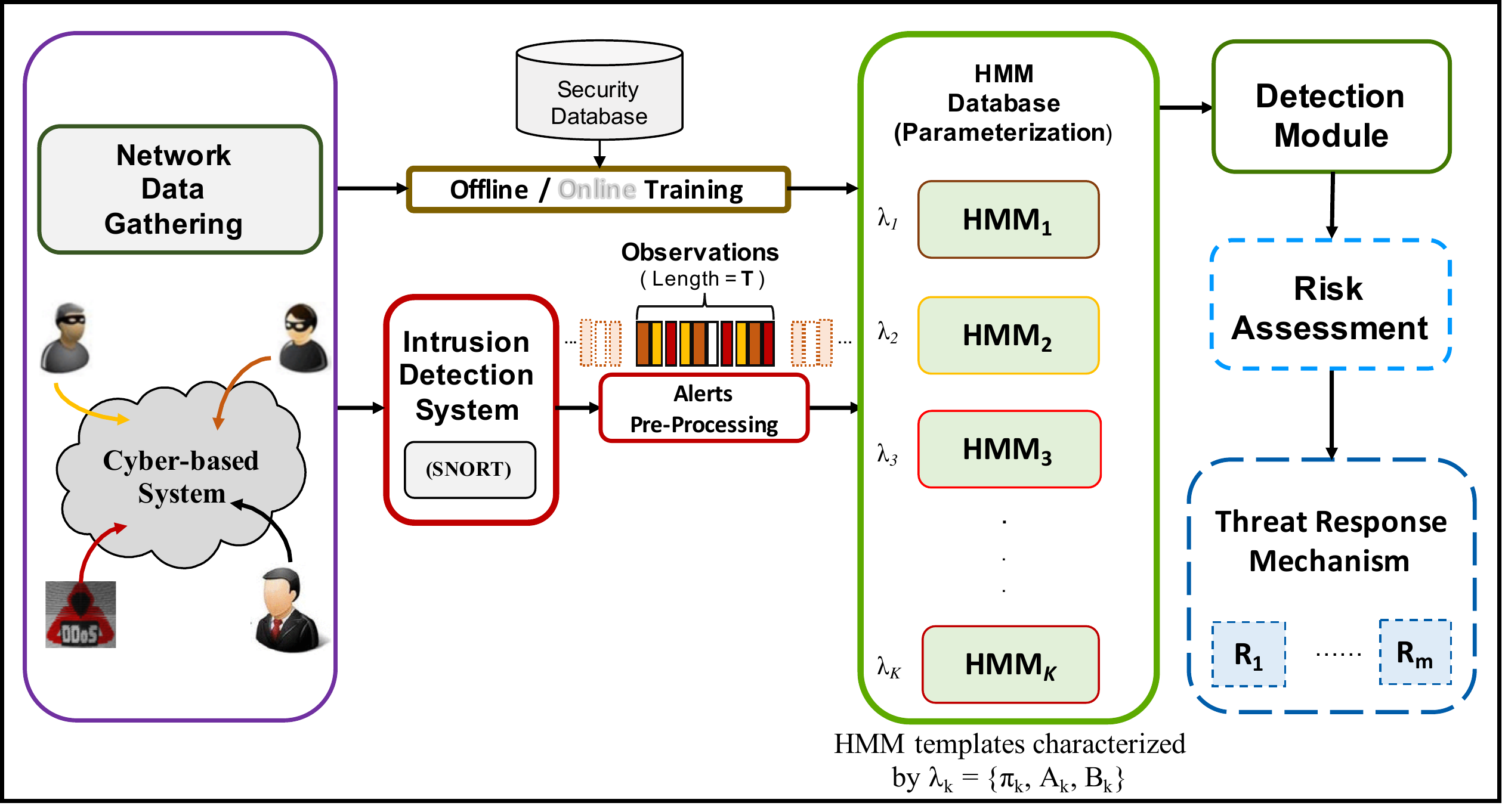}
% 2-Cols
% \includegraphics[width= 8.75cm, height= 5cm]{generalarchitecture2.pdf}
\caption{A Generic Architecture for Multiple Multi-stage Attack Detection using an HMM Database}
\label{fig:1}
\end{figure}

%Snort alert model
% some features are grouped as will 
In order to detect multiple multi-stage attacks, say $K$ attacks, one can generalize the existing single attack architecture by building a database of $K$ HMM templates. In Fig. \ref{fig:1}, we present a generic architecture for the threat detection process that uses such a database. Here, each HMM-based template is designed to detect a specific type of multi-stage attack. The goal of this generic architecture is to detect $K$ multi-stage attacks originated from a single or multiple attackers. Note, each of the $K$ HMM templates is trained to detect an individual multi-stage attack. As mentioned earlier, each template encompasses the HMM structure including all its parameters.

The second major component of this architecture is the Intrusion Detection System (IDS), (e.g., Snort software \cite{snort, barnyard2}), which generates the attack related alerts in real time from the network traffic according to a predefined set of rules. Typically, an IDS generates a stream of alerts which are temporally ordered based on their timestamps. 
The online processing of this stream of alerts can potentially require a large amount of memory \cite{Babcock2002}. 
Such memory requirements can be improved by implementing Snort rules using deterministic or nondeterministic forms of finite 
automata \cite{Becchi2007, Meiners2010, Yu2014, Yu2016}.
The selection of IDS rules can help to reduce the large volume of alerts and false positives by tuning these rules\cite{Tjhai2008}.
The interleaved alerts generated by Snort can belong to one or multiple attacks. These alerts can be preprocessed to generate observations in a suitable format that can be forwarded to the HMM database. Based on the information from Snort, the preprocessing module can assign different severity levels for the incoming alerts. The higher the level is, the more severe the alert which indicates an ongoing multi-stage attack is progressing towards an advanced stage.
In this paper, we assume a window-based technique which is needed in order to buffer a finite number of observations so that these can be  processed by the HMM templates. For this purpose, the incoming alerts to the system are grouped together to form an observation sequence of window size (observation length) ($T$). We assume no overlap occurs between two consecutive windows.
Note, the risk of progressing multi-stage attacks can be assessed in real time by the risk assessment component. Prioritized response actions can be taken based on detected states and the risk of the active attacks \cite{Jajodia2017}. 

%We are using 
%Why it is needed we ned to buffer a finite number of observations to be processed by the hhmm templates. For this peurpose observationtr grouped to and there is no overlap between teo two windows

\subsection{Modeling Interleaved Attacks}
Note, in general, $K$ distinct multi-stage attacks can be launched simultaneously in a network, and their related alerts, generated by the IDS (Snort), are forwarded to the HMM database in the form of a single stream of interleaved alerts. These alerts can be the result of a systematic interleaving of multiple multi-stage attacks initiated by a single attacker or can be generated randomly by different attackers. Note, for each observation length ($T$), we assume $T$ alerts are processed by the HMM templates sub-system. In particular, at any time, it is possible that these $T$ alerts can result from one attack or a mix of at most $K$ attacks. Some possible interleaved attack scenarios that can be orchestrated by an attacker include:
\begin{description}
\item [$\bullet$] An attacker starts and finishes an attack (Attack 2) in the middle of another ongoing attack (Attack 1) as shown in Scenario A in Fig. \ref{fig:2222}.
\item [$\bullet$] Multiple attacks start and finish at different times in the presence of one or multiple ongoing attacks.
\item [$\bullet$] Stages of attack(s) can be embedded at different times of an ongoing attack(s).
\item [$\bullet$] Systematic interleaving among multiple multi-stage attacks can be launched based on interleaving groups of alerts (see; for example, Scenario C in Fig. \ref{fig:2222}).

\end{description}

The existing datasets which feature multi-stage attacks and are publicly available,  do not consider these complex attack scenarios. The DARPA2000 alerts dataset, for instance, contains two distributed denial-of-service (DDoS) multi-stage attacks that happened at different times in which the attacker used multiple distributed compromised hosts to launch DoS attacks on a specific target \cite{darpa2000}. 
To address the challenge of generating the aforementioned interleaved attack scenarios, we generate interleaved alerts by altering timestamps and IP addresses of the DARPA2000 dataset. 

In order to detect the aforementioned attack scenarios, we propose two architectures based on the generic architecture shown in Fig. \ref{fig:1}. The design of the first architecture, Architecture I, is based on modifying the HMM model parameters so that they can deal with the interleaved alerts. 
The design of Architecture II improves attack detection capability by separating alerts from the various attacks prior to routing the alerts to HMM templates sub-system. 
%%%%%%%%Move this part to the end%

%%%%%%%%%%%%%%%%%%%%%%%%%%%%%%%%%%%%%%%%%%%%%%%%%%%%%%%%

\section{Proposed Architectures}
\subsection{Proposed Architecture I}
\begin{figure}%[htp]
 \centering
\includegraphics[width= 8.80cm, height=4.5cm]{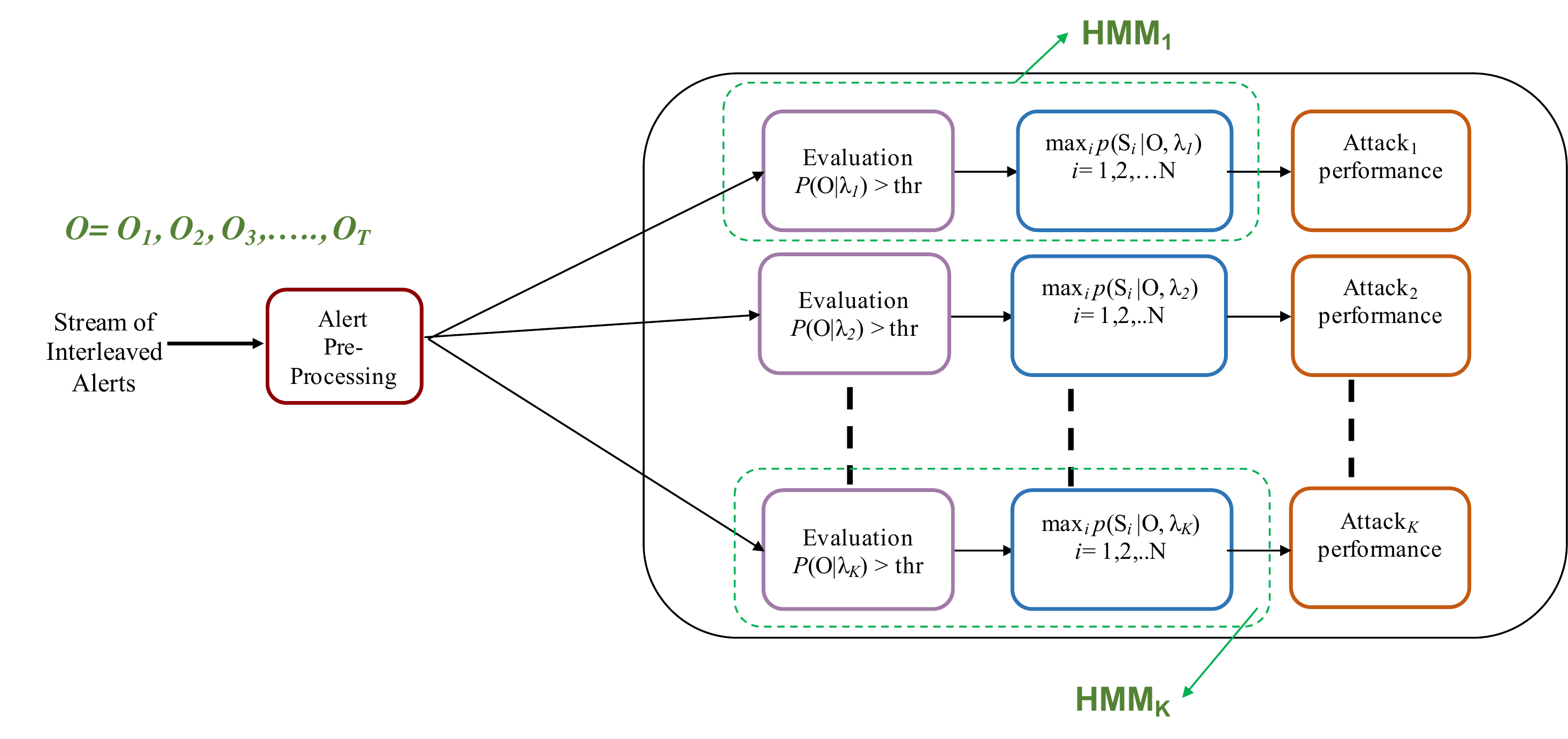}
\caption{ \small Architecture I }
\label{fig:2}
\end{figure}

As mentioned earlier, in order to deal with interleaved traffic alerts from different attacks, we modify the HMM of the generic architecture to accommodate observations from different attacks. The modified architecture is shown in Fig. \ref{fig:2}.
The stream of alerts generated by the IDS contains alerts that belong to one or more concurrent attacks. That is, for each observation length $T$, there are $T$ observations ($o_1, o_2,\dots, o_t,\dots, o_T$) processed by the HMM detection system, as shown in Fig. \ref{fig:2}. Arrival of these alerts represents the interleaved attacks mentioned in Section 3.2. The $\text{HMM}_k$ template is trained for Attack $k$. Therefore, out of $T$ observations, $\text{HMM}_k$ is expected to distinguish and process only those observations that belong to its attack, for which this HMM has been designed. Note, among $T$ observations, there are $L_{k}$ observations (i.e., $\{o_{1_k}, o_{2_k},\dots, o_{L_k} \}$) belonging to Attack $k$, and the $\text{HMM}_k$ consider the remaining $T-L_k$ observations to be unrelated (interfering) alerts. 
We introduce a common state that encompasses all of the unrelated alerts in $\text{HMM}_k$.

Regarding HMM structure, we focus on the alerts generated by the IDS which may consist of True Positive (TP) and False Positive (FP) observations. For each template, we consider State 1 as the most likely state that can be inferred by observing $T-L_k$ unrelated observations using $\text{HMM}_k$. In other words, the occurrence of these interfering (unrelated) observations leads to the lowest security state (State 1) in the $\text{HMM}_k$. To deal with these unrelated observations in parameterizing $\text{HMM}_k$, we introduce a new symbol, $\{o_t \notin V_k \}$, that represents all unrelated observations for Attack $k$. This requires modifying the HMM parameters, (i.e., matrices $A_k$ and $B_k$). This modification can be obtained by considering an observation $o_t$, such that  $\{o_t \notin V_k \}$. Therefore, we add an extra column in the emission probability matrix, $B_k$, to account for this new symbol, as follows:
\begin{align*}
B_k= \begin{bmatrix}
b_{11}&b_{12}&\cdots &b_{1M_k} & \epsilon_1 \\
b_{21}&b_{22}&\cdots &b_{2M_k} & 0\\
\vdots & \vdots & \ddots & \vdots\\
b_{N_k1}&b_{N_k2}&\cdots &b_{N_kM_k} & 0
\end{bmatrix}
\end{align*}
\noindent Note, transition to State 1, in the presence of unrelated observation $o_t$, occurs with probability $\epsilon_1$ which has a very small value (such as $ < 1\times 10^{-6}$) chosen such that $\sum_{j=1}^{M} b_{1j} =1$. Accordingly, almost no change is made to the other observation probabilities in the first row of the emission probability matrix. In addition, setting the probability to zero in the rest of the last column increases the probability that observing $\{o_t \notin V_k \}$ leads to State 1. A second modification is needed for the transition probability matrix ($A_k$) to ensure that whenever $\text{HMM}_k$  observes the $T-L_k$ alerts from attacks other than Attack $k$, transition to State 1 occurs. This transition can be achieved by introducing transition probability ($\epsilon_2$) in the first column of the $A_k$ matrix. Although our initial assumption is based on a left-right model, in this architecture, instead of adding a new state to the model we let all other states return only to State 1 whenever alerts from unrelated attacks occur. An important advantage of modeling unrelated alerts in this way is that the training of each HMM is simplified. Subsequently, by introducing $\epsilon_2$, the matrix $A_k$ becomes: 
\begin{align*}
A_k = \begin{bmatrix}
a_{11}&a_{12}&\cdots &a_{1N_k}  \\
\epsilon_2 &a_{22}&\cdots &a_{2N_k}\\
\vdots & \vdots & \ddots & \vdots\\
\epsilon_2 & 0&\cdots &a_{N_kN_k} 
\end{bmatrix}
\end{align*}

Based on this modification and training of the HMM template ($\lambda_k$), the evaluation module determines whether Attack $k$ is active or not, as shown in Fig. \ref{fig:2}, according to the criteria $Pr(O |\lambda_k) \geq thr$. Note, $thr$ is a threshold used to avoid unnecessary computations of the Viterbi algorithm module in case the attack is not active. The $thr$ value can be chosen within a range of $0$ to $0.5$. However, with the larger the value of $thr$, the HMM template ($\lambda_k$) estimates only the states of the high probability sequences. In this paper, we take a conservative approach in choosing $thr=0$. The evaluation probability can be computed using the forward algorithm \cite{rabiner1989tutorial}. In case the Attack $k$ is active, then $\text{HMM}_k$ ($\lambda_k$) runs the Viterbi algorithm to decode the most probable hidden states that correspond to the given observation sequence $O=\{o_1, o_2,\dots,o_t, \dots,o_T\}$, as follows:
\begin{equation}
\begin{aligned}
&&& x_t = \underset{{1 \leq i \leq N_k}}{\text{max}} \gamma_t(i) \\
&&& \gamma_t(i) = Pr(x_t = s_i |O, \lambda_k)\\
&&& t = 1,\dots, T
\end{aligned}
\label{eq:7}
\end{equation}
\noindent where $\gamma_t(i)$ represents the probability of being in state $s_i$ at time $t$ based on the observation sequence. In Architecture I, each HMM template in Fig. \ref{fig:2} uses the Viterbi algorithm to find the best state sequence, $X = \{x_1, \dots, x_t, \dots,x_T \}$. For a given observation sequence, the Viterbi algorithm finds the highest probability along a single path for every $o_t$ ($t \leq T$) such that:
\begin{equation}
\delta_{t} (i) = \underset{{s_1,\dots, s_{t-1}}}{\text{argmax}} Pr(s_1, \dots,s_t,  o_1,\dots,o_t  | \lambda_k)
\label{eq:8}
\end{equation}
\noindent Using induction, the algorithm determines the rest of the state sequence, as follows:
\begin{equation}
\delta_{t+1} (j) = \underset{{1 \leq i \leq N_k}}{\text{argmax}} \{\delta_{t}(i) a_{ij}(k) \} . b_i(o_{t+1}(k))
\label{eq:9}
\end{equation}

\noindent This computation for a given sequence is repeated by all HMM templates in the Architecture I (Fig. \ref{fig:2}). Table \ref{table1} shows the overall processing of alerts based on Architecture I. 

Architecture I has the limitation of a high probability of high false negatives in states detection, especially when the attacks are highly interleaved as shown in Section 5. The reason for this limitation in such scenarios is that each HMM template processes an observation sequence that contains interfering observations belonging to other attacks. However, the low performance of Architecture I is observed only in special attack scenarios. Nevertheless, Architecture I has a low computation complexity in terms of observations preprocessing (as discussed later).
More details about the detection performance of Architecture I is given in Section 5. 

To achieve better performance, we propose another variation of the generic architecture of Fig. \ref{fig:1}. Termed as Architecture II, this new architecture, is depicted in Fig. \ref{fig:222} and is discussed below.
\begin{table}[htp]
	\vspace{.05in}
\caption{\small Detection process for Architecture I  } % title of Table
\centering % used for centering table
\begin{tabular}{c l } % centered columns (4 columns)
\hline\hline %inserts double horizontal lines
  &   \\
 & \textbf{Input}: interleaved alerts: $O=\{o_1, o_2,\dots,o_t, \dots,o_T\}$,  \\
 & $\pi_k$: $\lambda_k, k=1,2,\dots,K$,\\ 
& \textbf{Output}: $X=\{x_1, x_2, \dots, x_T \}$ \\
1 &  \textbf{While} (${O}$ is not empty)\\
2 & \,\,\,\,\, \textbf{for}  $k=1:K$ \\
3 & \,\,\,\,\,\,\,\, \textbf{if} ($Pr(O | \lambda_k \geq thr$))\\
4 & \,\,\,\,\,\,\,\,\,\,  \textbf{for}  $t=1:T$\\
5&  \,\,\,\,\,\,\,\,\,\,\,\,          Compute $\gamma_t(i)$, \,\,\, $i= 1,2,\dots,N_k$ from equation (\ref{eq:7})\\
6 & \,\,\,\,\,\,\,\,\,\,\,\,         $ x_t = \underset{{1 \leq i \leq N_k}}{\text{max}} \gamma_t(i) $ \\
7 & \,\,\,\,\,\,\,\,\,\, \textbf{endfor}\\
8 & \,\,\,\,\,\,\,\, \textbf{endif}\\
9 & \,\,\,\,\, \textbf{endfor}\\
10 &  \textbf{endWhile}\\
   &  \\ [1ex] % [1ex] adds vertical space
\hline %inserts single line
\end{tabular}
\label{table1} % is used to refer this table in the text
\end{table}

%%%%%%%%%%%%%%%%%%%%%%%%%%%%%%%%%%%%%%%%%%%%%%%%%

\subsection{Proposed Architecture II}

\begin{figure}[htp]
 \centering
\includegraphics[width= 8.80cm, height=5.0cm]{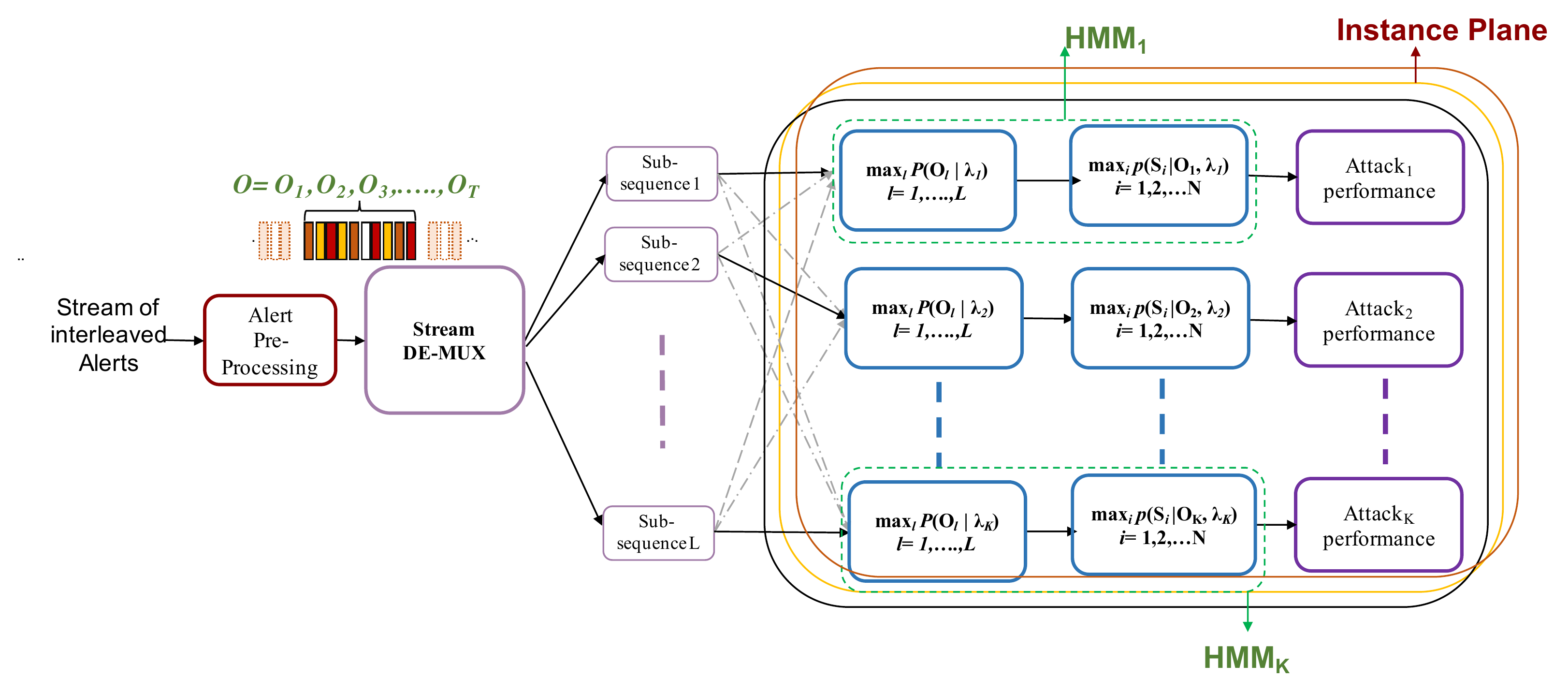}
\caption{ \small Architecture II exhibiting L Instance Planes with Substream Routing }
\label{fig:222}
\end{figure}

Again, we consider $K$ interleaved multi-stage attacks that can be simultaneously launched in the network. The IDS system, based on Snort, generates alerts from these attacks. Every alert is generated with a set of features, which includes alert ID, source/destination IP address, source/destination port number, and timestamp. In Architecture II, we use these features to improve detection efficiency of the HMM templates. In particular, unrelated observations that do not belong to the $k^{th}$ attack are separated and passed to their respective HMMs. Note, the major design philosophy behind Architecture II is to use aforementioned features to preprocess the online network traffic stream and demultiplex it into multiple substreams, as shown in Fig. \ref{fig:222}. Note, each substream is routed to individual instances (planes) where each instance plane contains templates of all attack types. We refer to this preprocessing step as a demultiplexing step. It is an important step as it helps in eliminating the number of interfering observations from other attacks that are not detectable by a particular HMM.

Note, alerts triggered by the same attack scenario are correlated based on some features, (e.g., the source and destination IP addresses). We define alert ($o_i$) as a 7-tuple features (timestamp, ID, srcIP, srcPort, desIP, desPort, priority) according to the IDMEF \cite{curry2002intrusion}. We refer to this tuple as a feature set $F=\{f_1, f_2, f_3, f_4, f_5, f_6, f_7\}$. The timestamp represents date and time of an alert generated by the IDS. ID is the identification of the alert, srcIP and srcPort indicate the source IP address and source port number, respectively. Also, desIP and desPort represent the destination IP address and port number, respectively, and priority indicates the alert's rank \cite{Valeur2004}. Note, these features are used to distinguish between attacks as to whether their instances are from the same or different types of attacks.

A subset, $S$, from the feature set $F$ ($S \subset F$) can be used for the demultiplexing operation. The simplest way in which we can demultiplex interleaved alerts is by grouping the alerts that are triggered by the same multi-stage attack into one subsequence based on their IP addresses relationships, i.e., $S=\{ f_3, f_5\}$. Note, the IP addresses of the alerts, which are triggered by the same attack scenario, are generally related in form a single substream. Consider there are two alerts, $o_i$ and $o_j$. The demultiplexer searches for their addresses to check if they have the same srcIP, or the same desIP. Moreover, it also checks whether destIP of the previous alert is the same as the srcIP of the next one, as in a multi-stage attack scenario, as when the destination node of an earlier alert is the source node of the next alert. Based on the IP address search, the demultiplexer either inserts $o_i$ and $o_j$ in the same subsequence or in different ones. 

 %to reduce computation overhead of, one can use hashing table
%, as illustrated in Fig. \ref{fig:4}
%\begin{figure}
% \centering
%\includegraphics[width= 7.5cm, height=3.5cm]{IP_Demux.pdf}
%\caption{ \small Alert sequence demultiplexed into multiple subsequences based on IP addresses only}
%\label{fig:4}
%\end{figure}

In essence, the demultiplexer module demultiplexes the alert streams into L subsreams ($1 \leq L  \leq K$). The demultiplexing process is based on one or more of the aforementioned distinguishing feature(s) of the incoming alerts and/or based on the correlation of IP addresses. Therefore, from the incoming stream of alerts, $O= \{o_1, o_2, \dots, o_t, \dots, o_T\}$, the demultiplexer generates $L$ substreams each of which belongs to a distinct multi-stage attack. These substreams are a subset of $O$, which are represented as, $O_1= \{o_{1_1}, o_{2_1}, \dots, o_{T_1} \}$, $O_k= \{o_{1_k}, o_{2_k}, \dots, o_{T_k} \}$, and so on till $O_L= \{o_{1_L}, o_{2_L}, \dots, o_{T_L} \}$, where $L \leq K$ and $T_k \leq T$. Note, the larger the feature subset we consider in stream demultiplexing, the more distinct substreams we obtain and, in turn, the more processing is entailed.
Note, within one observation sequence, alerts can belong to $L$ attacks where $1\leq L  \leq K$.  We assume that the demultiplexer does not cause any error in generating substreams.

The demultiplexer module does not distinguish among types of attacks, therefore, it cannot route a substream to its corresponding HMM template. To address this issue in Architecture II, each HMM can have $L$ instances, each of which can process one single substream. Thus, all the $L$ generated substreams pass through each HMM to find which subsequence matches a certain HMM. The computation by each instance is performed based on the posterior probability given in (\ref{eq:7.16}).
\begin{equation}
\begin{aligned}
& O^*_k = \underset{{1 \leq l \leq L}}{\text{max}}
& & Pr(O_l  | \lambda_l) , \,\,\,\,\, L \leq K
\end{aligned}
\label{eq:7.16}
\end{equation}
Note, this probability computation is performed $L \times K$ times. The next stage of the HMM process is to estimate the state probabilities for its corresponding subsequence, $O^*_k$, found from (\ref{eq:7.16}) using the Viterbi decoding algorithm, as follows:
\begin{equation}
\begin{aligned}
&&& x_t = \underset{{1 \leq i \leq N_k}}{\text{max}} \gamma_t(i) \\
&&& \gamma_t(i) = Pr(x_t = s_i |O_k, \lambda_k)\\
&&& t = 1,\dots, T_k\\
\end{aligned}
\label{eq:77}
\end{equation}

Unlike Architecture I, the first stage of every HMM in Architecture II has a maximum of $K$ instances of the forward algorithm and one instance of the Viterbi algorithm. In addition, the HMM in Architecture II deals with subsequences of length $T_k$, where $T_k \leq T$. Table \ref{table11} shows the overall processing of alerts based on Architecture II.
\begin{table}[t]
	\vspace{.05in}
\caption{\small Detection process for Architecture II } % title of Table
\centering % used for centering table
\scalebox{0.96}
{
\begin{tabular}{c l } % centered columns (4 columns)
\hline\hline %inserts double horizontal lines
  &   \\
 & \textbf{Input}: interleaved alerts: $O=\{o_1, o_2,\dots,o_t, \dots,o_T\}$, \\
 &  $\pi_k$: $\lambda_k, k=1,2,\dots,K$,\\ 
& \textbf{Output}: $X=\{x_1, x_2, \dots, x_T \}$ \\
1 &  \textbf{While} (${O}$ is not empty)\\
2 & \,\,\,\,Demultiplex sequence $O$ into $L$ subsequences,  $O_1, O_2, \dots, O_L$, \\
 & \,\,\,\,\,\,\,\,\,\,\,\,\, using features and address correlation \\
3 & \,\,\,\,\, \textbf{for}  $k=1:K$ \\
4 & \,\,\,\,\,\,\,\,\,\, \textbf{for}  $l=1:L$ \\
5 & \,\,\,\,\,\,\,\,\,\,\,\,\, Compute  ($Pr(O_l | \lambda_k $))\\
6 & \,\,\,\,\,\,\,\,\,\, \textbf{endfor}(\textit{l})\\
7 &\,\,\,\,\,\,\,\,\,\,  Find $O^*_k = \underset{{1 \leq l \leq L}}{\text{max}} Pr(O_l  | \lambda_k)$\\
8 & \,\,\,\,\,\,\,\,\,\,\,  \textbf{for}  $t=1:T_k$, $T_k$ is the length of sequence $O^*_k$ \\
9&  \,\,\,\,\,\,\,\,\,\,\,\,\,\,\,  Compute $\gamma_t(i) = Pr(x_t = s_i |O^*_k, \lambda_k)$, \,\, from equation (\ref{eq:77})\\
10  & \,\,\,\,\,\,\,\,\,\,\,\,\,\,\,         $ x_t = \underset{{1 \leq i \leq N_k}}{\text{max}} \gamma_t(i) $ \\
11 & \,\,\,\,\,\,\,\,\,\,\, \textbf{endfor}(\textit{t})\\
12   & \,\,\,\,\, \textbf{endfor}(\textit{k})\\
13 &  \textbf{endWhile}\\
   &  \\ [1ex] % [1ex] adds vertical space
\hline %inserts single line
\end{tabular}
}
\label{table11} % is used to refer this table in the text
\end{table}
\subsection{Complexity Comparison of the Proposed Architectures}
Note, in Architectures I and II in Figs. \ref{fig:2} and \ref{fig:222}, the main modules that contribute to their computational complexity are the alert preprocessing module for assigning alert severity, the stream demultiplexing module, and the HMM parallel branch modules. The first preprocessing module is the same for both architectures. However, the demultiplexing module exists only in Architecture II, which demultiplexes all $T$ alerts based on a subset ($S$) of alert features considered in the demultiplexing operation. The larger the $T$ and $S$ sets are, the more complex computation is performed by this module. In other words, as a result of the demultiplexing operation, Architecture II has $T \times |S|$ additional computational steps as compared with Architecture I.

Next, we consider the HMM database component of the architectures. Note, two algorithms need to be executed in each branch of the HMM database, the forward algorithm (FW) to compute posterior probability for the evaluation purpose and the Viterbi algorithm (VA) to estimate the best state sequence. In Architecture I, each incoming sequence of $T$ alerts is processed by all of the $K$ branches. In other words, $K$ computations of the FW algorithm plus $K$ computations of the Viterbi algorithm are performed. On the other hand, in Architecture II, each HMM template processes, on the average, with a shorter sequence length compared to the sequence lengths in Architecture I. In the first module of each branch, the FW algorithm is executed $L$ times, and in the second module of the branch, the Viterbi algorithm is executed only once. Therefore, in Architecture II, $L \times K$ computations of the FW algorithm plus $L$ computations of the Viterbi algorithm are performed. It is important to note that although Architecture II seems to perform a greater number of computations in the HMM database, the length of sequences processed by both the FW and the Viterbi algorithms is, on the average, shorter than the sequences in Architecture I. The primary shortcoming of Architecture II is the computation overhead needed for the demultiplexing operation. This overhead can be high especially in cases where $T$ has a very large value and a large number of features. 
%%%%%%%%%%%%%%%%%%%%%%%%%%%%%%%%%%%%%%%%%%%%%%%%%%%%%%%%
\begin{figure*}
  \centering
  \begin{tabular}{cccc}
  \begin{subfigure}[b]{0.22\textwidth}
  \includegraphics[width= 4.5cm, height=4.5cm]{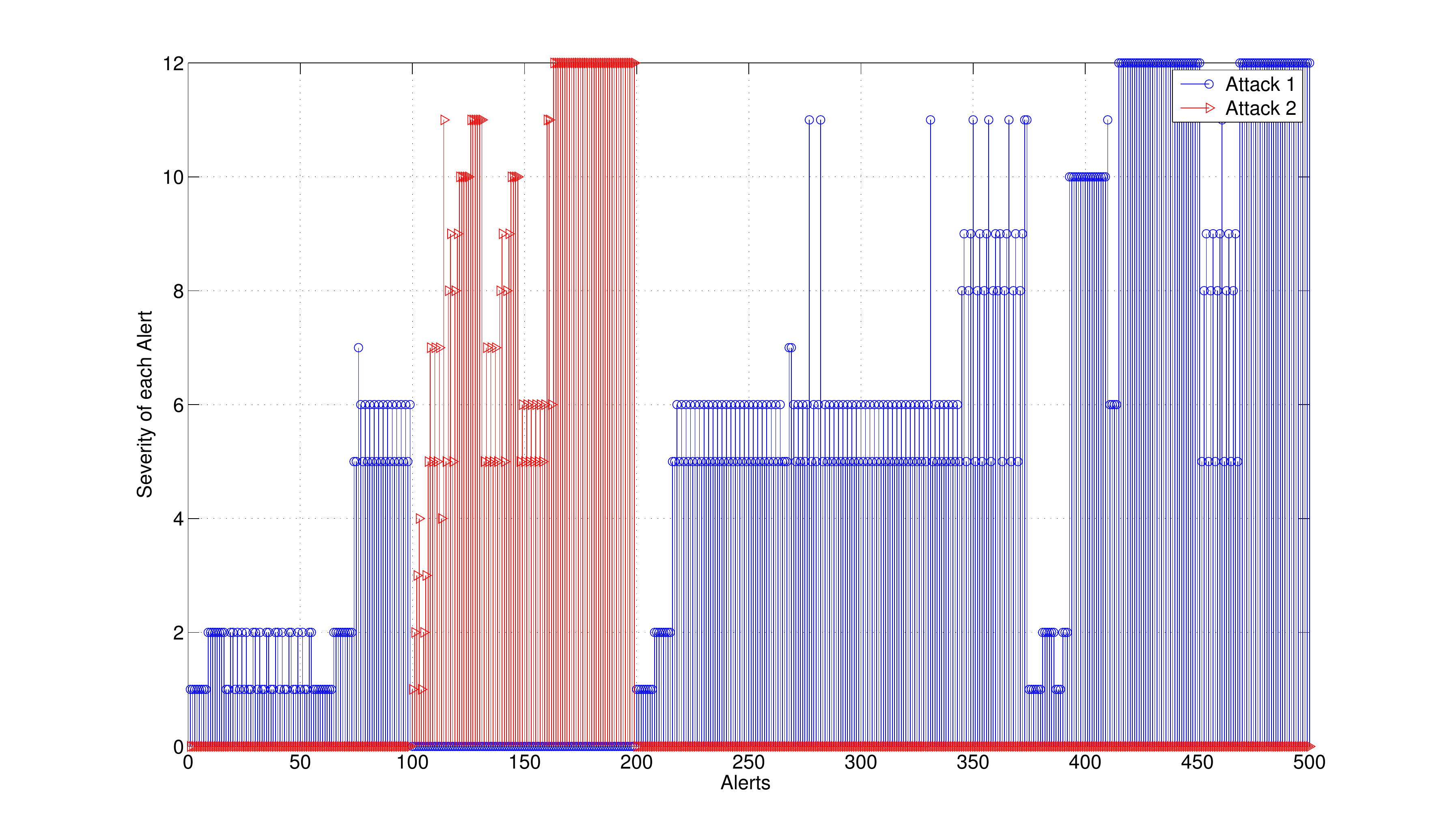}
  \caption{Scenario 1}
  \end{subfigure} &
  \begin{subfigure}[b]{0.22\textwidth}
    \includegraphics[width= 4.5cm, height=4.5cm]{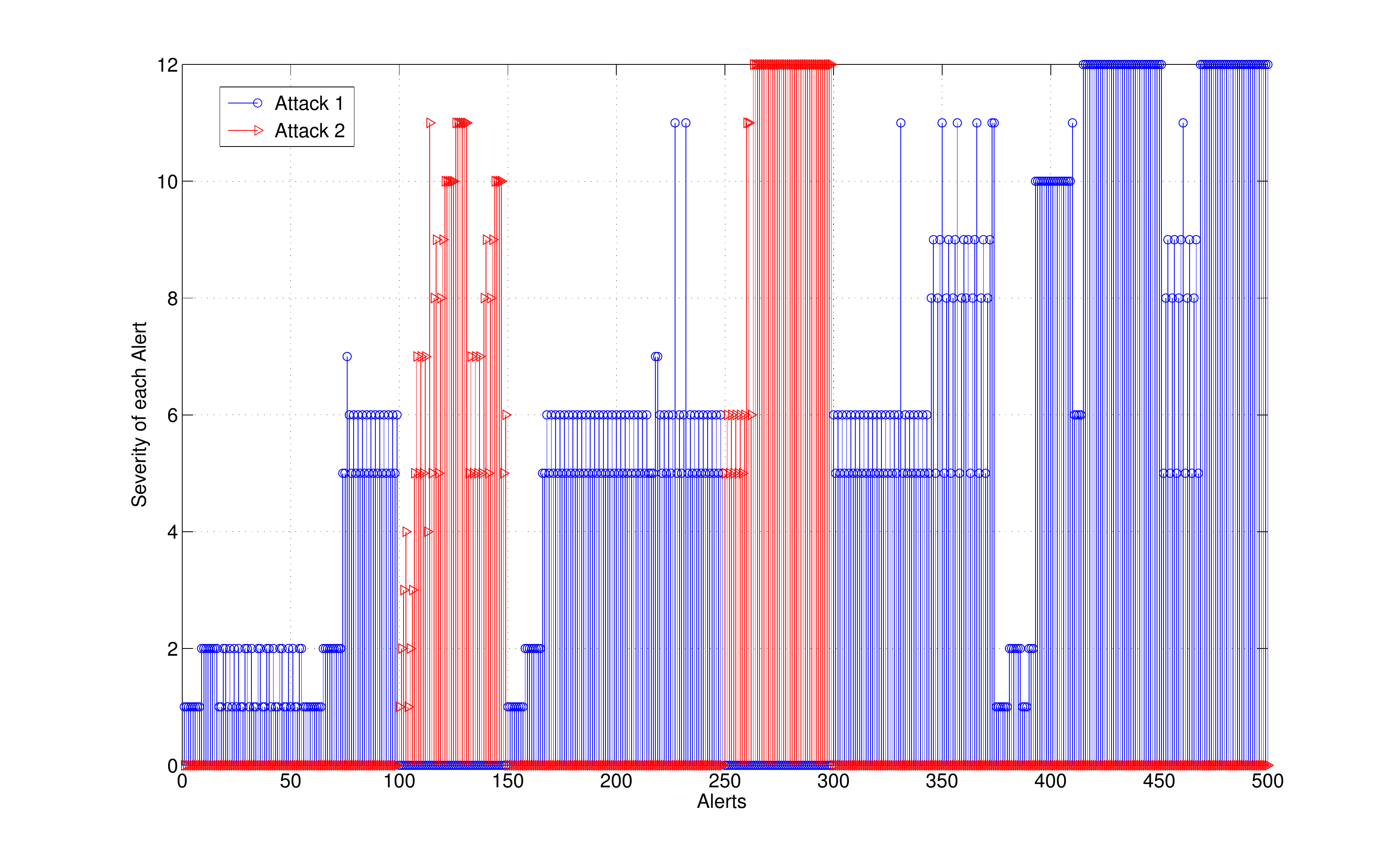}
    \caption{Scenario 2}
  \end{subfigure} &
    \begin{subfigure}[b]{0.22\textwidth}
    \includegraphics[width= 4.5cm, height=4.5cm]{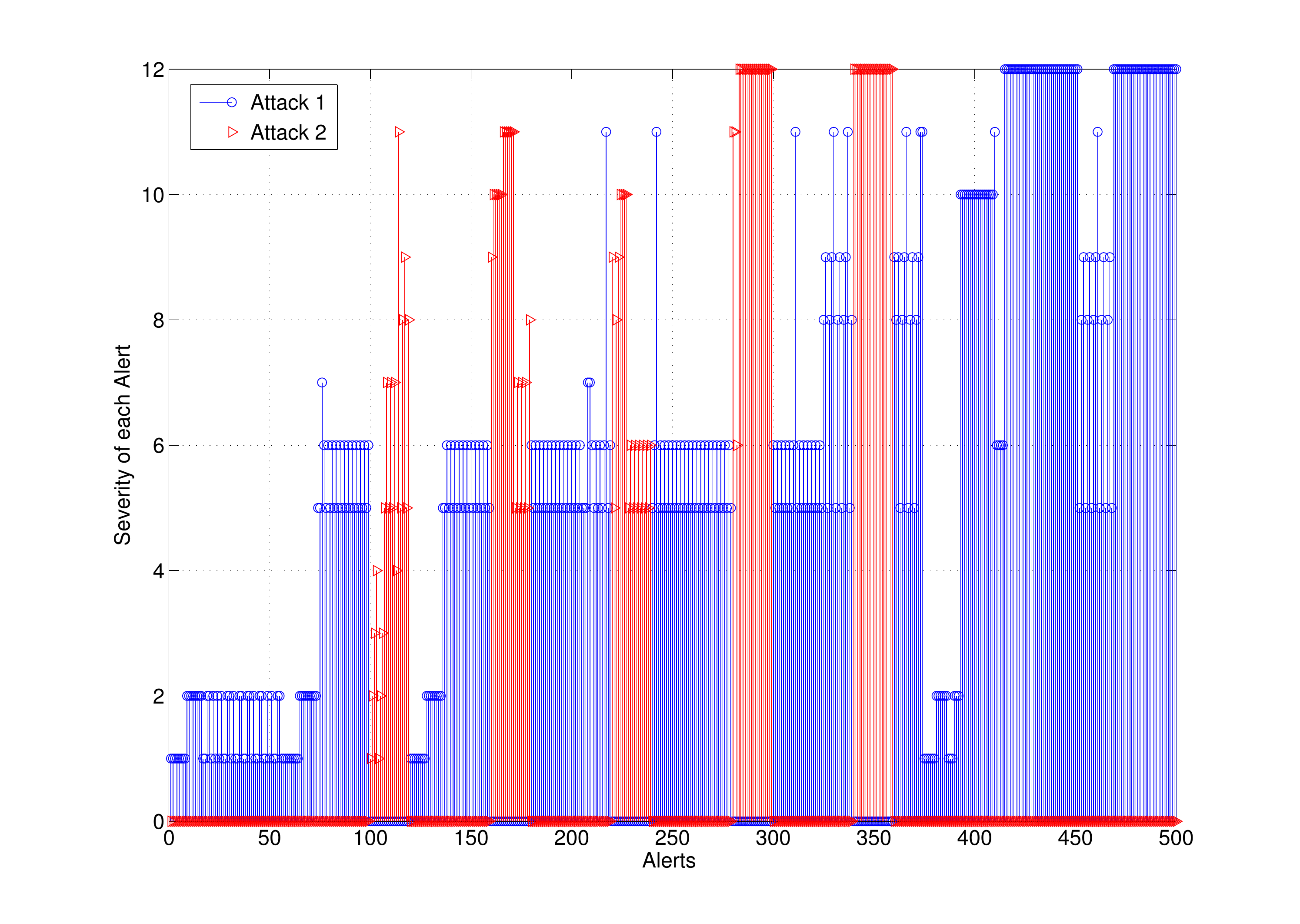}
    \caption{Scenario 3}
  \end{subfigure} &
    \begin{subfigure}[b]{0.22\textwidth}
    \includegraphics[width= 4.5cm, height=4.5cm]{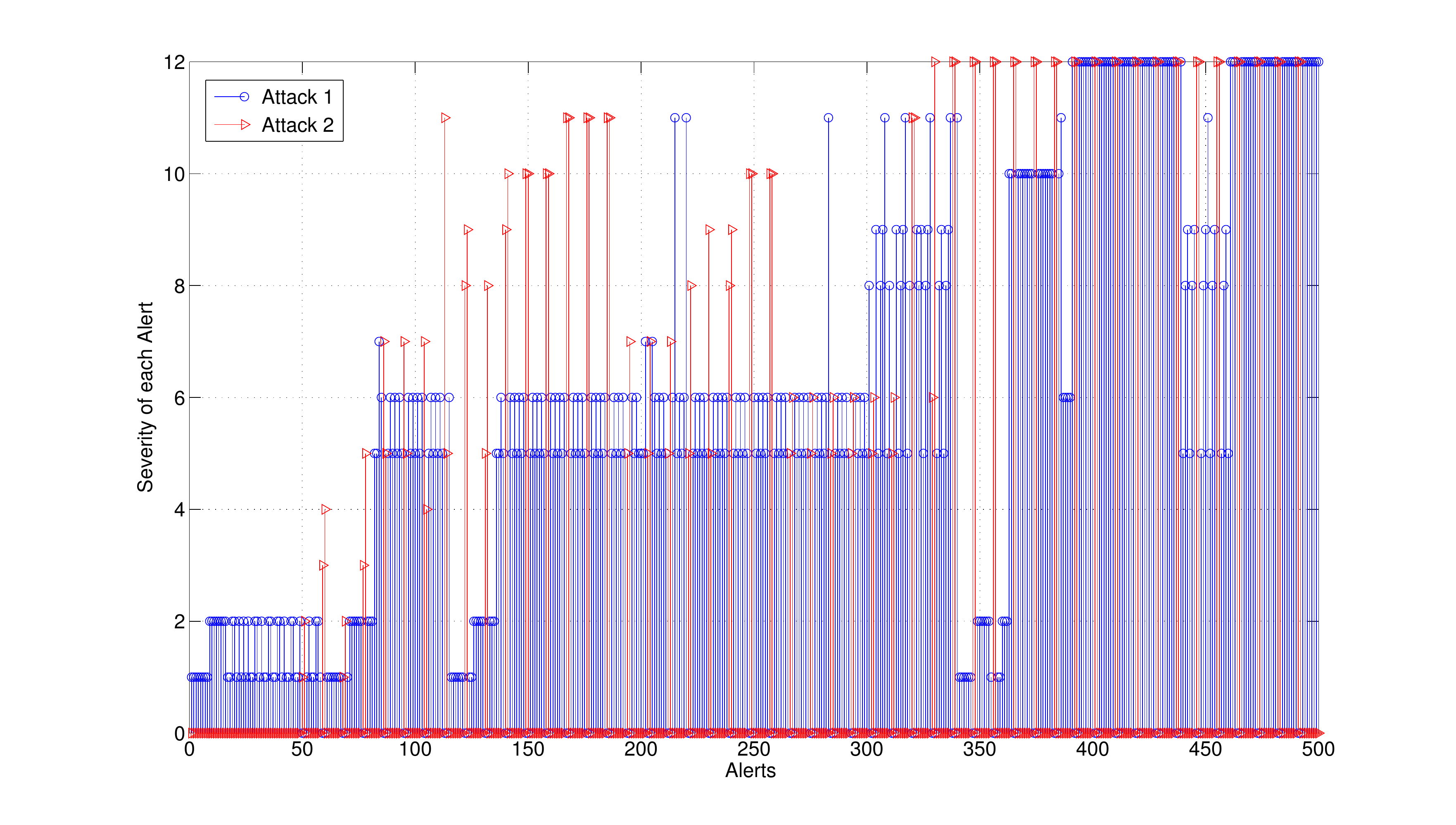}
    \caption{Scenario 4}
  \end{subfigure} \\
  \end{tabular}
  \caption{ \small Interleaved Alerts Scenarios from LLDDOS 1.0 and LLDDOS 2.0.2 Attacks}
  \label{figur11}
\end{figure*}

\section{Performance Evaluation}
In this section, we discuss the experimental results based on the DARPA2000 dataset \cite{darpa2000}, since limited datasets are available for this particular evaluation. The DARPA2000 dataset contains two DDoS multi-stage attacks labeled as LLDDOS 1.0 and LLDDOS 2.0.2. Each of these attacks has five stages: 1) IP sweeping, 2) Sadmind probing, 3) Sadmind exploitation, 4) DDoS software installation, and 5) Launching the DDoS attack. In our experiment, DARPA2000 raw network packets were processed by Snort IDS \cite{snort} to generate alerts. The total number of alerts resulting from this process is 3500 and 2000, for LLDDOS 1.0 and LLDDOS 2.0.2 attacks, respectively.  These alerts are clustered into 12 distinct symbols, therefore, $M_k=12, k=1,2$. The preprocessing module assigns a severity level to these alerts based on their relation to the stages of the multi-stage attacks. In the case of more than one alert which leads to a state, the higher severity level is given to the alert which indicates that the attack is progressing. The training of the two HMMs is conducted based on a five-state model ($N_k=5, k=1,2$), which corresponds to the five stages in each attack. A discussion on the training and the parameterization of each HMM is given in Appendix A.  

%we present the detection performance of the two multi-stage attacks using their respective HMMs when these attacks occur one at a time. Subsequently, 
For the completeness of our evaluation, several scenarios of the two simultaneous attacks are generated with a varying degree of interleaving to test the performance of the proposed architectures. For some cases, we compare the three architectures of Figs. \ref{fig:1}, \ref{fig:2}, and \ref{fig:222}. The reason for using the generic architecture of Fig. \ref{fig:1} for the comparison is that no evaluation has been done in the literature for multiple multistage attacks. For all the results, the x-axis shows the running count of alerts as they are generated by Snort. For evaluation purposes, we propose three performance metrics, in addition to the widely used state probability metric \cite{sendi2012real}. The proposed metrics are: (1) the attack risk probability which provides insight to the speed of the attacks, (2) the detection error rate performance, which measures how much error is generated by an architecture in estimating states, and (3) the number of correctly detected stages. The justification of these metrics is given in the following subsections.

\subsection{Generating Alert Interleaving Scenarios}
Based on the two multi-stage attacks in the DARPA2000 dataset, we alter the timestamp of some of these alerts in both attacks so that we can generate a single sequence of alerts that is composed of a mix of the two attacks without altering the temporal order of the original alerts. In addition, the IP addresses of the hosts attacked by Attack 2 (LLDDOS 2.0.2) are also changed. The reason for this modification is to simulate two simultaneous attacks intruding into a network. Fig. \ref{figur11} shows several scenarios of interleaved alerts for two simultaneous DDoS attacks. Note, the degree of interleaving increases from Scenario 1 to Scenario 4 indicating an increase in the sophistication of actions and complexity of attacks. Since Attack 2 takes a shorter time to compromise the target and launch DDoS, we manipulate the timestamps of Attack 2 so that it spreads across different times of Attack 1. Note also, in this experiment Case Study 1, we only implement systematic interleaving scenarios and no random interleaving scenario is used.
The y-axis in Fig. \ref{figur11} represents the alert severity based on the preprocessing module. Based on these scenarios the performance results of the proposed architectures are given in the following subsections.

%%%%%%%%%%%%%%%%%%%%%%%%%%%%%%%%%%%%%%%%%%%%%%%%%%%%%%%%%%%%%%%%%%%%%%%%%%%%%%%%%%%%%%%%%%%

\subsection{Probability of State Estimation and the Effect of Interleaving}
\begin{figure*} [htp]
  \centering
  \begin{tabular}{cccc}
  \begin{subfigure}[b]{0.22\textwidth}
  \includegraphics[width= 4cm, height=4cm]{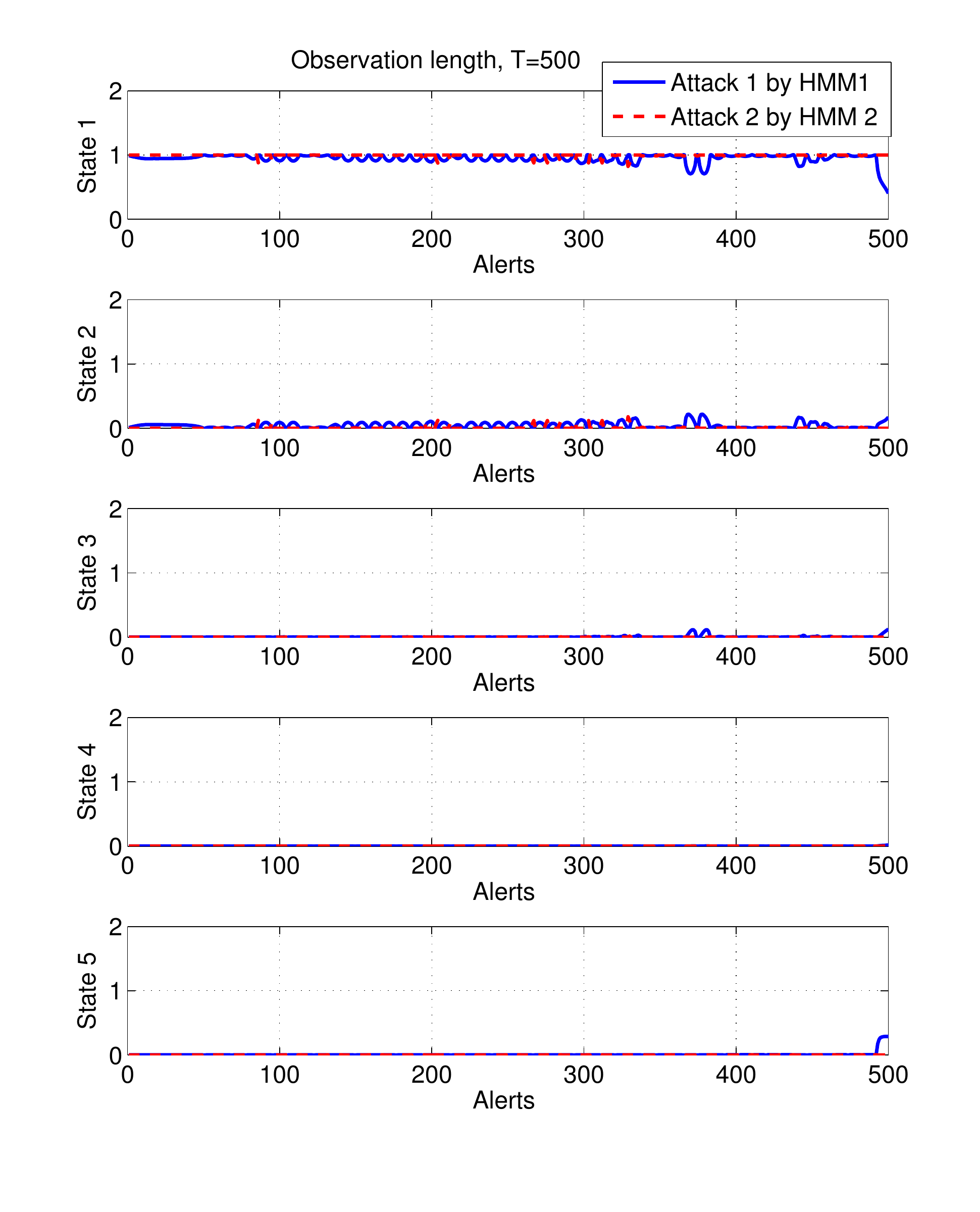}
  \caption{}
  \end{subfigure} &
  \begin{subfigure}[b]{0.22\textwidth}
    \includegraphics[width= 4cm, height=4cm]{Attak1a2_state_prob_alg1_w500_witheps-eps-converted-to.pdf}
    \caption{}
  \end{subfigure} &
    \begin{subfigure}[b]{0.22\textwidth}
    \includegraphics[width= 4cm, height=4cm]{Attak1a2_state_prob_alg1_w500_witheps-eps-converted-to.pdf}
    \caption{}
  \end{subfigure} &
    \begin{subfigure}[b]{0.22\textwidth}
    \includegraphics[width= 4cm, height=4cm]{Attak1a2_state_prob_alg1_w500_witheps-eps-converted-to.pdf}
    \caption{}
  \end{subfigure} \\
  \end{tabular}
  \caption{ \small State Probability of Attacks 1 and 2 Detected by HMM1 and HMM2 Based on Architecture I, T=500}
  \label{figur12}
\end{figure*}

\begin{figure*} [htp]
  \centering
  \begin{tabular}{cccc}
  \begin{subfigure}[b]{0.22\textwidth}
  \includegraphics[width= 4cm, height=4cm]{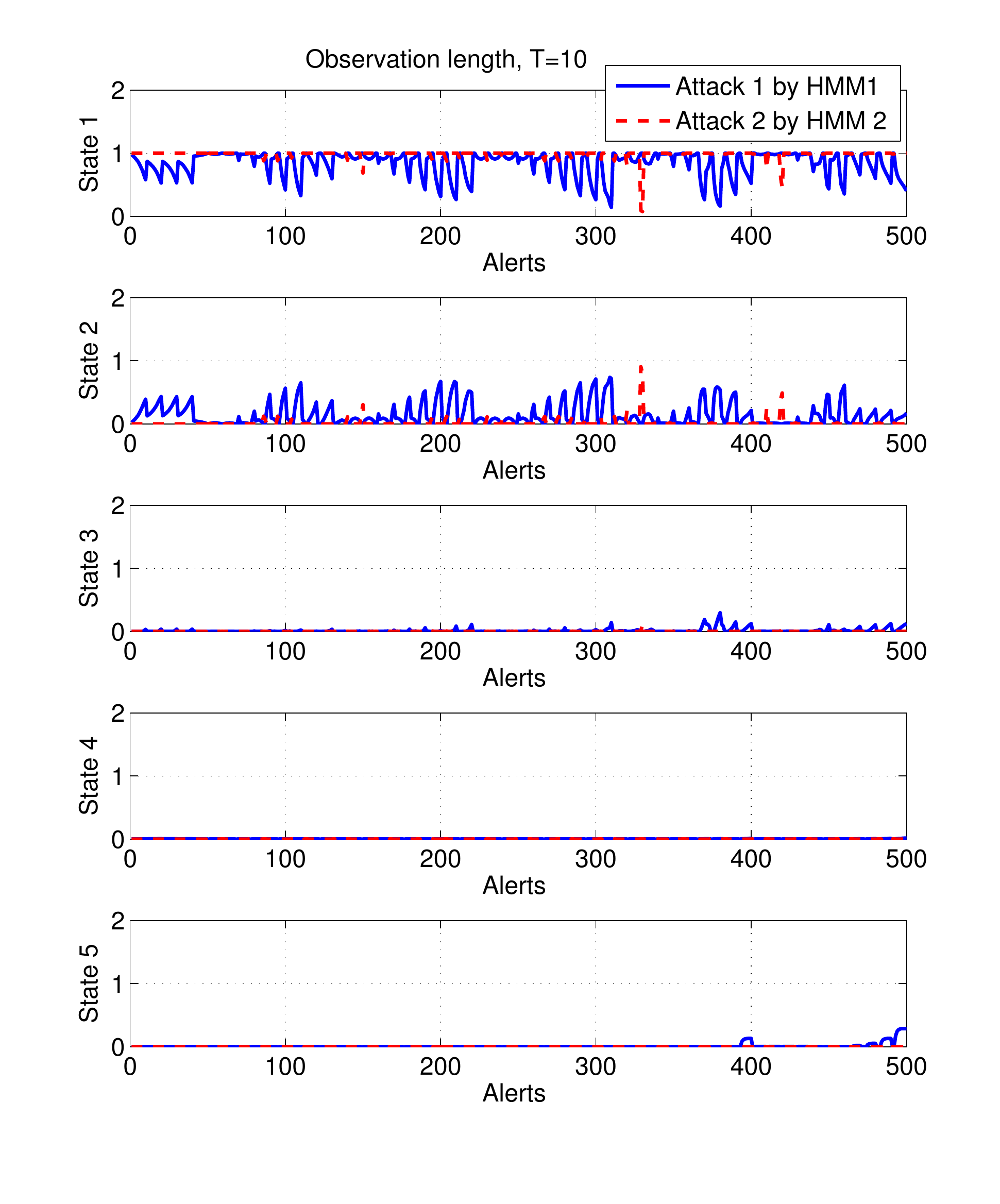}
  \caption{}
  \end{subfigure} &
  \begin{subfigure}[b]{0.22\textwidth}
    \includegraphics[width= 4cm, height=4cm]{Attak1a2_state_prob_alg1_w10_witheps-eps-converted-to.pdf}
    \caption{}
  \end{subfigure} &
    \begin{subfigure}[b]{0.22\textwidth}
    \includegraphics[width= 4cm, height=4cm]{Attak1a2_state_prob_alg1_w10_witheps-eps-converted-to.pdf}
    \caption{}
  \end{subfigure} &
    \begin{subfigure}[b]{0.22\textwidth}
    \includegraphics[width= 4cm, height=4cm]{Attak1a2_state_prob_alg1_w10_witheps-eps-converted-to.pdf}
    \caption{}
  \end{subfigure} \\
  \end{tabular}
  \caption{\small State Probability of Attacks 1 and 2 Detected by HMM1 and HMM2 Based on Architecture I, T=10}
  \label{figur14}
\end{figure*}

In this subsection, we present the state probability, $\gamma_t(i)$, from (\ref{eq:7}) and (\ref{eq:77}) for $i=1, \dots, 5$ with two observation lengths, $T=10$ and $T=500$. Regarding $T=500$,
it can be seen from Figs. \ref{figur12}a and \ref{figur12}b that Architecture I can estimate\footnote{Note: The terms detecting a state  and estimating a state are used interchangeably in this paper} the states of both attacks with a high probability, especially for States 1, 2, 4, and 5. However, as the degree of interleaving increases from Scenario 1 to Scenario 4, Architecture I fails to detect many states. For example, for Scenario 3, States 3 and 4 of Attack 2 are not detected, as depicted in Fig. \ref{figur12}c. For Scenario 4, Architecture I performs very poorly as it fails to detect all the states of both attacks, as depicted in Fig. \ref{figur12}d. For $T=10$, Architecture I shows a small improvement in detecting States 3 and 4 for Scenarios 1 and 3, respectively, as can be seen from Fig. \ref{figur12} and Fig. \ref{figur14}. The reason for the poor performance of Architecture I is that the increasing  degree of interleaving between alerts allows for more interfering alerts to exist within a given sequence. These conditions cause the Viterbi algorithm to incorrectly determine the state probability of the non-interfering alerts. 

%%%%%%%%%
\begin{figure*} [htp]
  \centering
  \begin{tabular}{cccc}
  \begin{subfigure}[b]{0.22\textwidth}
  \includegraphics[width= 4cm, height=4cm]{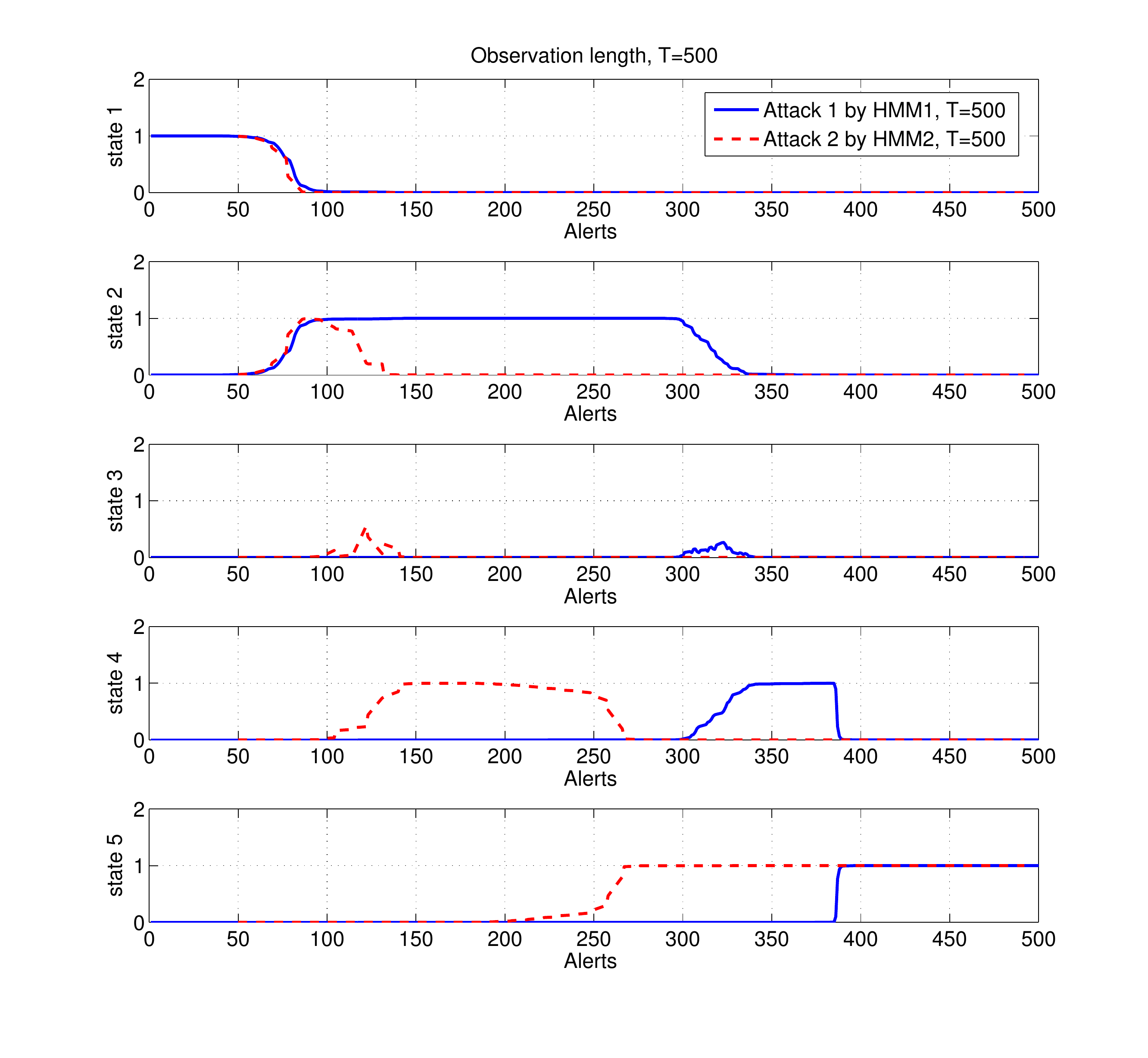}
  \caption{}
  \end{subfigure} &
  \begin{subfigure}[b]{0.22\textwidth}
      \includegraphics[width= 4cm, height=4cm]{Attak1a2_state_prob_alg2_w500-eps-converted-to.pdf}
    \caption{}
  \end{subfigure} &
    \begin{subfigure}[b]{0.22\textwidth}
    \includegraphics[width= 4cm, height=4cm]{Attak1a2_state_prob_alg2_w500-eps-converted-to.pdf}
    \caption{}
  \end{subfigure} &
    \begin{subfigure}[b]{0.22\textwidth}
    \includegraphics[width= 4cm, height=4cm]{Attak1a2_state_prob_alg2_w500-eps-converted-to.pdf}
    \caption{}
  \end{subfigure} \\
  \end{tabular}
  \caption{ \small State Probability of Attacks 1 and 2 Detected by HMM1 and HMM2 Based on Architecture II, T=500}
  \label{figur13}
\end{figure*}

%%%%%%%%%%%
%%%%%%%%%%
\begin{figure*} [htp]
  \centering
  \begin{tabular}{cccc}
  \begin{subfigure}[b]{0.22\textwidth}
  \includegraphics[width= 4cm, height=4cm]{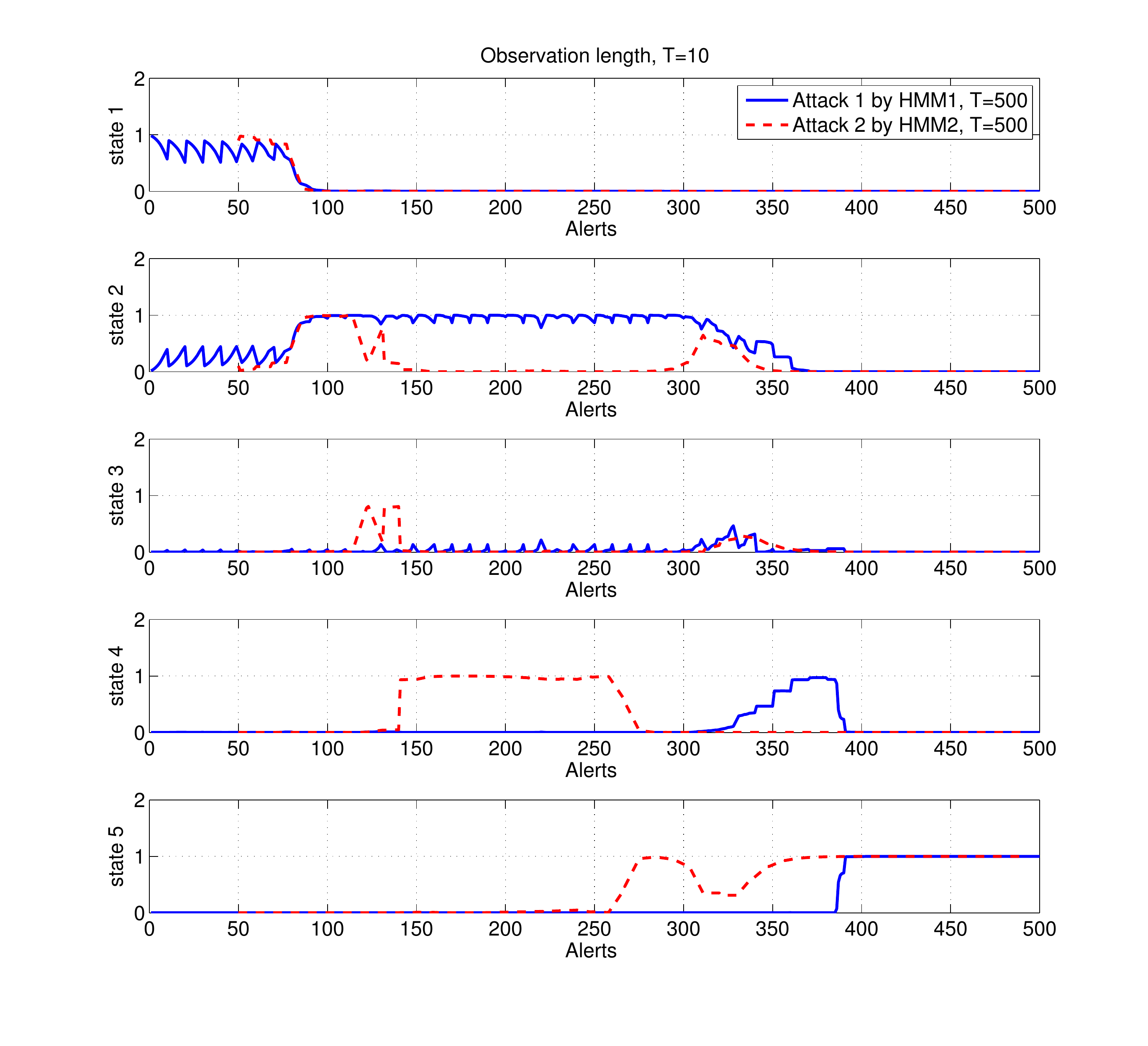}
  \caption{}
  \end{subfigure} &
  \begin{subfigure}[b]{0.22\textwidth}
    \includegraphics[width= 4cm, height=4cm]{Attak1a2_state_prob_alg2_w10-eps-converted-to.pdf}
    \caption{}
  \end{subfigure} &
    \begin{subfigure}[b]{0.22\textwidth}
    \includegraphics[width= 4cm, height=4cm]{Attak1a2_state_prob_alg2_w10-eps-converted-to.pdf}
    \caption{}
  \end{subfigure} &
    \begin{subfigure}[b]{0.22\textwidth}
    \includegraphics[width= 4cm, height=4cm]{Attak1a2_state_prob_alg2_w10-eps-converted-to.pdf}
    \caption{}
  \end{subfigure} \\
  \end{tabular}
 \caption{ \small State Probability of Attacks 1 and 2 Detected by HMM1 and HMM2 Based on Architecture II, T=10}
  \label{figur15}
\end{figure*}

 \begin{figure*} [htp]
  \centering
  \begin{tabular}{cccc}
  \begin{subfigure}[b]{0.22\textwidth}
  \includegraphics[width= 4cm, height=4cm]{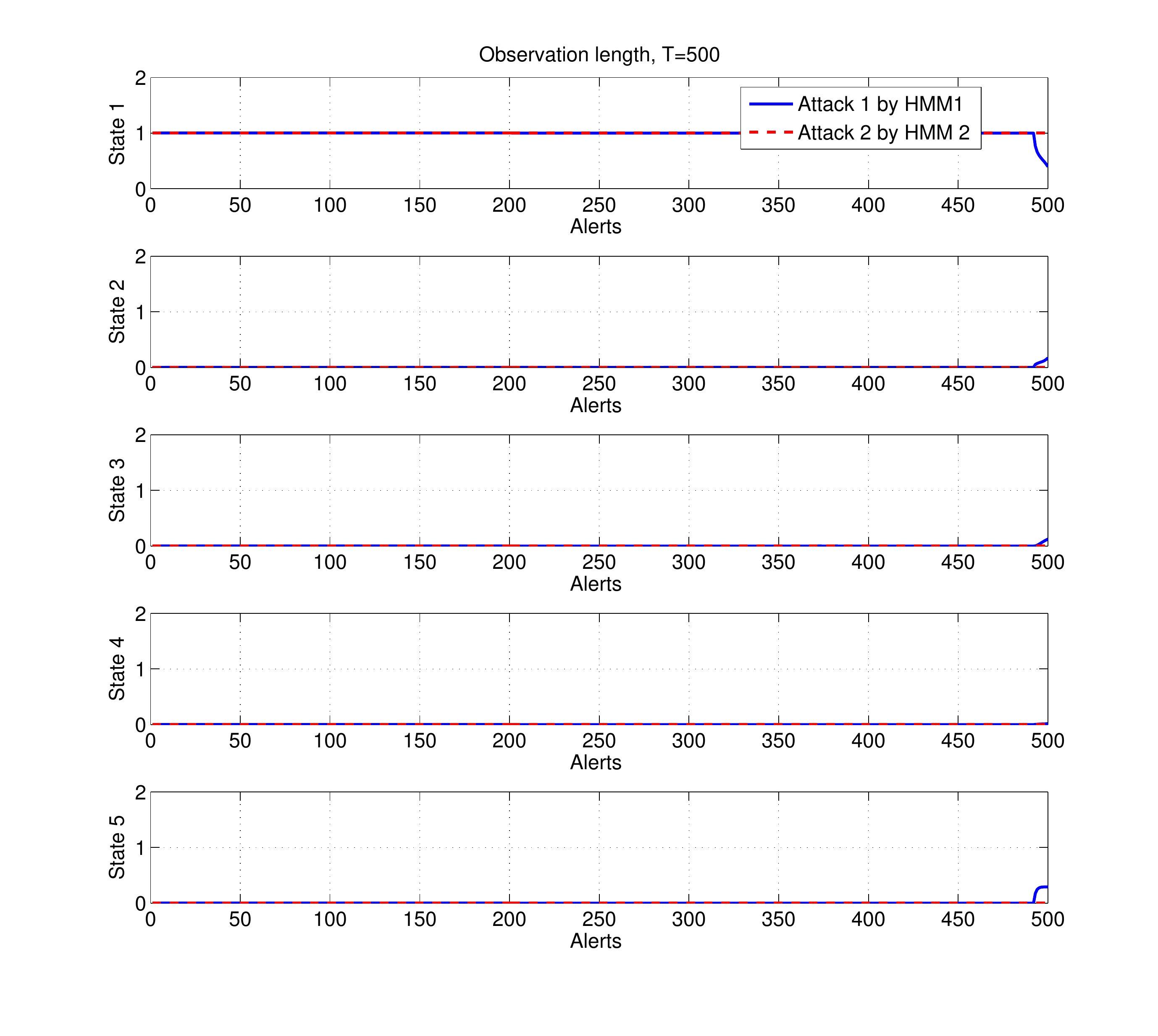}
  \caption{SC2, $\epsilon_2=0$}
  \end{subfigure} &
  \begin{subfigure}[b]{0.22\textwidth}
    \includegraphics[width= 4cm, height=4cm]{Attak1a2_state_prob_alg1_w500_witheps-eps-converted-to.pdf}
    \caption{SC2, $\epsilon_2=0.001$}
  \end{subfigure} &
    \begin{subfigure}[b]{0.22\textwidth}
    \includegraphics[width= 4cm, height=4cm]{Attak1a2_state_prob_alg1_w500-eps-converted-to.pdf}
    \caption{SC3, $\epsilon_2=0$}
  \end{subfigure} &
    \begin{subfigure}[b]{0.22\textwidth}
    \includegraphics[width= 4cm, height=4cm]{Attak1a2_state_prob_alg1_w500_witheps-eps-converted-to.pdf}
    \caption{SC3, $\epsilon_2=0.001$}
  \end{subfigure} \\
  \end{tabular}
  \caption{ \small Effect of $\epsilon_2$ on State Probability Based on Architecture I}
  \label{figur16}
\end{figure*}
Architecture II, on the other hand, performs better as compared to Architecture I in terms of estimating correct states of all incoming alerts for both $T=10$ and $T=500$. This performance improvement, even for higher degrees of interleaving, can be observed from Figs. \ref{figur12}c, \ref{figur12}d, \ref{figur13}c, \ref{figur13}d, \ref{figur14}c, \ref{figur14}d, \ref{figur15}c, and \ref{figur15}d. The reason behind this performance improvement for Architecture II is the presence of the demultiplexing module that helps each HMM to process only relevant attack alerts. 
Note, for both architectures the state probability for State 3 of Attack 1 is very low for both values of $T$ because Snort does not produce enough alerts for this stage. 

We observe discontinuity in the state probability plot of Architecture I in Figs. \ref{figur12} and \ref{figur14}, a condition that results when both of the HMMs return to State 1 whenever interfering alerts exist from other attacks. However, in Architecture II, as the alerts from different attacks are demultiplexed prior to their processing by the HMMs, the states of the HMMs are not interrupted. Fig. \ref{figur16} shows the importance of considering $\epsilon_2$ in designing the HMM used by Architecture I. In this experiment, we choose $\epsilon_2=0.001$, which is a small value that does not significantly affect the transition probability matrices, $A_1$ and $A_2$, obtained from training. Note, $\epsilon_2=0$ represents the case of generic architecture, for which returning to State 1 is not allowed when HMM1 receives alerts belonging to Attack 2 or when HMM2 receives alerts belonging to Attack 1. Setting $\epsilon_2=0$ reduces the accuracy of state detection for the two attacks, as can be seen in Fig. \ref{figur16}. For example, for interleaving Scenario 2, Fig. \ref{figur16}a provides no estimate for state probability of States 2 and 4 for the first 200 alerts when $\epsilon_2=0$, as compared to Fig. \ref{figur16}b when $\epsilon_2=0.001$. A similar observation can be made by comparing Figs. \ref{figur16}c and \ref{figur16}d for the first 350 alerts of State 2.

In summary, Figs. \ref{figur12}, \ref{figur13}, \ref{figur14}, and \ref{figur15} show that no significant change occurs in the state detection performance of each architecture as the observation length changes from $10$ to $500$ alerts. Moreover, Architecture II is more robust in terms of having a better state probability estimation metric at a higher degree of interleaving as compared to Architecture I.
%%%%%%%%%%%%%%%%%%%%%%%%%%%%%%%%%%%%%%%%%%%%%%%%%%%%%%%%%%%%%%%%%%%%%%%%%%%%%%%%%%%%%%%%%%%

\subsection{Attack Risk Probability}
We define the first proposed performance metric as the attack risk probability, which is the probability of how far an attack is from compromising the target, i.e., reaching the final state. The calculation of this attack probability depends on the estimated state probability ($\gamma_t(i)$) averaged over the total number of states. Its value gets updated at every observation length in a non-decreasing manner, as shown below in (\ref{eq:20}):
\begin{equation}
\begin{aligned}
& Pr_{attack_k}(t) =  \frac{{\sum_{i=1}^{N_k}  \gamma_t(i)  s_i}}{N_k}\\
& t=1,\dots, T \,\,\,  i = 1,\dots, N_k, \,\,\,  k = 1,\dots, 2\\
\end{aligned}
\label{eq:20}
\end{equation}
This performance measure can help in  tracking the progress of each attack, especially when there are multiple organized attacks. It can be noted that, the rate at which the attack risk probability changes with respect to alerts gives an indication of how fast or slow an attack is progressing. Consequently, this measure can help in prioritizing response actions for each ongoing attack. 

\begin{figure*} [htp]
  \centering
  \begin{tabular}{cccc}
  \begin{subfigure}[b]{0.22\textwidth}
  \includegraphics[width= 4cm, height=4cm]{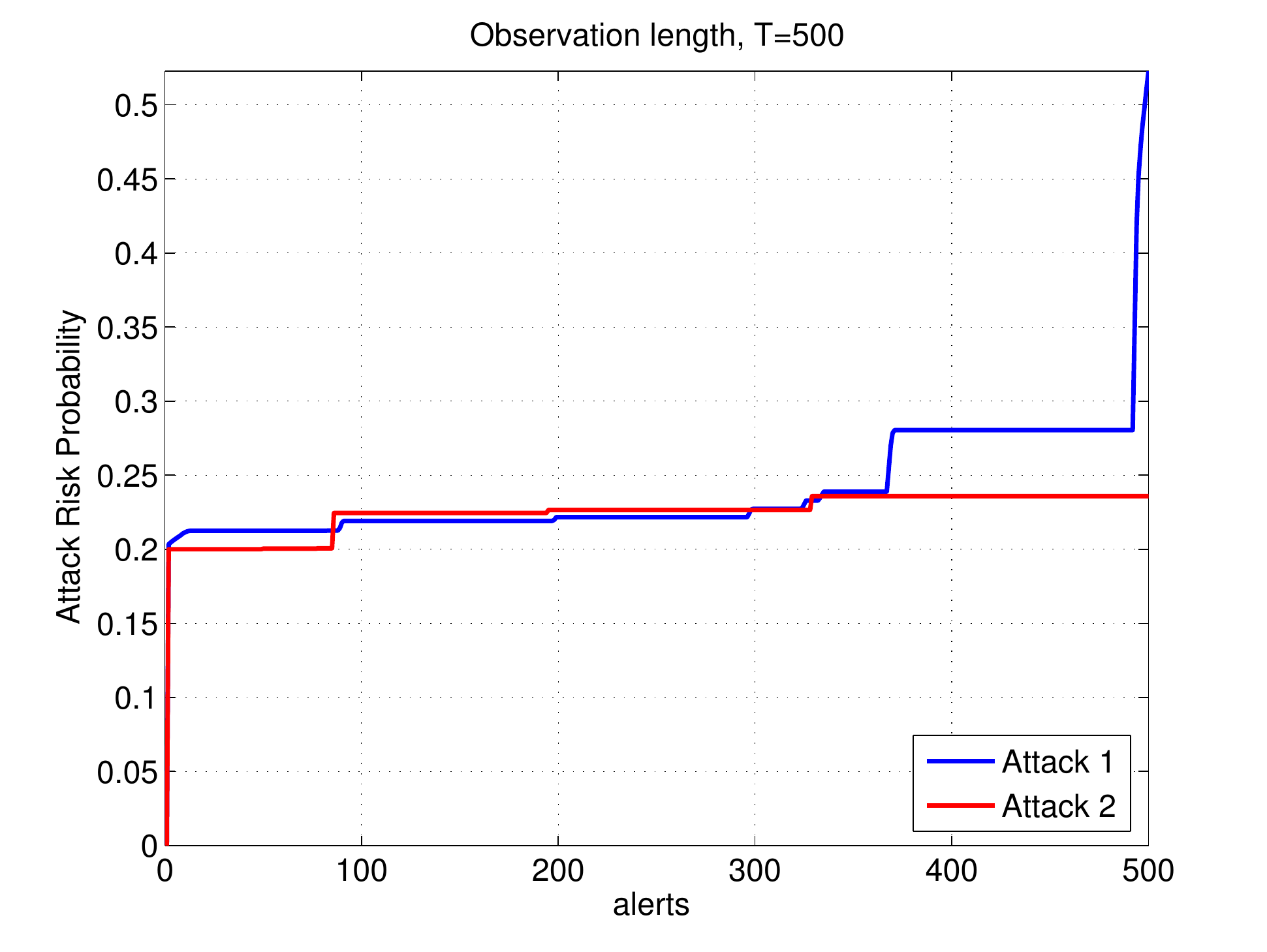}
  \caption{Scenario 3}
  \end{subfigure} &
  \begin{subfigure}[b]{0.22\textwidth}
    \includegraphics[width= 4cm, height=4cm]{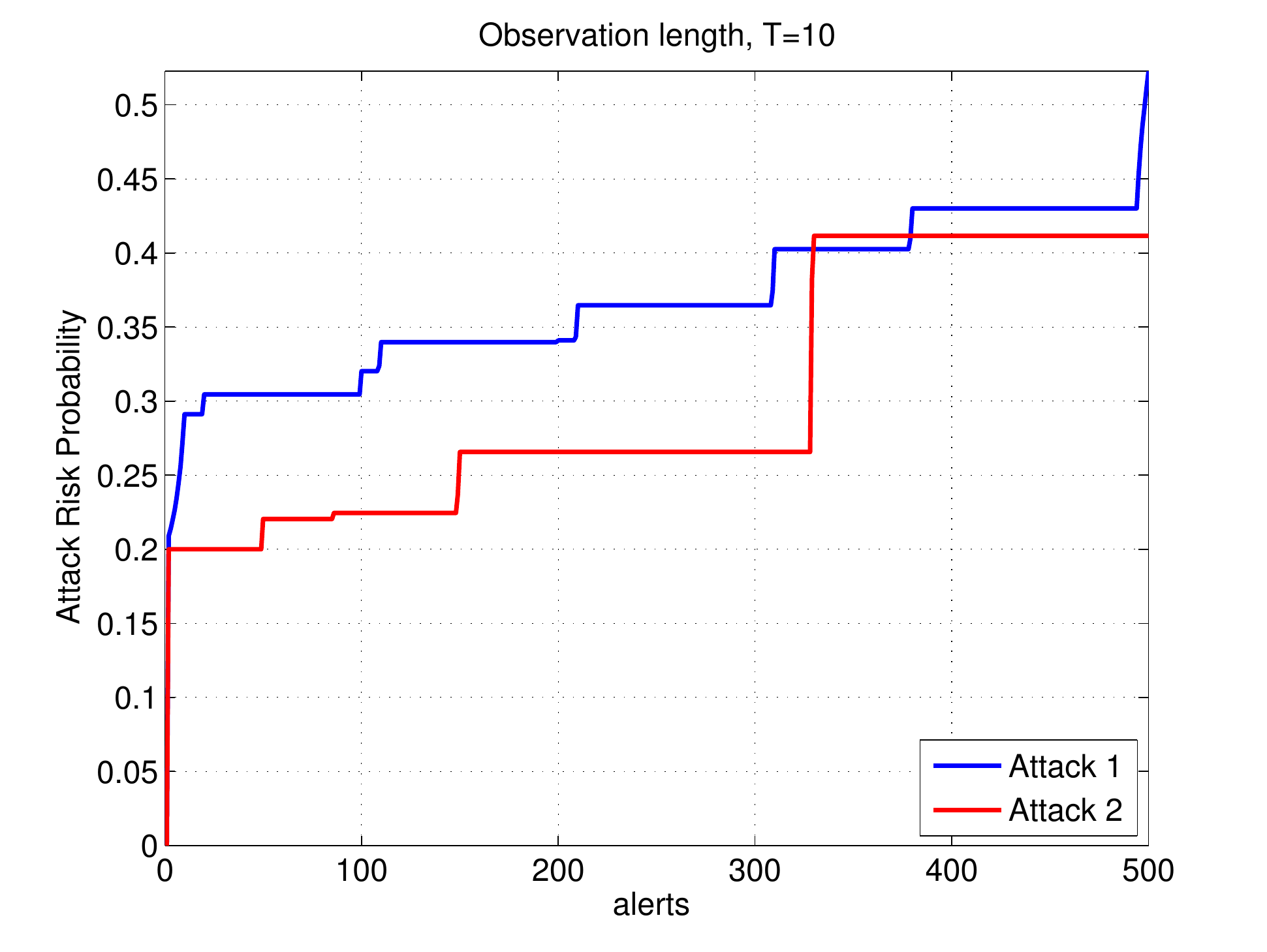}
    \caption{Scenario 3}
  \end{subfigure} &
    \begin{subfigure}[b]{0.22\textwidth}
    \includegraphics[width= 4cm, height=4cm]{Attak_probability_alg1_w500_epslon-eps-converted-to.pdf}
    \caption{Scenario 4}
  \end{subfigure} &
    \begin{subfigure}[b]{0.22\textwidth}
    \includegraphics[width= 4cm, height=4cm]{Attak_probability_alg1_w10_epslon-eps-converted-to.pdf}
    \caption{Scenario 4}
  \end{subfigure} \\
  \end{tabular}
  \caption{ \small Attack Risk Probability Based on Architecture I}
  \label{figur223}
\end{figure*}

\begin{figure*} [htp]
  \centering
  \begin{tabular}{cccc}
  \begin{subfigure}[b]{0.22\textwidth}
  \includegraphics[width= 4cm, height=4cm]{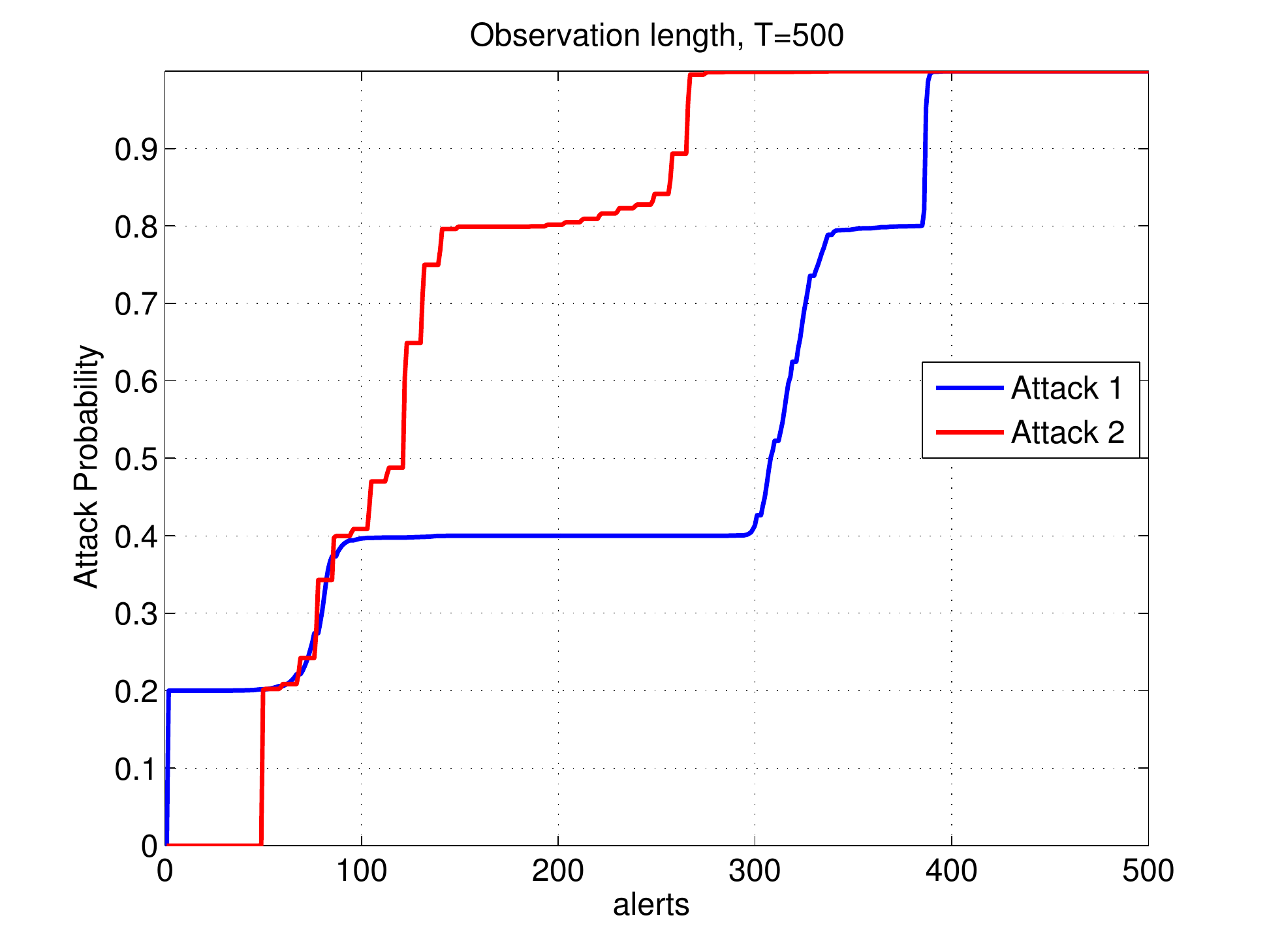}
  \caption{Scenario 3}
  \end{subfigure} &
  \begin{subfigure}[b]{0.22\textwidth}
    \includegraphics[width= 4cm, height=4cm]{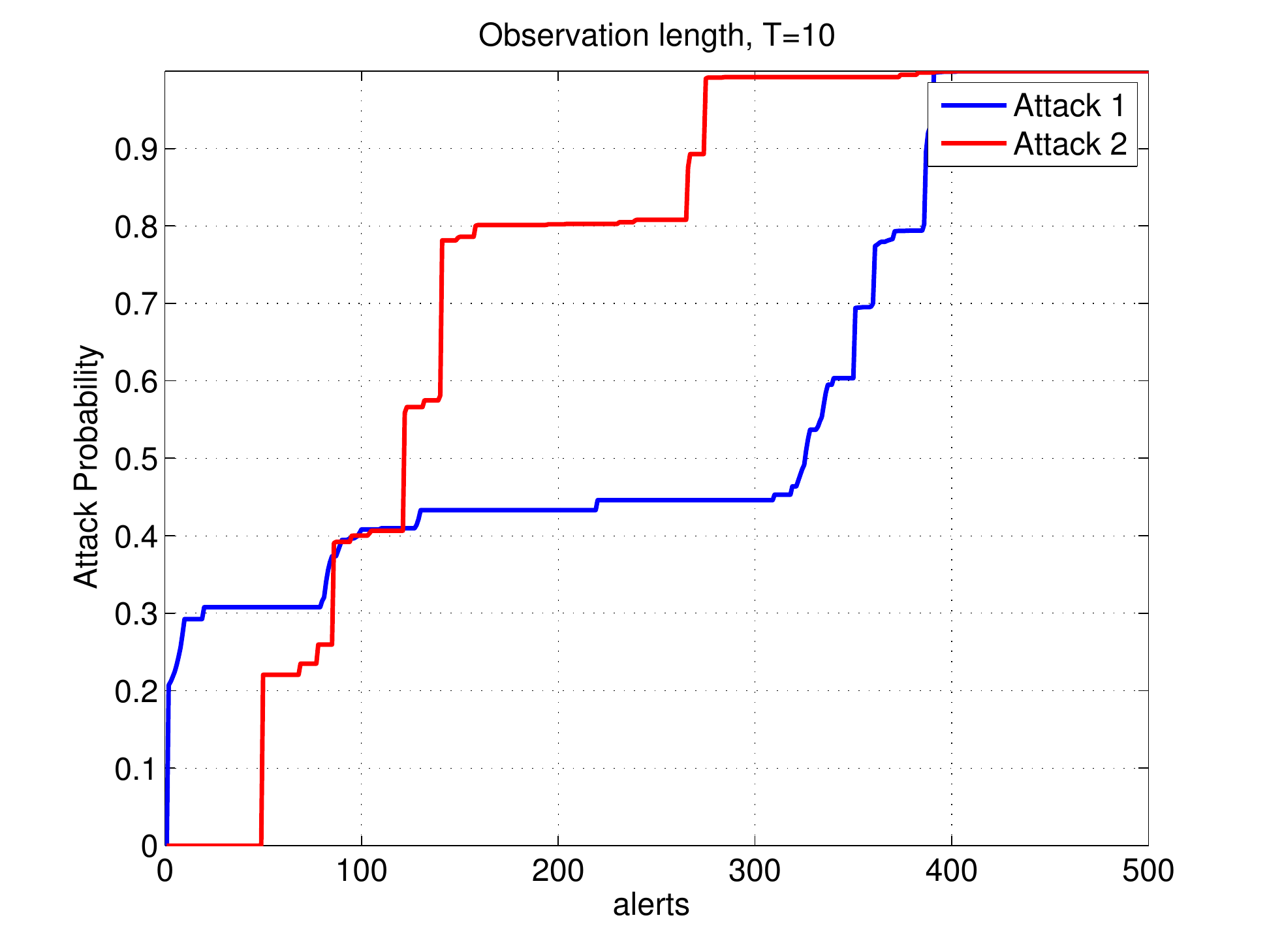}
    \caption{Scenario 3}
  \end{subfigure} &
    \begin{subfigure}[b]{0.22\textwidth}
    \includegraphics[width= 4cm, height=4cm]{Attak_probability_alg2_w500-eps-converted-to.pdf}
    \caption{Scenario 4}
  \end{subfigure} &
    \begin{subfigure}[b]{0.22\textwidth}
    \includegraphics[width= 4cm, height=4cm]{Attak_probability_alg2_w10-eps-converted-to.pdf}
    \caption{Scenario 4}
  \end{subfigure} \\
  \end{tabular}
 \caption{ \small Attack Risk Probability Based on Architecture II}
  \label{figur334}
\end{figure*}

Figs. \ref{figur223} and \ref{figur334} show the attack risk probability for both DARPA multi-stage attacks using Architecture I and II for different interleaving scenarios and for the two observation lengths, $T=10$ and $T=500$. The results shown are for Scenarios 3 and 4, as they are relatively more complex to detect. Fig. \ref{figur334} shows that Architecture II can track the progress of Attacks 1 and 2 for both interleaving scenarios accurately based on the knowledge of the generated input alerts shown in Figs. \ref{figur11}c and \ref{figur11}d. Note, there is no significant difference is shown between the case of $T=10$, and $T=500$. Also note, after 100 alerts, Attack 2 progresses relatively quickly, and reaches the compromised state well before Attack 1. In contrast, however, Architecture I underestimates the progress of Attacks 1 and 2, as shown in Figs. \ref{figur223}c and \ref{figur223}d, because Architecture I fails to detect some of the states for Scenarios 3 and 4, as illustrated in the previous subsection. Fig. \ref{figur223}c shows that both attacks progress at a slow pace. However, this discrepancy is not true as depicted in Fig. \ref{figur334}c where Architecture II shows both attacks progress quickly at different rates. For instance, Attack 2 reaches $80 \%$ after 100 alerts, while Attack 1 reaches $80 \%$ after 300 alerts. Moreover, Fig. \ref{figur223}d shows that Attack 1 progresses faster than Attack 2, which is also not true, as depicted in Fig. \ref{figur334}d which indicates Attack 2 is faster than Attack 1. This inaccurate detection of attacks can adversely affect response decisions, especially, when a priority-based mechanism is employed \cite{Jajodia2017}.

\subsection{Error Rate Performance}
The next performance measure we propose is the error rate ($ER$), which is the ratio of the number of errors resulting from the inconsistency between the type of an alert and the corresponding estimated state relative to the total number of incoming alerts. Formally, $ER$ is given by the following equation:

\normalsize

\scriptsize
\begin{equation}
ER =  \frac{{ \text{Number of incorrect detected states of the incoming Alerts}}}{\text{Total number of Alerts }} \times 100
\label{eq: 21}
\end{equation}
\normalsize
Note, the exact state corresponding to every incoming alert is considered based on the knowledge of the input alerts and their corresponding states. The reasons for inconsistency between the type of alerts and their detected states are: (1) the presence of interfering alerts, and (2) the state estimation error resulting from the enforcement of the left-right HMM model along with some of the observations which may be out of sequence due to the packets generated by the attacker.  In Subsection 5.6, we analyze the effect of False Positives (FPs) and False Negatives (FNs) introduced by the Snort alert generation system on the state estimation error.%This error is assumed to be the way the attack sequence ..... according to Tables \ref{table2: alert and states} and \ref{table3: alert and states}.
\begin{figure*} [htp]
  \centering
  \begin{tabular}{cccc}
  \begin{subfigure}[b]{0.22\textwidth}
  \includegraphics[width= 4cm, height=4cm]{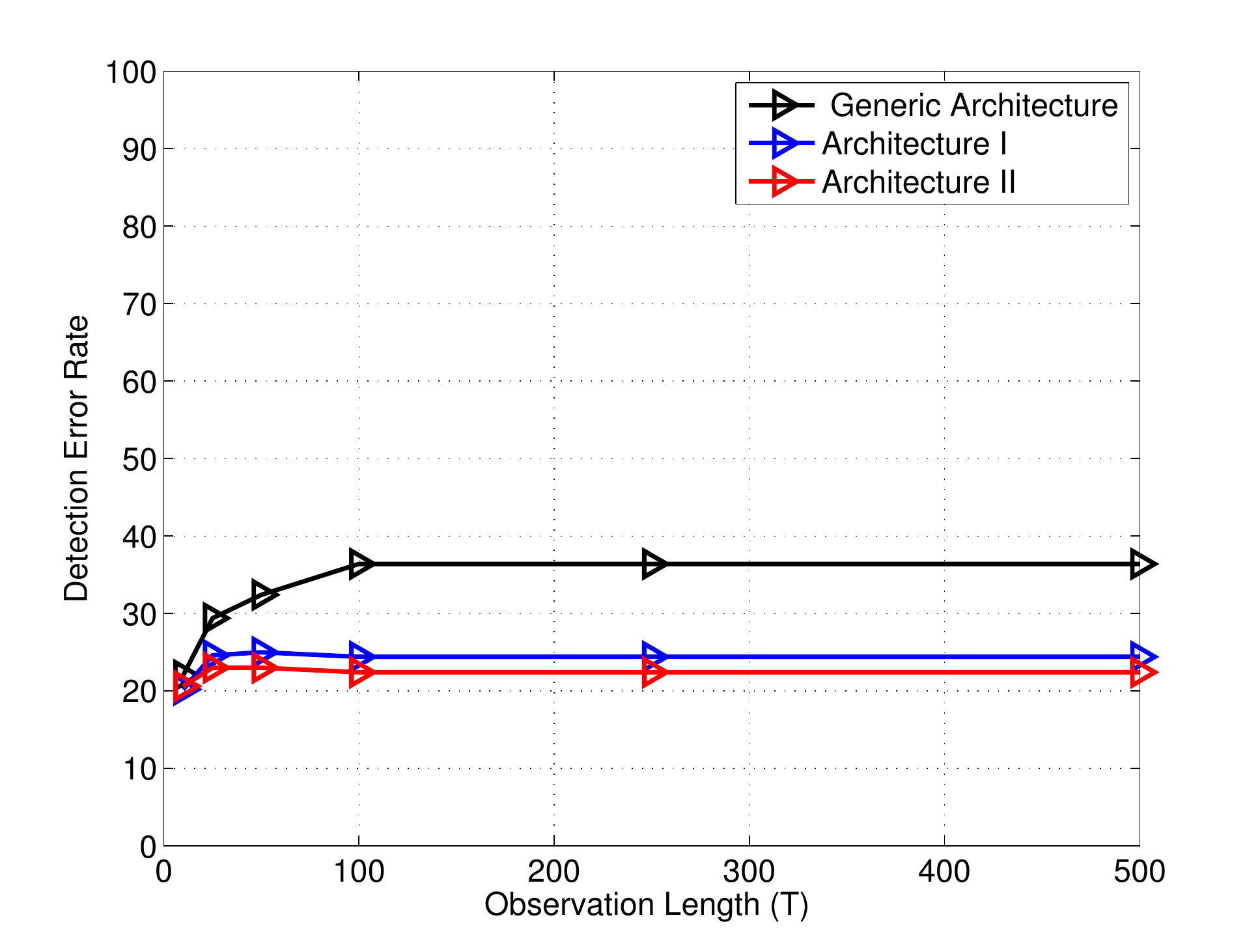}
  \caption{Scenario 1}
  \end{subfigure} &
  \begin{subfigure}[b]{0.22\textwidth}
    \includegraphics[width= 4cm, height=4cm]{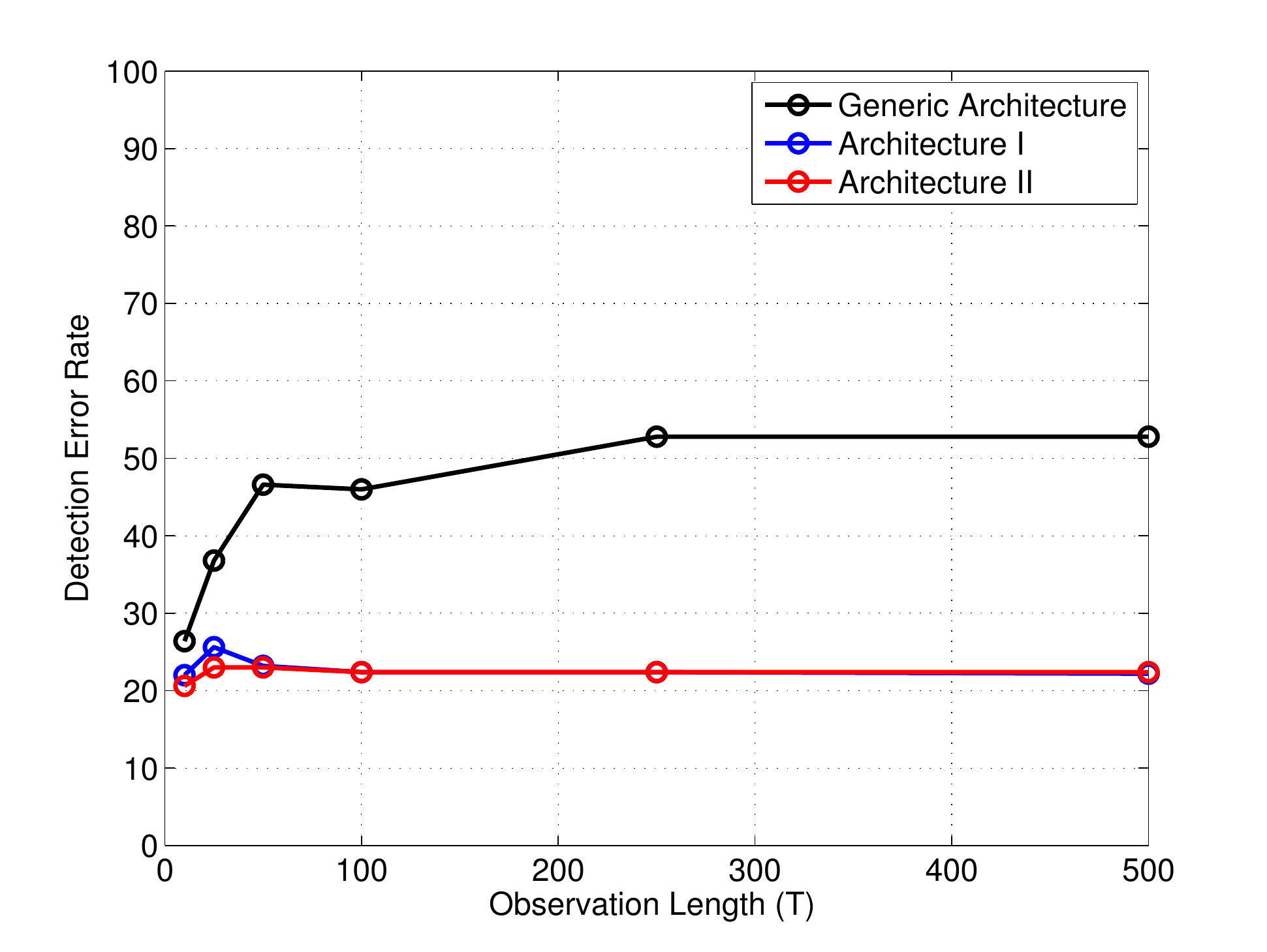}
    \caption{Scenario 2}
  \end{subfigure} &
    \begin{subfigure}[b]{0.22\textwidth}
    \includegraphics[width= 4cm, height=4cm]{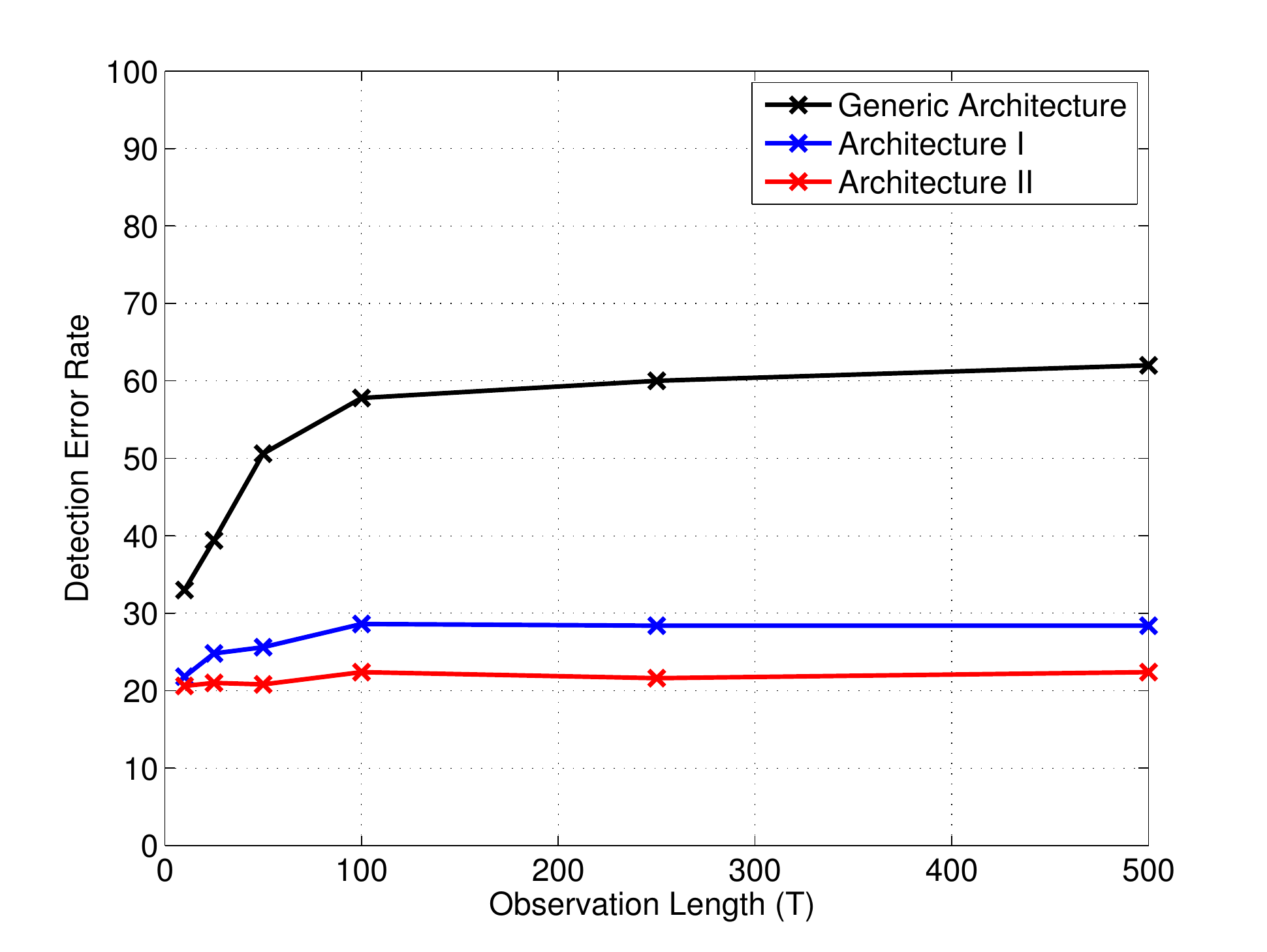}
    \caption{Scenario 3}
  \end{subfigure} &
    \begin{subfigure}[b]{0.22\textwidth}
    \includegraphics[width= 4cm, height=4cm]{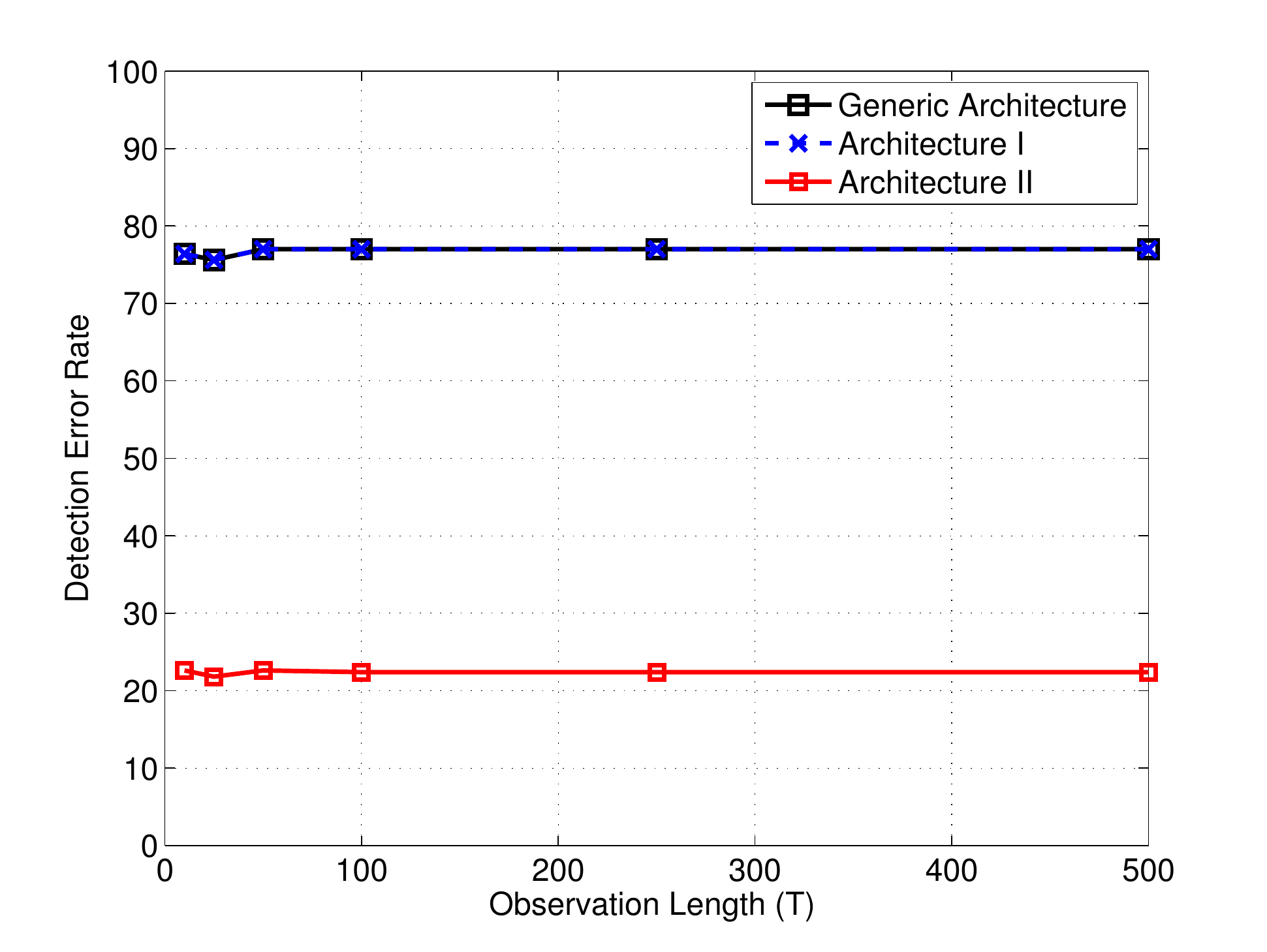}
    \caption{Scenario 4}
  \end{subfigure} \\
  \end{tabular}
  \caption{ \small State Detection Error Rate at Various interleaving Scenarios}
  \label{figur33444}
\end{figure*}

\begin{table*}[ht]
\centering

\caption{\small Number of Correctly Detected Stages per Attack at Various Interleaving Scenarios}
\label{table4}
\scalebox{0.90}
{
\begin{tabular}{|l|l|l|l|l|l|}
\hline
\small \textbf{Interleaving Scenario}       & \small \textbf{Architecture}        & \small \textbf{Attack}   & $T=10$ & $T=100$ & $T=500$ \\ \hline
\multirow{4}{*}{Scenario 2} & \multirow{2}{*}{I}  & Attack 1 & 4    & 4     & 4     \\ \cline{3-6} 
                            &                     & Attack 2 & 5    & 5     & 5     \\ \cline{2-6} 
                            & \multirow{2}{*}{II} & Attack 1 & 4    & 4     & 4     \\ \cline{3-6} 
                            &                     & Attack 2 & 5    & 5     & 5     \\ \hline
\multirow{4}{*}{Scenario 3} & \multirow{2}{*}{I}  & Attack 1 & 4    & 4     & 5     \\ \cline{3-6} 
                            &                     & Attack 2 & 4    & 4     & 3     \\ \cline{2-6} 
                            & \multirow{2}{*}{II} & Attack 1 & 4    & 5     & 5     \\ \cline{3-6} 
                            &                     & Attack 2 & 5    & 5     & 5     \\ \hline
\multirow{4}{*}{Scenario 4}  & \multirow{2}{*}{I}  & Attack 1 & 1    & 1     & 1     \\ \cline{3-6} 
                            &                     & Attack 2 & 1    & 1     & 1     \\ \cline{2-6} 
                            & \multirow{2}{*}{II} & Attack 1 & 4    & 5     & 5     \\ \cline{3-6} 
                            &                     & Attack 2 & 5    & 5     & 5     \\ \hline
\end{tabular}
}
\end{table*}

\begin{figure*} [htp]
  \centering
  \begin{tabular}{cccc}
  \begin{subfigure}[b]{0.27\textwidth}
  \includegraphics[width= 5cm, height=4cm]{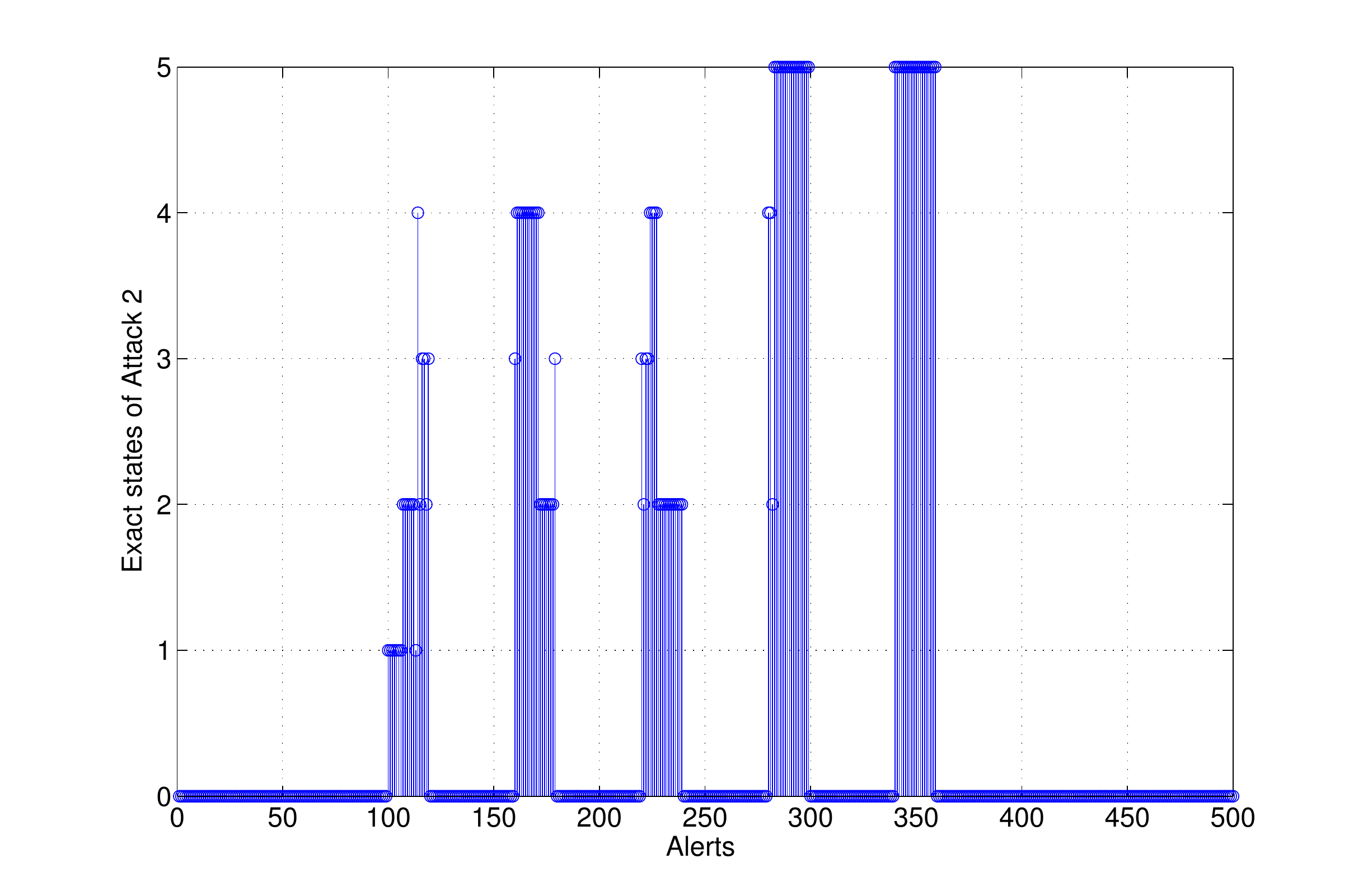}
  \caption{\small Exact States}
  \end{subfigure} &
  \begin{subfigure}[b]{0.27\textwidth}
    \includegraphics[width= 5cm, height=4cm]{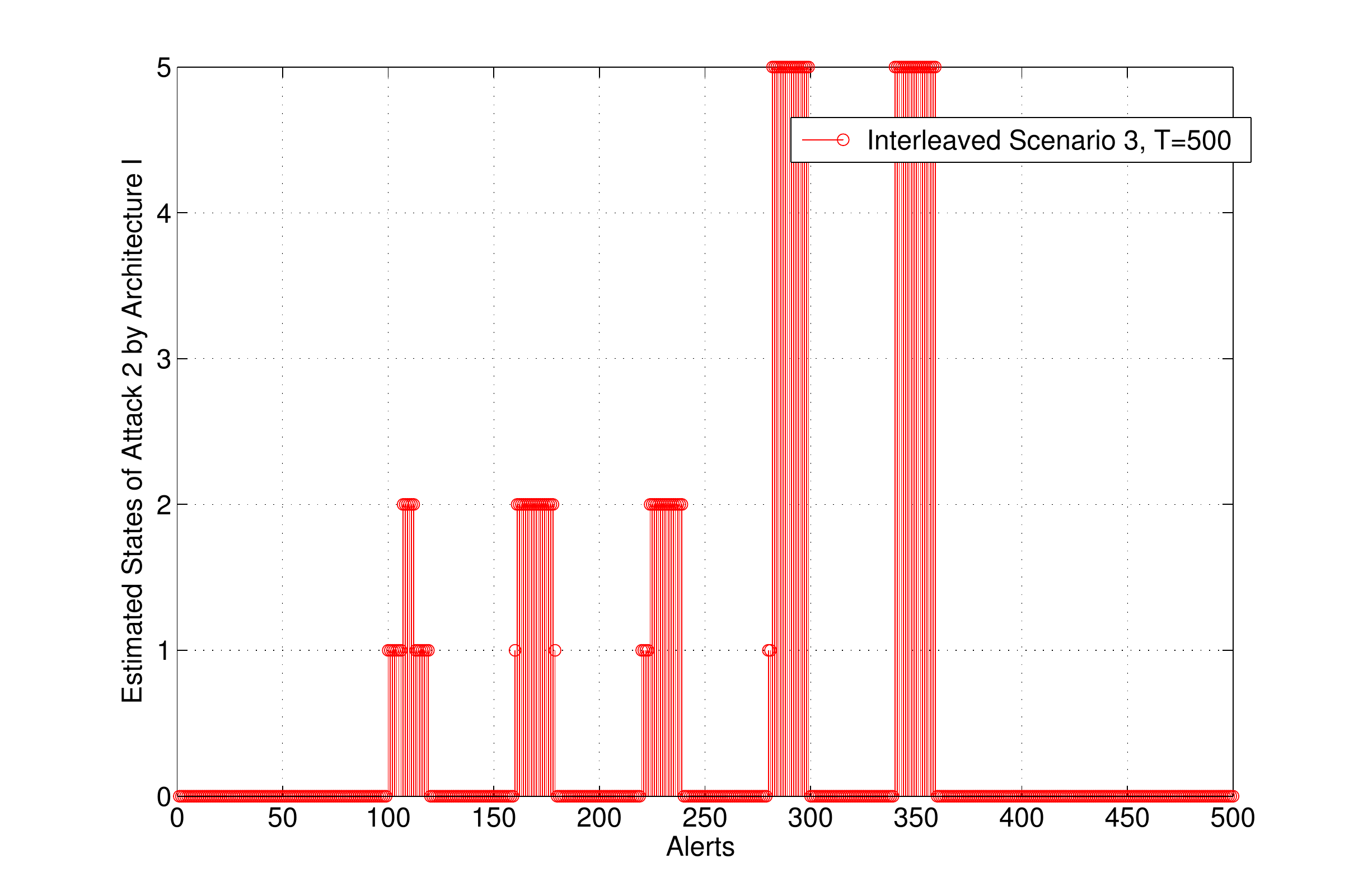}
    \caption{\small Architecture I}
  \end{subfigure} &
    \begin{subfigure}[b]{0.27\textwidth}
    \includegraphics[width= 5cm, height=4cm]{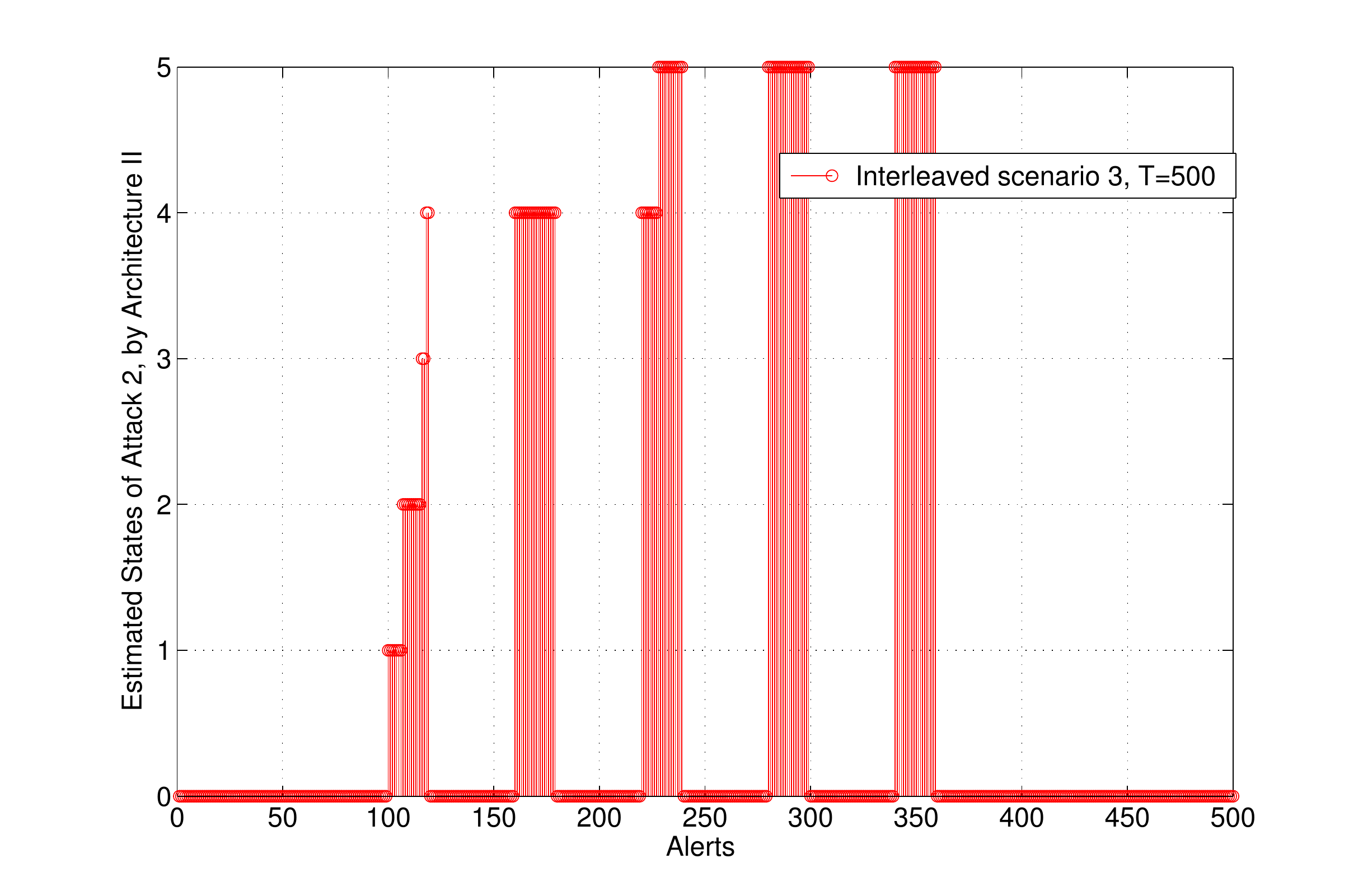}
    \caption{\small Architecture II}
  \end{subfigure} \\
  \end{tabular}
  \caption{ \small Comparison between Architectures I and II in detecting stages of Attack 2 for Scenario 3}
  \label{figur335}
\end{figure*}

Fig. \ref{figur33444} shows the plot of $ER$ for different interleaving scenarios and for several values of $T$. Note, the error for Architecture I is due to the aforementioned reasons (1) and (2), while the error for Architecture 2, is due to only reason (2). It can be seen from the figure that the proposed architectures outperform generic architecture (Fig. (\ref{fig:1})) for interleaving Scenarios 1,2, and 3. However, for Scenario 4, both Architecture I and the generic architecture have similar $ER$, which is higher than Architecture II.
It can also be noted that the $ER$ for Architecture II remains almost constant with respect to $T$ and is also the same for all scenarios. Similarly, Architecture I has an almost constant $ER$ with respect to $T$; however, its $ER$ performance gets worse as the degree of interleaving increases as compared to Architecture II. For instance, for Scenario 4, the $ER$ for Architecture I is as high as $77\%$ as compared to Architecture II which has a value of $22\%$. Note, for the generic architecture, the $ER$ generally increases with $T$ and saturates to a value. The main reason for this trend is the same as aforementioned reason (1) and the lack of capability of this architecture to distinguish between alerts from two different attacks. In addition, due to the same reason, the $ER$ of the generic architecture also increases from Scenario 1 through Scenario 4. 

%%%%%%%%%%%%%%%%%%%%%%%%%%%%%%%%%%%%%%%%%%%%%%%%%%%%%

\subsection{Number of Correctly Detected Stages per Attack}
The third performance measure we propose is the number of correctly detected stages per attack, which allows us to analyze the security impact due to missing or incorrectly detecting stages in a multi-stage attack, especially in consideration of response actions. We compare between architectures in terms of the number of detected stages per attack. 
\begin{figure*} [htp]
  \centering
  \begin{tabular}{cccc}
  \begin{subfigure}[b]{0.22\textwidth}
  \includegraphics[width= 4cm, height=4cm]{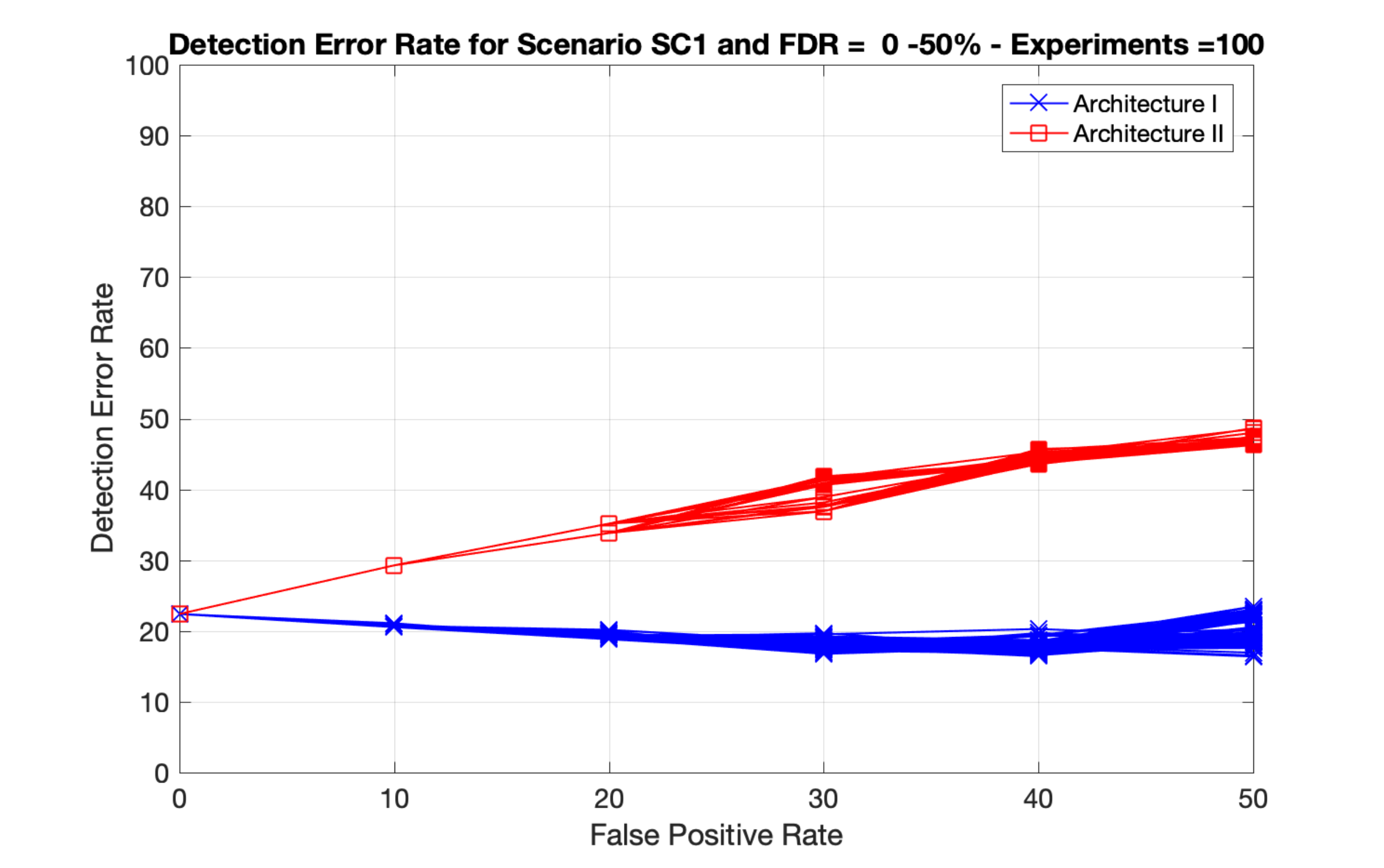}
  \caption{Scenario 1}
  \end{subfigure} &
  \begin{subfigure}[b]{0.22\textwidth}
    \includegraphics[width= 4cm, height=4cm]{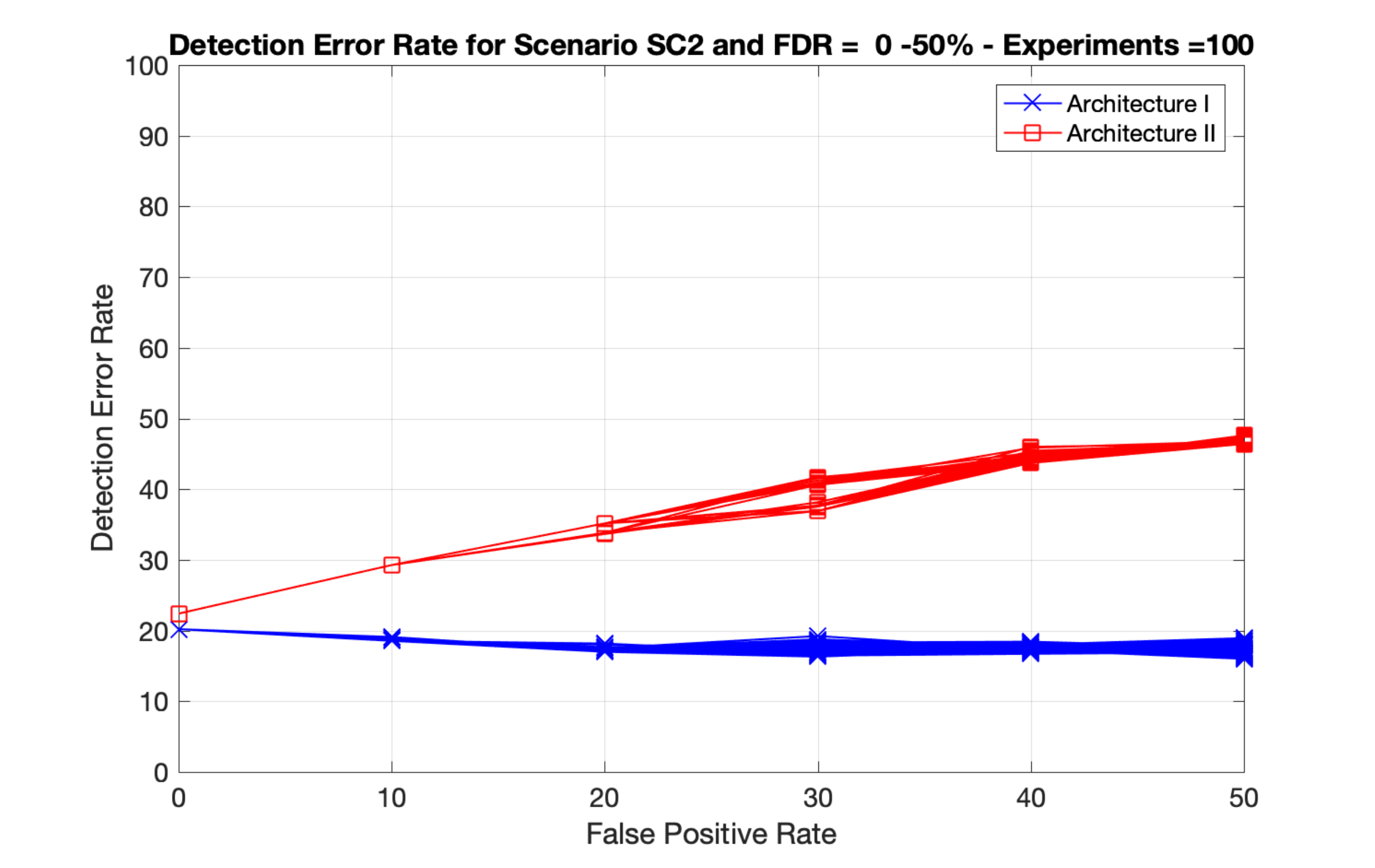}
    \caption{Scenario 2}
  \end{subfigure} &
    \begin{subfigure}[b]{0.22\textwidth}
    \includegraphics[width= 4cm, height=4cm]{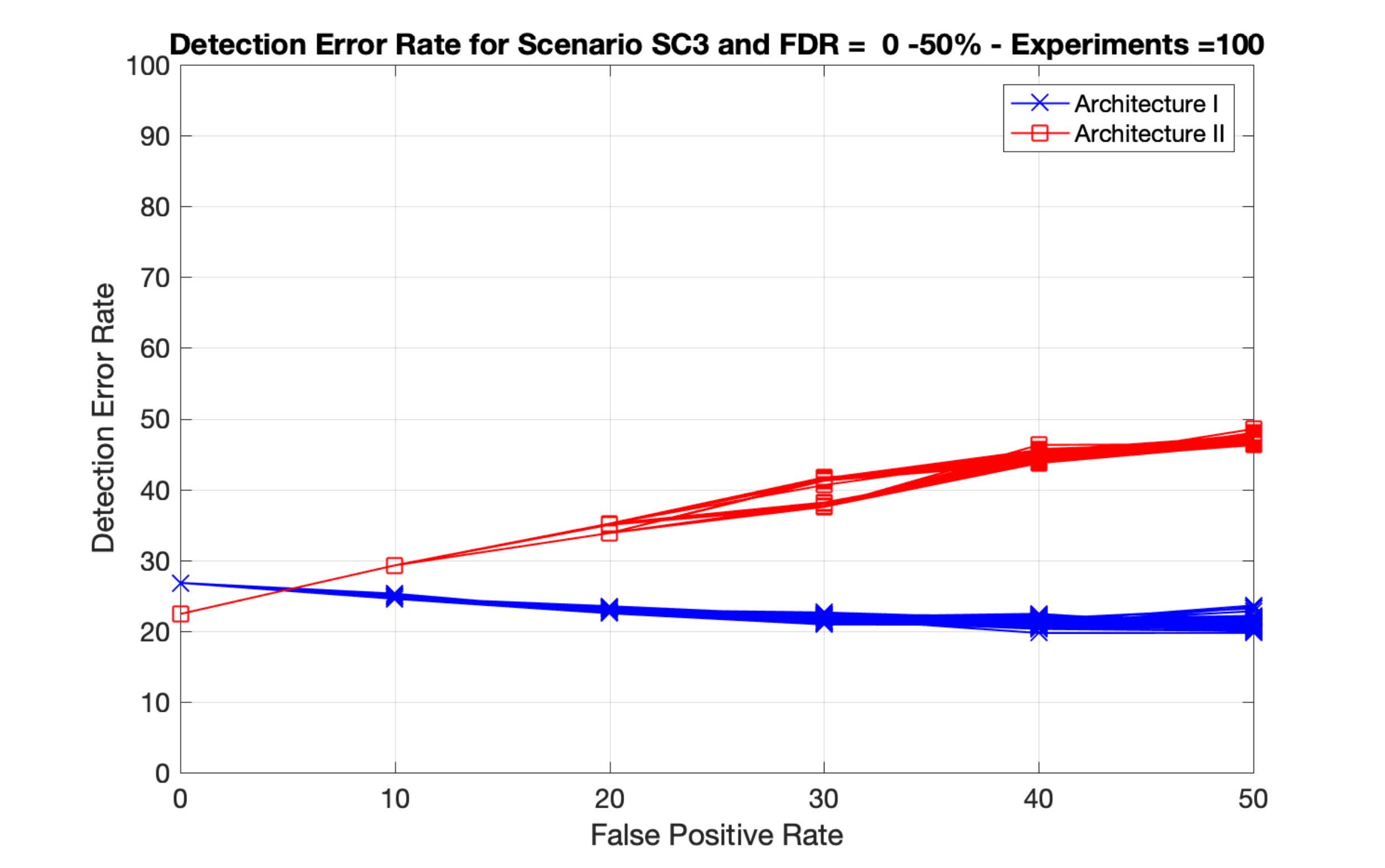}
    \caption{Scenario 3}
  \end{subfigure} &
    \begin{subfigure}[b]{0.22\textwidth}
    \includegraphics[width= 4cm, height=4cm]{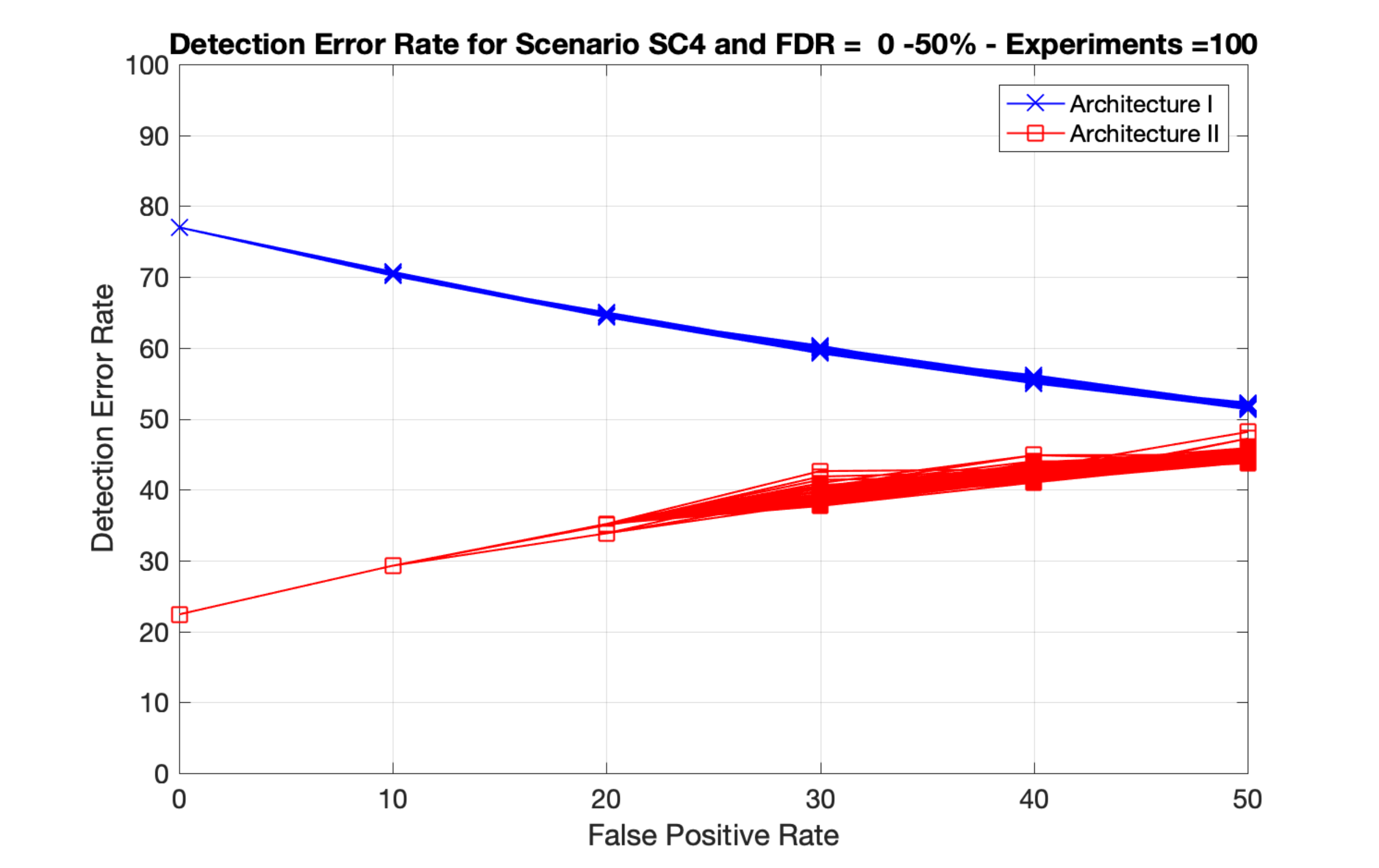}
    \caption{Scenario 4}
  \end{subfigure} \\
  \end{tabular}
  \caption{ \small The Impact of False Positives on the State Detection Error Rate of Architectures I \& II in Scenarios 1-4 Using Various False Discovery Rates ($FDR = 0\% - 50\%$) - The Observation Window Size = 500 and the Number of Experiments = 100}
  \label{figur44999}
\end{figure*}

\begin{figure*} [htp]
  \centering
  \begin{tabular}{cccc}
  \begin{subfigure}[b]{0.22\textwidth}
  \includegraphics[width= 4cm, height=4cm]{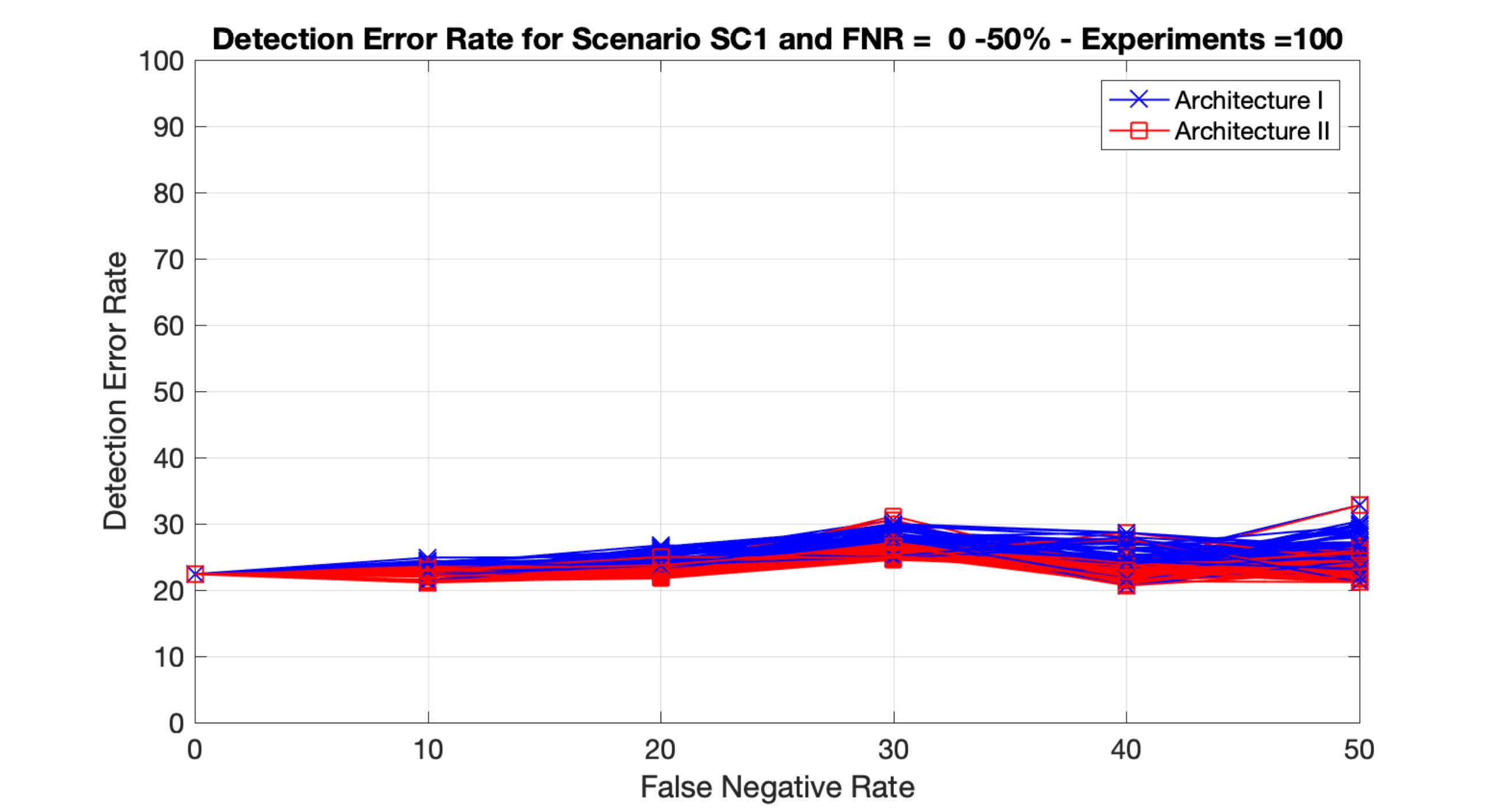}
  \caption{Scenario 1}
  \end{subfigure} &
  \begin{subfigure}[b]{0.22\textwidth}
    \includegraphics[width= 4cm, height=4cm]{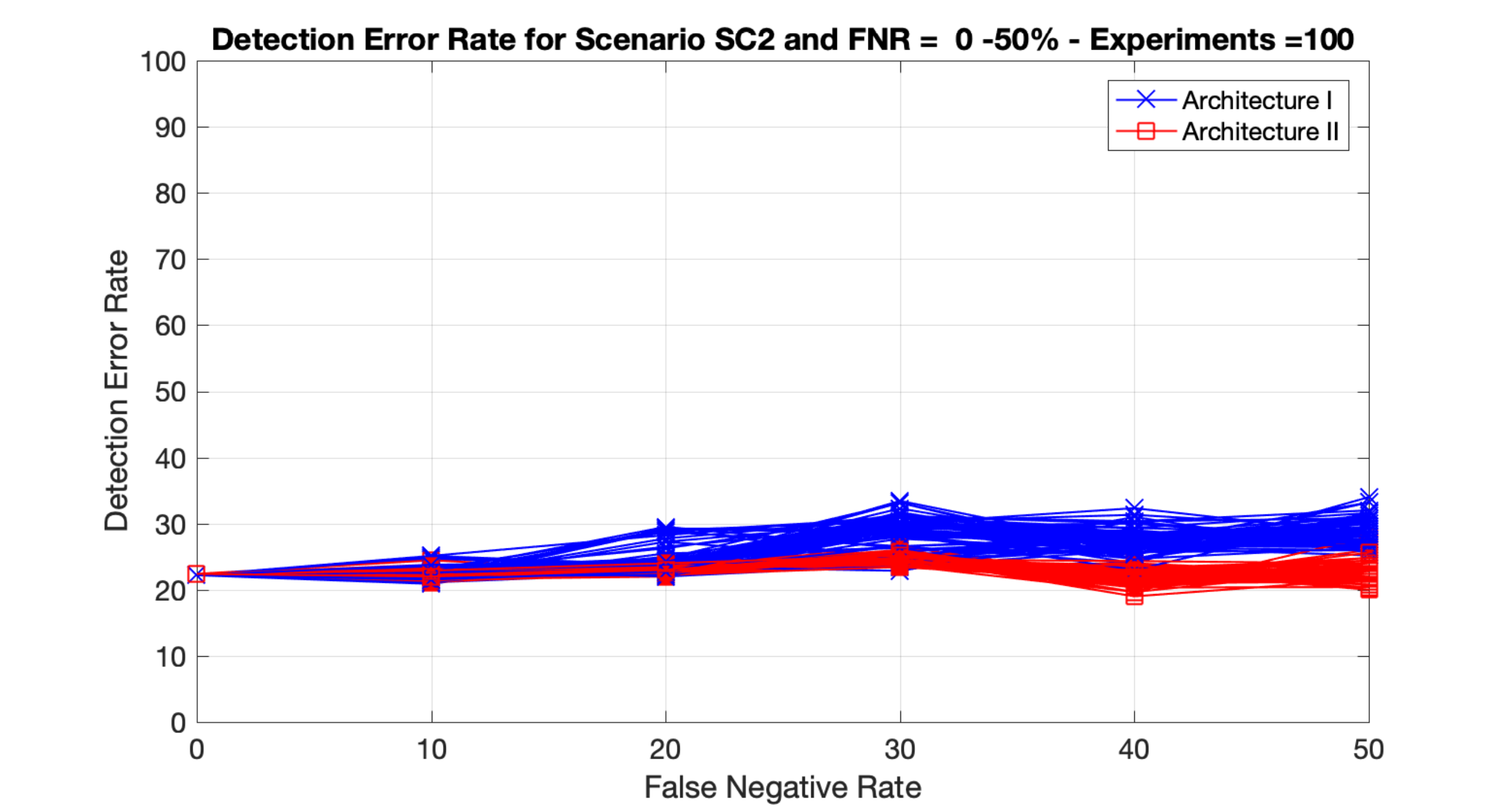}
    \caption{Scenario 2}
  \end{subfigure} &
    \begin{subfigure}[b]{0.22\textwidth}
    \includegraphics[width= 4cm, height=4cm]{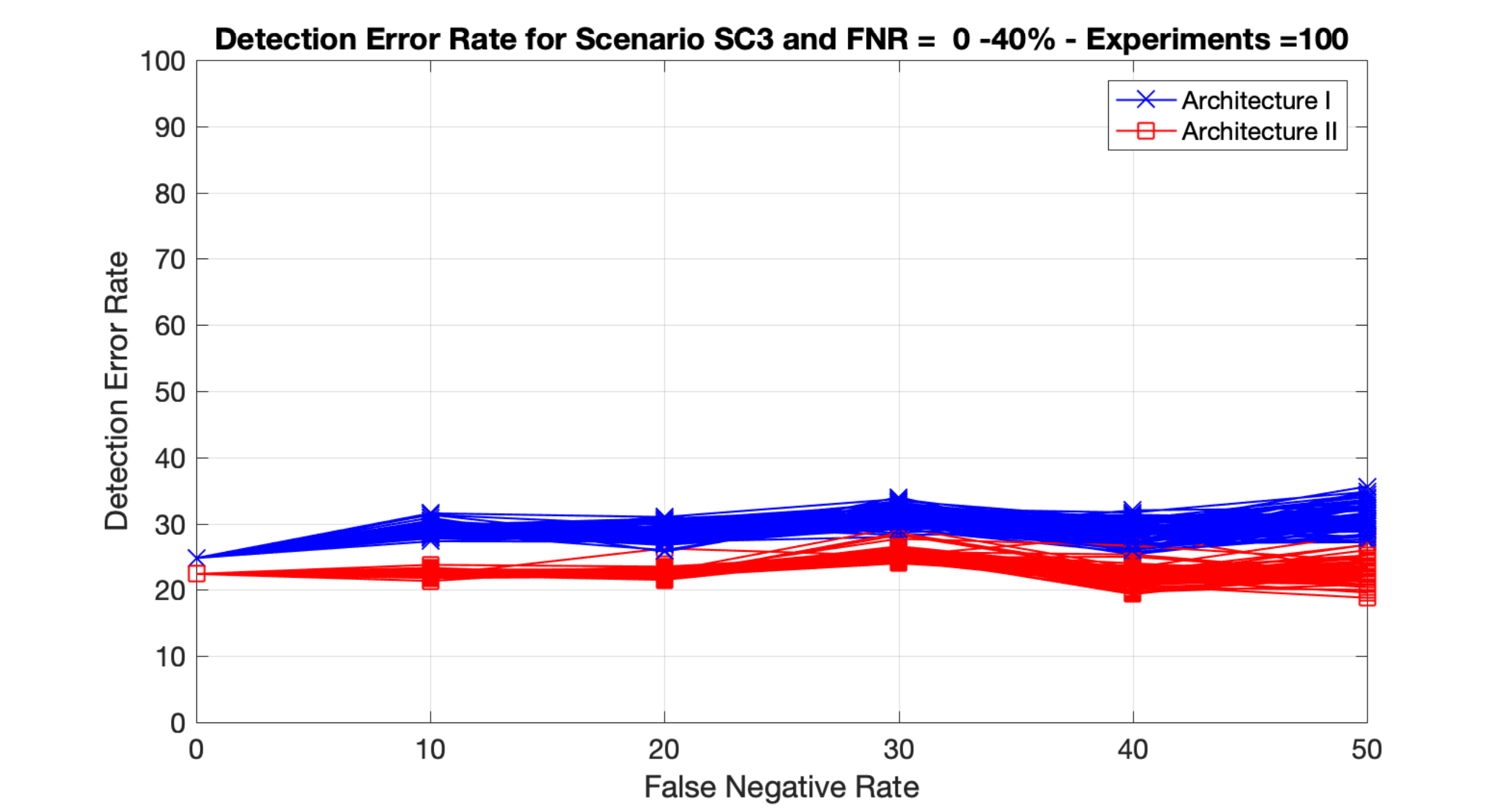}
    \caption{Scenario 3}
  \end{subfigure} &
    \begin{subfigure}[b]{0.22\textwidth}
    \includegraphics[width= 4cm, height=4cm]{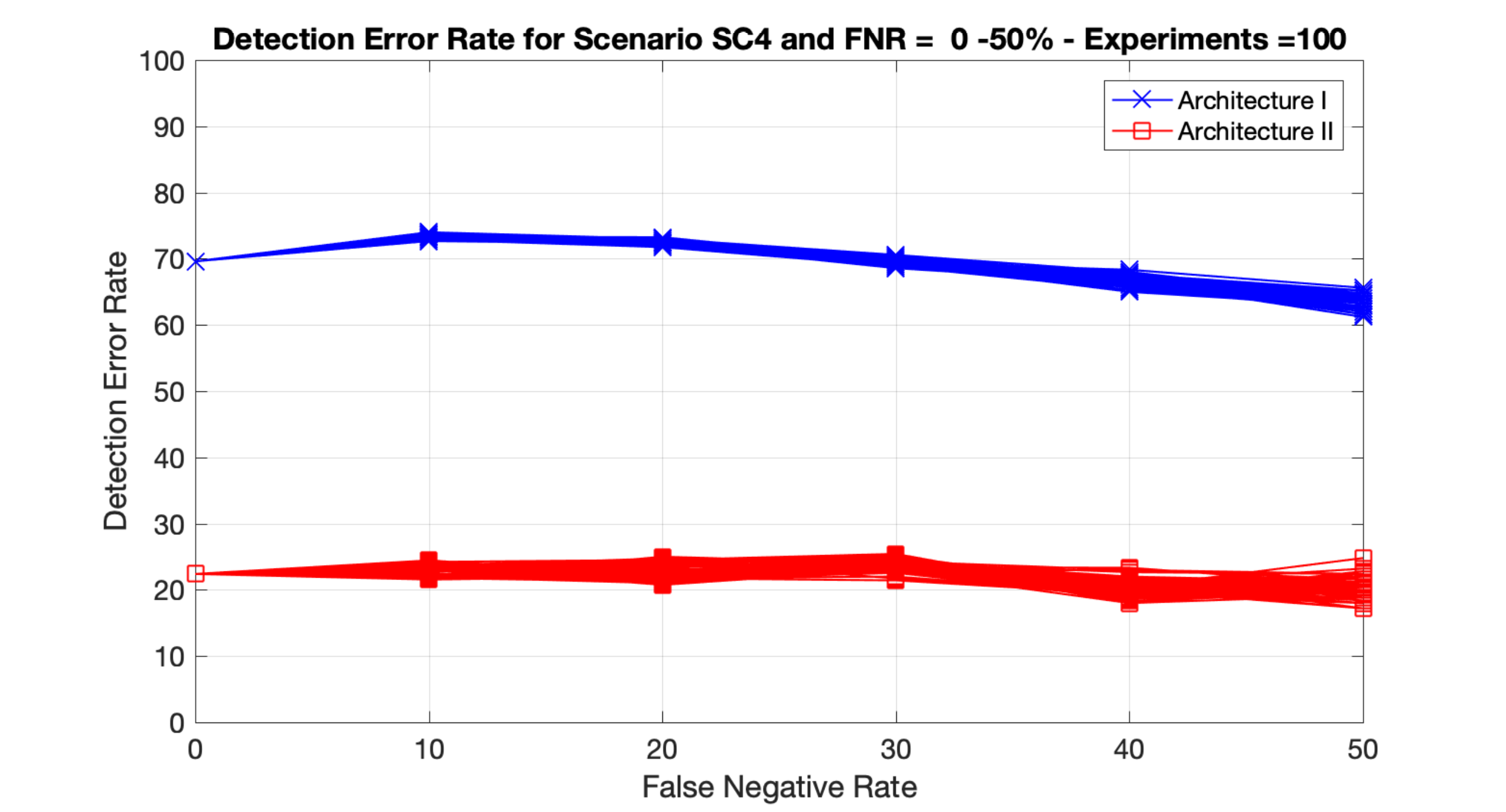}
    \caption{Scenario 4}
  \end{subfigure} \\
  \end{tabular}
  \caption{ \small The Impact of False Negatives on the State Detection Error Rate of Architectures I \& II in Scenarios 1-4 Using Various False Negative Rates ($FNR = 0\% - 50\%$) - The Observation Window Size = 500 and the Number of Experiments = 100}
  \label{figur45999}
\end{figure*}

This measure is computed as follows. As we know the correspondence between alerts and stages (or states) of the attacks based on the knowledge of the DARPA2000 dataset, we compare the estimated states from each HMM with the exact states. Table \ref{table4} provides the results for three different values of $T$. It can be observed that Architecture II outperforms Architecture I in correctly detecting more stages for both attacks. The performance of the two architectures is the same for the interleaving Scenario 2, as both of them can detect stages 1, 2, 4, and 5 but not 3. This can be seen in Figs. \ref{figur12}b, \ref{figur13}b, \ref{figur14}b, and \ref{figur15}b. 

For Scenarios 3 and 4, Architecture II detects more stages than Architecture I. For example, all five stages of Attack 2 are detected in Scenario 3 using Architecture II for $T=500$, while Architecture I only detects Stages 1 and 5. Fig. \ref{figur335} shows the estimated states by HMM2 for Attack 2 using Scenario 3. It can be observed that with Architecture I, Stages 3 and 4 are not detected. The advantage of Architecture II is more apparent in Table \ref{table4} when the interleaving Scenario 4 is used.

Fig. \ref{figur335} and Table \ref{table4} show the importance of this performance metric in terms of how much lead time is available to respond to ongoing attacks. For instance, Fig. \ref{figur335}b shows that Architecture I detects State 2 then immediately detects State 5 of Attack 2, which implies that not enough lead time is available to respond to a progressing attack. While Fig. \ref{figur335}c shows, on the other hand, that Architecture II detects all states of Attack 2 in a correct sequence similar to the synthesized input states of Attack 2 shown in Fig. \ref{figur335}a. This performance metric establishes the importance of considering a large number of states in modeling the HMM in essence that the effect of missing a few number of states does not drastically impact lead time while making real time response decisions to an ongoing multi-stage attack.

\subsection{Impact of False Positives (FPs) and False Negatives (FNs)}
The security alerts generated by an IDS are, in general, noisy and suffer from both FPs and FNs. In the former case, the IDS (e.g. Snort) generates false alerts when no attack attempts are happening in the network, and in the latter case, the IDS fails to detect exploit attempts and does not generate alerts \cite{Tjhai2008}.
%In general, the performance of the existing IDS techniques suffers due to both FPs and FNs. %\cite{Tjhai2008}. 
In our evaluation of the proposed architectures, similar effects are observed, as discussed below. 

We have conducted several experiments to study the impact of FPs and FNs for the proposed HMM architectures. In our experiments, we synthesize the dataset by eliminating some of the True Positive alerts (TPs), in order to mimic FNs, and inject some FPs into the observation sequence in a randomized fashion.
%\footnote{The distances between consecutive FPs are based on Normal distribution $X \sim \mathcal{N}(1.5, 0.2)\,.  $}. 
In our experiments, we vary the False Discovery Rate (FDR) and the False Negative Rate (FNR) from 0\% to 50\% for the alert generation system (Snort). In addition, due to randomized injection and elimination, we conduct 100 experiments for each interleaving scenario and for each value of FDR and FNR to identify any potential outliers. Note, in our experiments, we assume that the FP error is uniformly induced by all of the alert generation rules employed by Snort. In other words, the effect of the FPs is uniformly distributed over generated alerts by Snort.

The results for the impact of FDR for both architectures are shown in Fig. \ref{figur44999}. It can be noticed that the performance of Architecture II degrades with the increase of FDR. This lowered performance is expected as some of FPs are also "demultiplexed" and affect the TPs in their respective substreams when each of these substreams is processed by the associated HMM template.

However, for Architecture I, the general trend observed is an improvement in the detection error rate performance which is more noticeable for Scenario 4 (Fig. \ref{figur44999}d). A plausible explanation for this trend is that the FPs in the whole stream either maintain the current state or allow for a transition to a subsequent state in the HMM. The DARPA2000 dataset contains a high number of observations related to State 1 as compared to other states. Therefore, under the assumption of a uniform injection of FPs in the alert stream, the likelihood of the HMM staying in State 1 increases with the increase in FDR. 
Note, an HMM template for Architecture I always leads to State 1 for unrelated observations. Therefore, as the percentage of FDR increases, this tendency of staying in State 1 also increases with the high degree of interleaving due to the increase in the number of unrelated observations.
 
%owever, some FP related alerts in a given state can allow state transition to move forward which can improve state estimation probability. Note, such state estimation can increase with increase FDR thereby improving the performance of the Architecture I. In other words, the performance curve of Fig13d can improve with increasing of $FDR$.
\begin{figure*} [htp]
  \centering
  \begin{tabular}{cccc}
  \begin{subfigure}[b]{0.27\textwidth}
  \includegraphics[width= 5cm, height=4cm]{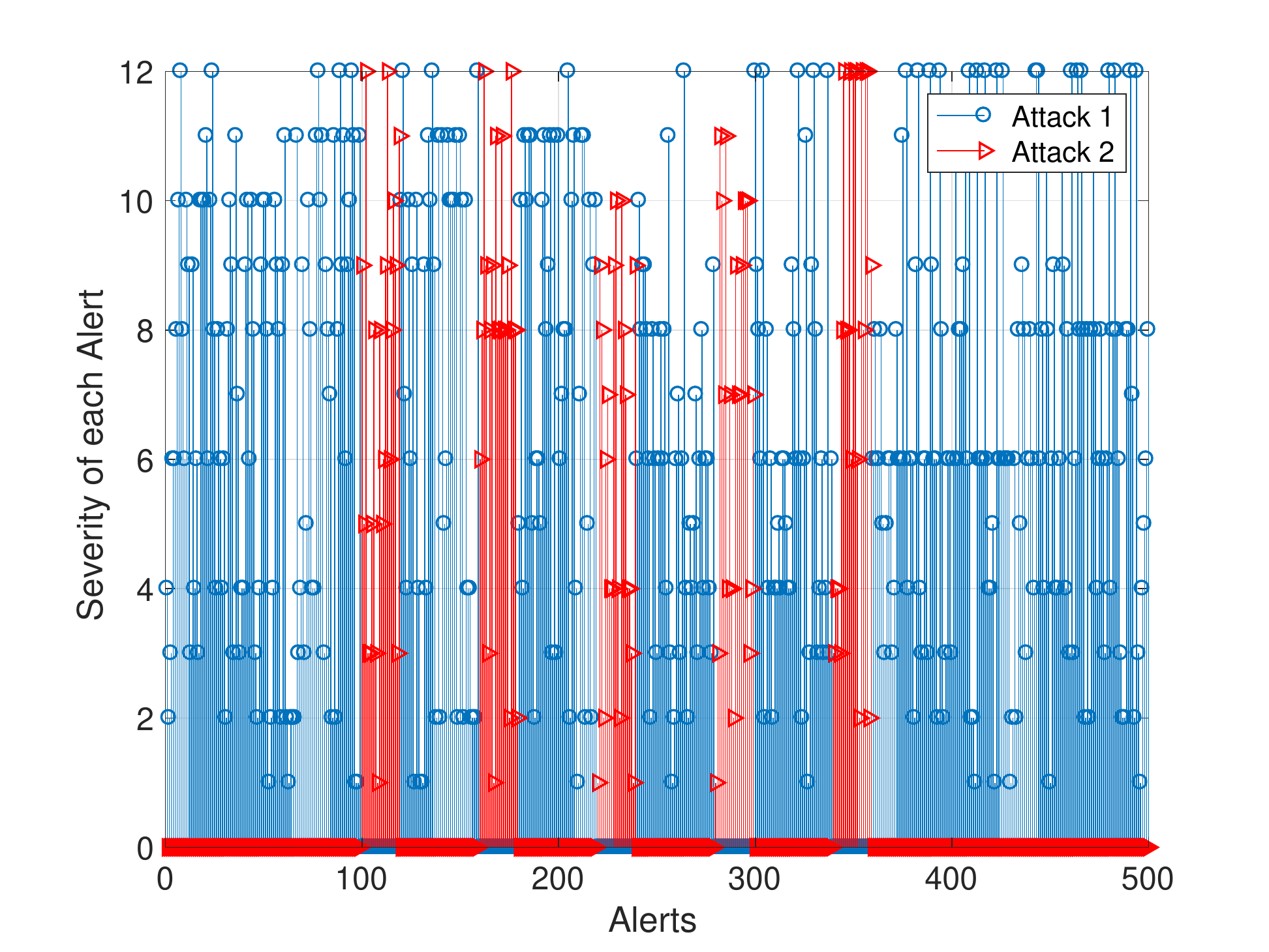}
  \caption{\small Exact States}
  \end{subfigure} &
  \begin{subfigure}[b]{0.27\textwidth}
    \includegraphics[width= 5cm, height=4cm]{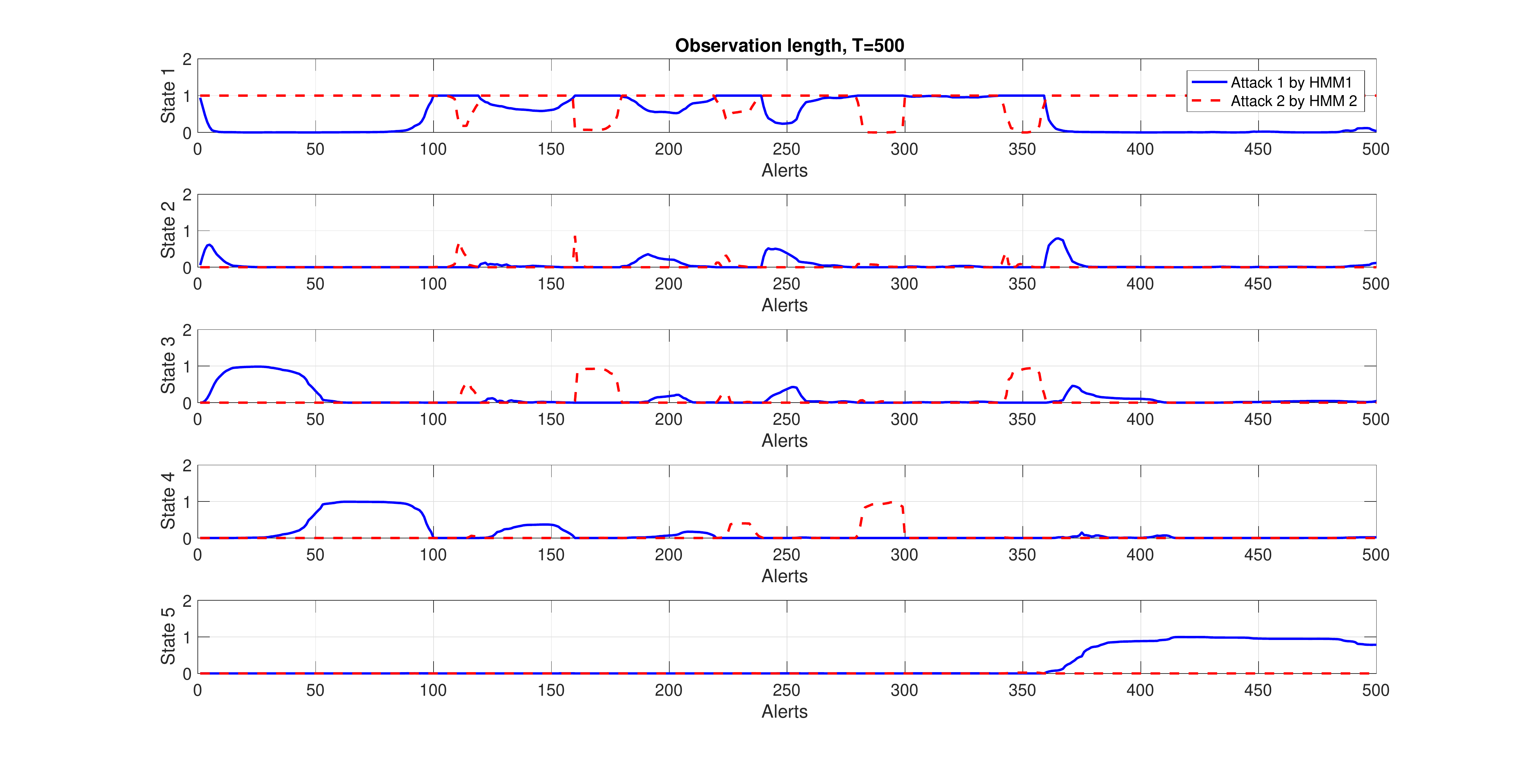}
    \caption{\small Architecture I}
  \end{subfigure} &
    \begin{subfigure}[b]{0.27\textwidth}
    \includegraphics[width= 5cm, height=4cm]{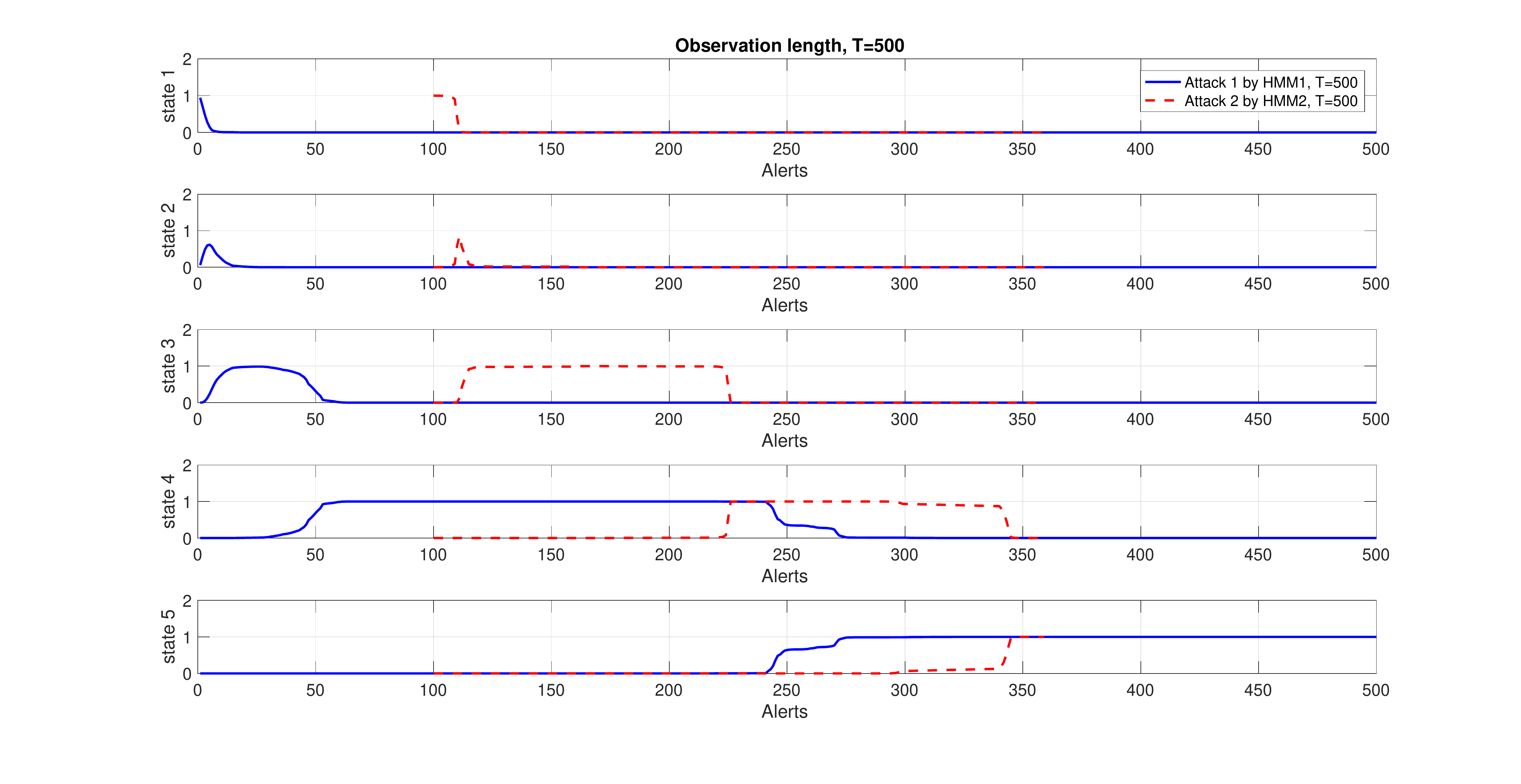}
    \caption{\small Architecture II}
  \end{subfigure} \\
  \end{tabular}
  \caption{ \small State Probability of Synthesized Attacks 1 and 2 for the Interleaving Scenario 3 Detected by HMM1 and HMM2 Based on both Architectures, T=500}
  \label{figur9935}
\end{figure*}

The effect of FNs of the IDS system (Snort) on both architectures is shown in Fig. \ref{figur45999}. The effect of the FNR on the performance of Architecture II is shown in Fig. \ref{figur45999}. 
Note, some of the TPs which are eliminated by FNs may belong to the erratic behavior of the attacker (which is reason (2) mentioned in Subsection 5.4), while some other TPs eliminated by FNs are legitimate (i.e. correctly sequenced) alerts. A positive effect is shown on performance in the case of erratic behavior results in improved performance of detection error rate, while for the case of legitimate alerts, the performance degrades. We can notice such improvement and degradation in the performance for different values of FDR and FNR as shown in Figs. \ref{figur44999} and \ref{figur45999}.

For Architecture I, in addition to reason (2), the reason (1) (mentioned in Subsection 5.4) also comes into play, whereby FNs can eliminate some unrelated alerts and thereby reduce the possibility of transition to State 1 and increasing the possibility of transition from a given state to the next state. This improvement in the performance is more noticeable for Scenario 4 where the prospects of making such forward transitions are higher. %why higher - clarify
 %which reduces the performance degradation which is more noticeable based on reason 2 which is related to the attacker's behavior as mentioned in Section 5.4. impact the performance in two ways based on reason 1 mentioned in Section 5.2 which are related to reason 2 related to the attacker's behavior and TPs. 

\subsection{Performance Evaluation Using Synthesized Datasets - Case Study 2} 
The evaluation experiments in the previous case study (Subsections 5.1-5.6) have been implemented with DARPA2000 simultaneously interleaved attacks. Due to the limitation of this dataset in terms of number of scenarios and due to the lack of availability of datasets with a large number of attacks, in this case study, we generate synthesized datasets that contain different instances of the DARPA2000 multi-stage attacks. Note, it is important to point out that our objective is not to evaluate a specific dataset, but rather to evaluate the proposed architectures for detecting complex multi-stage attacks that are orchestrated by an adversary through interleaving. The design of the architectures is generic in the sense they can process any dataset with multiple attacks. Specifically, the goal of this case study is to study the effectiveness of the proposed architectures when tested on various datasets that have multi-stage attack instances which vary from from the trained HMM templates. The importance of this evaluation is that, in reality, the attacker(s) may not follow the exact same steps for the same multi-stage attack type, for example, in terms of the targeted nodes or the number of attempts. %and the vulnerability exploitation time.

In particular, an HMM generator \cite{mathworks}, which generates sequences for HMM, is used to orchestrate several instances of a multi-stage attack type. 
In this case study, the original DARPA2000 dataset is used for training, and the generated synthesized datasets is used for testing.
%Note, in this experiment, we generate FP observations randomly and with high rate to model the erroneous behavior of an IDS and to have a realistic synthesized dataset. 

\subsubsection{Performance Evaluation - Two Synthesized Multi-stage Attacks} 
Fig. \ref{figur9935} shows the state probability of synthesized Attacks 1 and 2 for the interleaving Scenario 3 detected by HMM1 and HMM2 for Architecture I, Fig. \ref{figur9935}b, and for Architecture II, (Fig. \ref{figur9935}c). 
It can be observed in Fig. \ref{figur9935} that the results of the case study of synthesized multi-stage attacks are consistent with the previous results for Scenario 3 from Case study 1, discussed in Subsection 5.2 (Figs. \ref{figur11}c, \ref{figur12}c , \ref{figur13}c), in terms of detection performance and state estimation. 
In particular, Architecture II detects all stages of the interleaved multi-stage attack scenario. In contrast, Architecture I fails to detect State 5 of Attack 2 as it estimated the state as State 2 and State 3 due to the noisy observations resulting from unrelated alerts.

\begin{figure*} [ht]
  \centering
  \begin{tabular}{cccc}
  \begin{subfigure}[b]{0.27\textwidth}
  \includegraphics[width= 5cm, height=4cm]{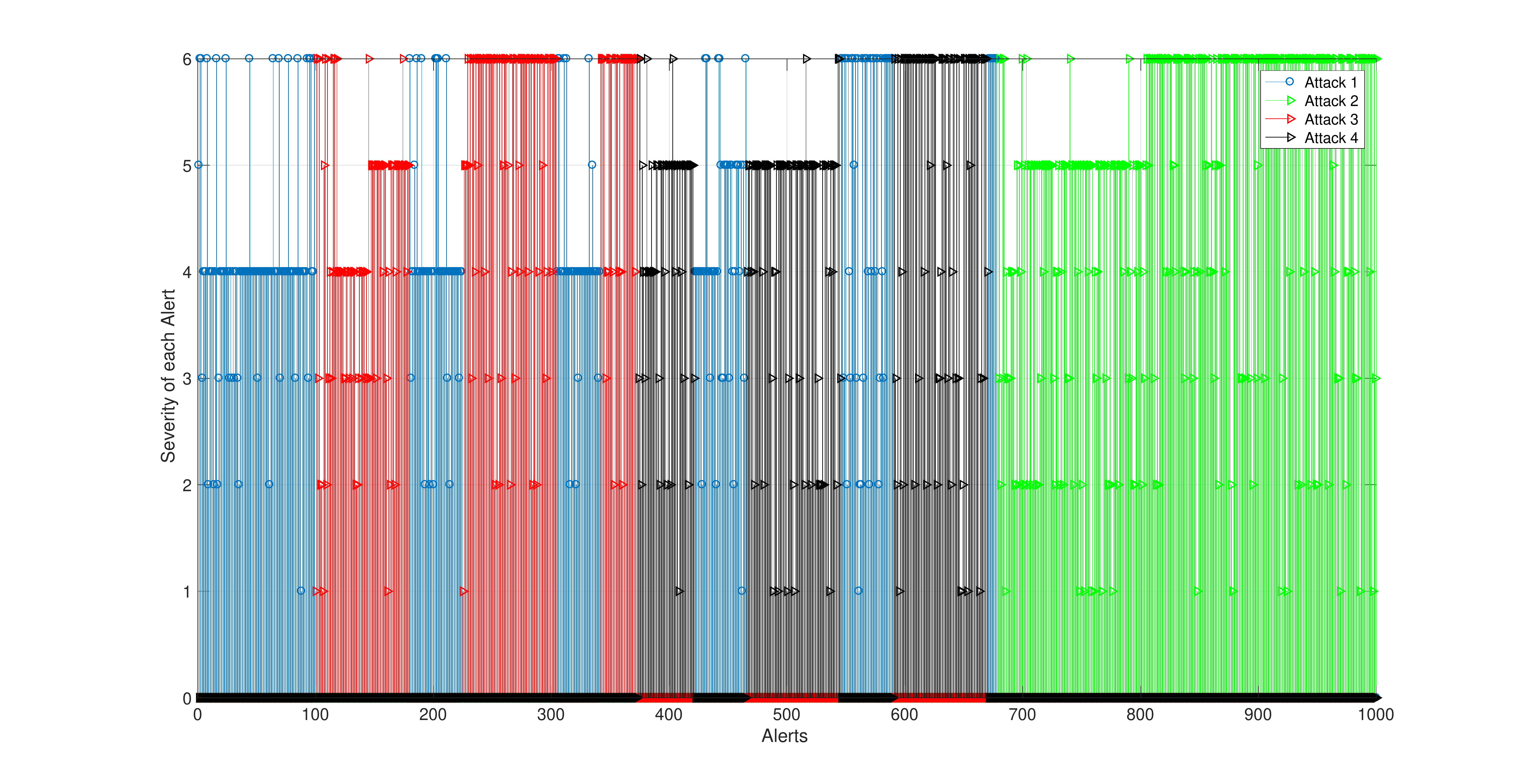}
  \caption{\small Exact States}
  \end{subfigure} &
  \begin{subfigure}[b]{0.27\textwidth}
    \includegraphics[width= 5cm, height=4cm]{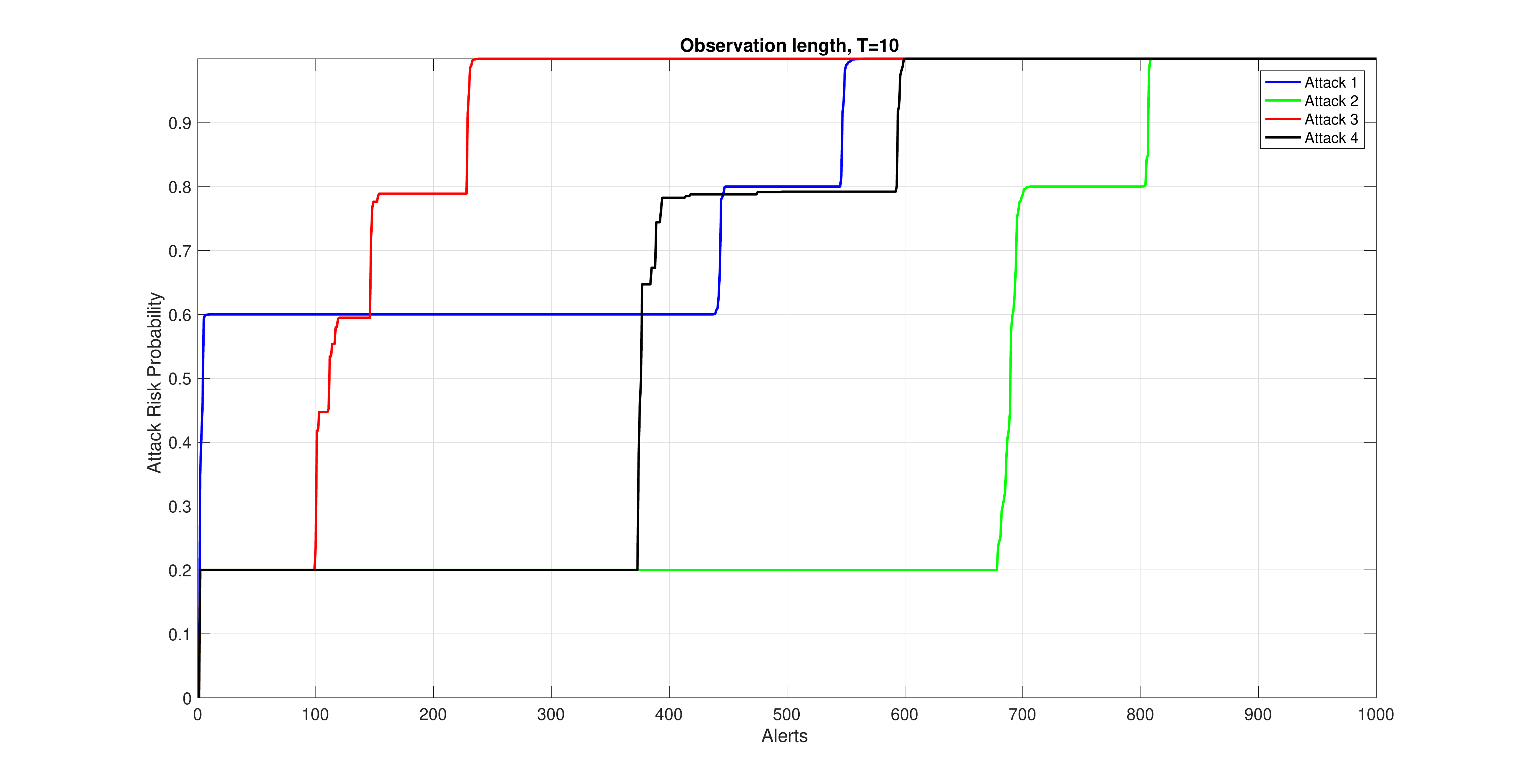}
    \caption{\small Attack Probability}
  \end{subfigure} &
    \begin{subfigure}[b]{0.27\textwidth}
    \includegraphics[width= 5cm, height=4cm]{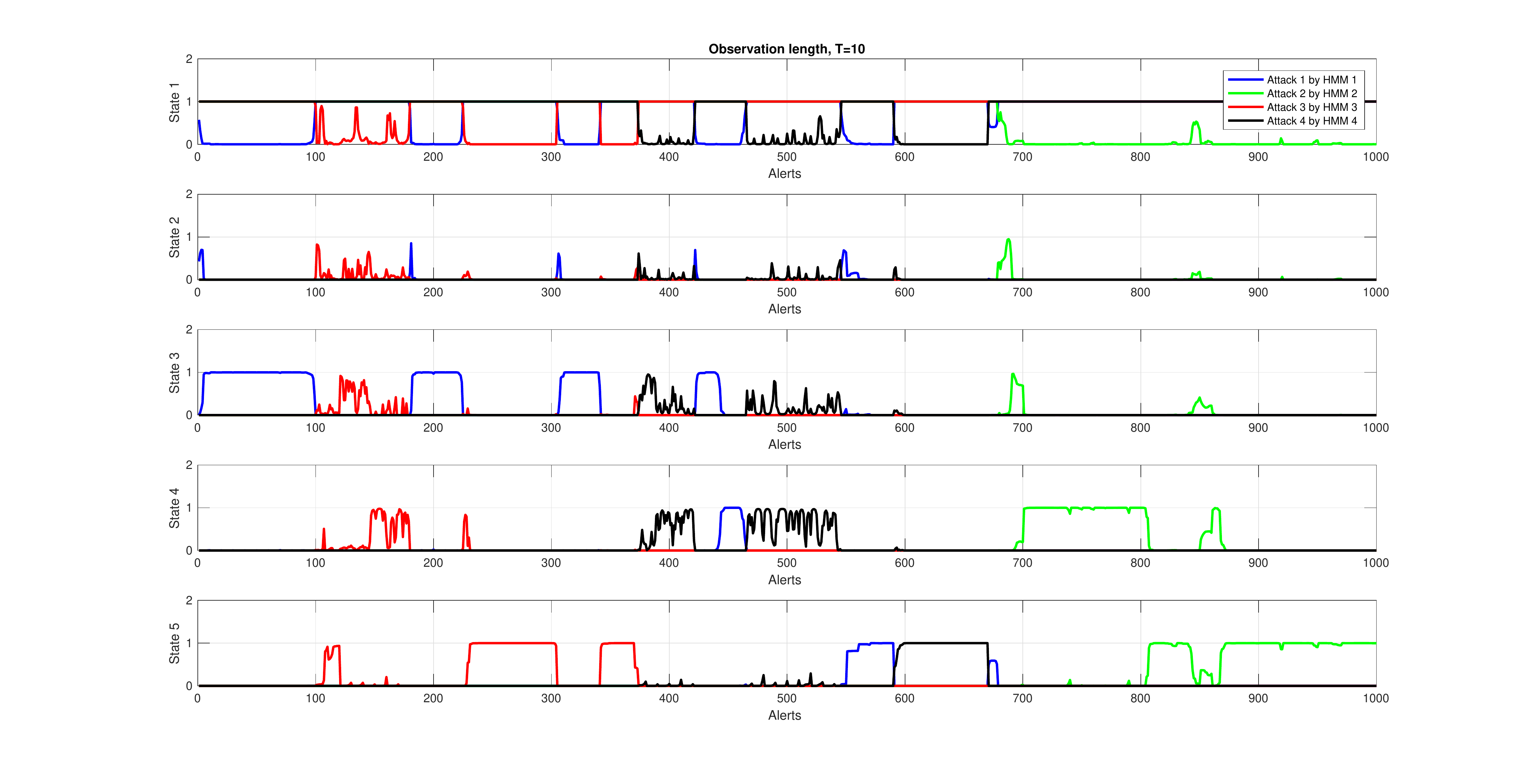}
    \caption{\small State Probability}
  \end{subfigure} \\
  \end{tabular}
  \caption{ \small State Probability of Synthesized Attacks 1 - 4 the Detected by HMM1 and HMM2 Based on Architecture I, T=10}
  \label{figur9955}
\end{figure*}

\subsubsection{Performance Evaluation - Four Synthesized Multi-stage Attacks}
The proposed architectures can be applied to more than two simultaneous attacks, as well as attacks with different instances. We have conducted several experiments, using synthetic datasets, to study the effect of having more than two simultaneous multi-stage attacks on the performance of the proposed architectures. 
In particular, Architecture II performs better than Architecture I since it depends essentially on the demultiplexing operation. However, with more than two multi-stage attacks, more computations are involved, especially in the demultiplexing module and also in the HMM database component. Consequently, the mean time to demultiplex the stream and to estimate the state for a window size of 100 increases from 0.46 milliseconds to 1.9 milliseconds.
Architecture I works well with more than two attacks, but its detection performance deteriorates significantly with a large number of attacks and a higher degree of interleaving.  %ER 
Fig. \ref{figur9955} shows the results for a scenario of four multi-stage attacks. In this scenario, the attacker(s) in Attack 3 and Attack 4 attempt to hide an attack with some of their previous attempts that he/she exploited successfully, which represent two quite different instances from the trained HMM templates for these attacks. Although the sequence of observations has unrelated observations from four different multi-stage attacks, especially in the window between 300 and 400, Fig. \ref{figur9955}b shows that the proposed architecture estimates the progress of the attack correctly. Due to page limit, we show the results for only one scenario of interleaving from the four multi-stage attacks.

\section{Conclusion}
This paper addresses the detection problem of interleaved multiple multi-stage attacks intruding into a computer network. We emphasize the importance of this problem by showing how interleaving and stealthy attacks can deceive the detection system. Therefore, we propose two architectures based on a well-known machine learning technique, i.e., the Hidden Markov Model, and provide their performance results and computational complexity. Both architectures can track interleaved attacks by detecting the correct states of the system for each incoming alert. However, as the degree of interleaving among attacks increases, Architecture II, which employs a demultiplexing mechanism, exhibits more robustness and better performance as compared to Architecture I. For the performance assessment of these architectures, we propose three performance metrics which include (1) attack risk probability, (2) detection error rate, and (3) the number of correctly detected stages. The DARPA2000 dataset is chosen to synthesize interleaved multi-stage attack scenarios and to demonstrate the efficacy of the proposed architectures. The proposed architectures are generic in terms of their capability to process any dataset that contains multiple interleaved multi-stage attacks.

%%%%%%%%%%%%%%%%%%%%%%%%%%%%%%%%%%%%%%%%%%%%%%%%%%%%%

\section{Acknowledgment}
This research was supported by the grants from Northrop Grumman Corporation and US National Science Foundation (NSF) Grant IIS-0964639 and was supported by King Abdulaziz University. Distribution Statement A: Approved for Public Release; Distribution is
Unlimited \#18-1658, Dated 10/30/18.

\bibliographystyle{IEEEtran}
\bibliography{references1}

\begin{IEEEbiography}[{\includegraphics[width=1in, height=1.25in,clip,keepaspectratio]{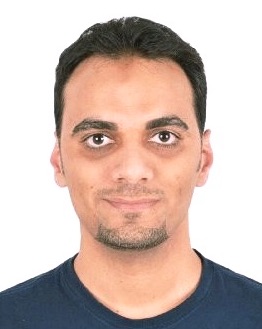}}]{Tawfeeq Shawly} received the M.S. and Ph.D. degrees in communications, networking, and signal processing from the School of Electrical and Computer Engineering, Purdue University, West Lafayette, IN, USA, in 2016 and 2019, respectively. % He received his B.S. and M.S. degrees in Computer Engineering from the Department of Electrical and Computer Engineering at King Abdulaziz University in Jeddah, Saudi Arabia. 
He is an Assistant Professor in the Electrical Engineering Department at the King Abdulaziz University in Rabigh, Saudi Arabia. His research interests include security of networked cyber-physical systems, artificial intelligence and machine learning.\end{IEEEbiography}

\begin{IEEEbiography}[{\includegraphics[width=1in, height=1.25in,clip,keepaspectratio]{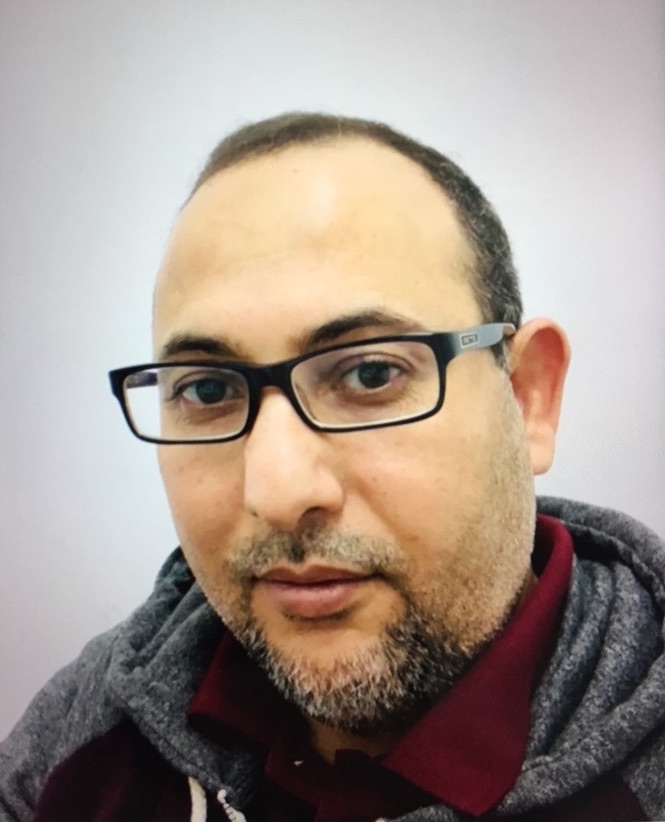}}]{Ali Elghariani} (S'12-M'14) received the B.S. and M.S. degrees in electrical and electronic engineering from the University of Tripoli, Tripoli, Libya. He obtained his Ph.D. degree in communications, networking, and signal processing from the School of Electrical and Computer Engineering, Purdue University, West Lafayette, IN, USA, in 2014. Currently, he is a research engineer at XCOM-Labs in San Diego CA. When this paper was submitted, he was a visiting researcher at Purdue University, West Lafayette IN. In 2015-2016, he was a Lecturer at the
Department of Electrical and Electronic Engineering, University of Tripoli, Tripoli, Libya. His research interests include signal detection and channel estimation in large-scale MIMO systems,  mmwave communication, optimization techniques in wireless communications, and network security.\end{IEEEbiography}

\begin{IEEEbiography}{Jason Kobes} works as a principal cyber architect and research scientist in Washington, DC for Northrop Grumman Corporation. He has over 20 years of experience concentrated in information systems design analytics, business/mission security architecture, enterprise risk management, information assurance research, and business consulting. He has an M.S. in information assurance (MSIA) and a B.S. in computer science from Iowa State University.\end{IEEEbiography}

\begin{IEEEbiography}[{\includegraphics[width=1in, height=1.25in,clip,keepaspectratio]{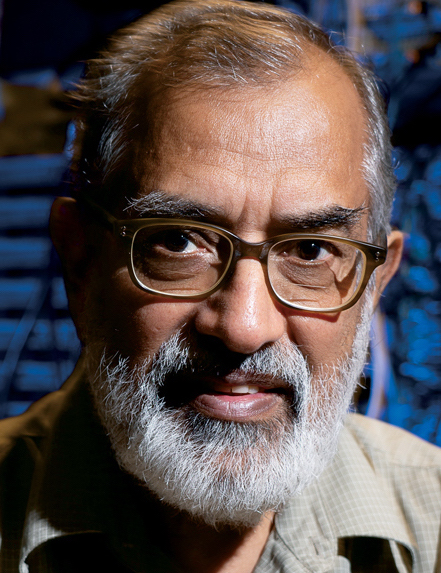}}]{Arif Ghafoor} is a professor in the School of Electrical and Computer Engineering at Purdue University. His research interests include multimedia information systems, database security, and distributed computing. He is a Fellow of the IEEE.\end{IEEEbiography}

\end{document}